\documentclass[fleqn,usenatbib]{mnras}

\usepackage{newtxtext,newtxmath}
\usepackage[T1]{fontenc}
\usepackage{ae,aecompl}
\usepackage{graphicx}	
\usepackage{amsmath}	
\usepackage{amssymb}	

\def\be{\begin{equation}} 
\def\ee{\end{equation}}

\def\kms{\,{\rm {km\, s^{-1}}}}

\def\gsim{\lower.5ex\hbox{\gtsima}} 
\def\lsim{\lower.5ex\hbox{\ltsima}} \def\gtsima{$\; \buildrel > \over 
\sim \;$} \def\ltsima{$\; \buildrel < \over \sim \;$} \def\prosima{$\; 
\buildrel \propto \over \sim \;$} \def\gsim{\lower.5ex\hbox{\gtsima}} 
\def\lsim{\lower.5ex\hbox{\ltsima}} 
\def\simgt{\lower.5ex\hbox{\gtsima}} 
\def\simlt{\lower.5ex\hbox{\ltsima}} 
\def\simpr{\lower.5ex\hbox{\prosima}} \def\la{\lsim} \def\ga{\gsim} 
  
\def\gtsima{$\; \buildrel > \over \sim \;$} 
\def\ltsima{$\; \buildrel < \over \sim \;$} 
\def\gsim{\lower.5ex\hbox{\gtsima}} 
\def\lsim{\lower.5ex\hbox{\ltsima}} 
\def\simgt{\lower.5ex\hbox{\gtsima}} 
\def\simlt{\lower.5ex\hbox{\ltsima}} 
\def\simpr{\lower.5ex\hbox{\prosima}} 
\def\la{\lsim} 
\def\ga{\gsim}

\def\Msun{M_{\odot}} 
\def\Zsun{Z_{\odot}}

\def\E3{{\cal E}_{\rm g}^{III}}

\def\ObOm{\Omega_b/\Omega_M}
\def\Om{\Omega_M}
\def\Ob{\Omega_b}
\def\MIII{\rm m_{\tt PISN}}
\def\MpopIII{${\rm m_{\tt popIII}}$}
\def\fratio{${\rm f_*/f_{dil}}$\;}
\def\teff{$T\rm_{eff}$}
\def\kms{$\mathrm{km\, s^{-1}}$}
\newcommand{\mygi}{MyGIsFOS}

\title{Probing the existence of very massive first stars}
\author[Salvadori, Bonifacio, Caffau et al.]
{S. Salvadori$^{1,2,3}$\thanks{E-mail:stefania.salvadori@unifi.it}, 
P. Bonifacio$^{3}$, 
E. Caffau$^{3}$, 
S. Korotin$^{4}$,
S. Andreevsky$^{5,3}$, 
\newauthor
M. Spite$^{3}$, 
and \'{A}. Sk\'{u}lad\'{o}ttir$^{6}$
\\
$^{1}$Dipartimento di Fisica e Astronomia, Universit\'a degli Studi di Firenze, Via G. Sansone 1, I-50019 Sesto Fiorentino, Italy\\
$^{2}$INAF/Osservatorio Astrofisico di Arcetri, Largo E. Fermi 5, I-50125
Firenze, Italy\\
$^{3}$GEPI, Observatoire de Paris, Universit\'e PSL, CNRS, Place Jules
Janssen, F-92195 Meudon, France\\
$^{4}$Crimean Astrophysical Observatory, Nauchny 298409, Republic of Crimea\\ 
$^{5}$Department of Astronomy and Astronomical Observatory, Odessa National University, Isaac Newton Institute of Chile, \\
Odessa Branch, Shevchenko Park, 65014, Odessa, Ukraine\\
$^{6}$Max-Planck-Institut f$\ddot{\text{u}}$r Astronomie, K$\ddot{\text{o}}$nigstuhl 17, D-69117 Heidelberg, Germany\\
}

\date{Accepted XXX. Received YYY; in original form ZZZ}
\pubyear{2019}
\begin{document}
\label{firstpage}
\pagerange{\pageref{firstpage}--\pageref{lastpage}}
\maketitle

\begin{abstract}
We present a novel approach aimed at identifying the key chemical elements to search for the (missing) descendants of very massive first stars exploding as Pair Instability Supernovae (PISN). Our simple and general method consists in a parametric study accounting for the unknowns related to early cosmic star-formation and metal-enrichment. Our approach allow us to define the most likely [Fe/H] and abundance ratios of long-lived stars born in inter-stellar media polluted by the nucleosynthetic products of PISN at a $> 90\%$, $70\%$, and $50\%$ level. In agreement with previous works, we show that the descendants of very massive first stars can be most likely found at [Fe/H]$\approx -2$. Further, we demonstrate that to search for an under-abundance of 
[(N, Cu, Zn)/Fe]$< 0$ is the key to identify these rare descendants. The ``killing elements" N, Zn, and Cu are not produced by PISN, so that their sub-Solar abundance with respect to iron persists in environments polluted by further generations of normal core-collapse supernovae up to a $50\%$ level. We show that the star BD +80$^\circ$ 245, which has [Fe/H]$= -2.2$, [N/Fe]$= -0.79$, [Cu/Fe]$=-0.75$, and [Zn/Fe]$= -0.12$ can be the smoking gun of the chemical imprint from very massive first stars. To this end we acquired new spectra for BD +80$^\circ$ 245 and re-analysed those available from the literature accounting for Non-Local Thermodynamic Equilibrium corrections for Cu. We discuss how to find more of these missing descendants in ongoing and future surveys to tightly constrain the mass distribution of the first stars.
\end{abstract}
\section{Introduction}
Primordial composition stars were very likely more massive than those that we observe 
in Local star-forming regions \citep[e.g.][]{bromm13}. Analytical calculations (from the 
pioneering work by \cite{silk77} to \cite{mckee08}), along with state-of-the-art 
numerical simulations that include radiative feedback \citep[e.g.][]{hosokawa11}, show that 
the mass of these Population III (Pop~III) stars is typically \MpopIII${\rm > 10\Msun}$, 
and can reach extreme values when a statistical sample of cosmological formation 
sites is considered, \MpopIII$\approx 1000\Msun$ \citep{hirano14,susa14,hirano15}. 
On the other hand, 3D simulations following gas fragmentation for 
$< 10^3$~yrs and resolving individual proto-stars, find that the proto-stellar 
disk can fragment in sub-solar clumps, which implies the existence of some 
stars such that \MpopIII${\rm < 1\Msun}$. 
On larger time-scales, however, most of these fragments are expected to merge 
to form more massive stars \citep[e.g.][]{greif12}. 
At the moment, therefore, we can only assert that the characteristic mass of Pop~III 
stars is much larger than normal Pop~II/I stars, ${\rm m^{ch}_{popIII}\approx 10 \Msun}$, 
and that their high-mass end possibly extends up to \MpopIII${\rm \approx 1000\Msun}$ 
\citep[e.g.][]{greif15,hirano15}. 

Among such a variety of stellar masses those with \MpopIII${\rm > 140\Msun}$ 
represent the dominant sources of ionizing photons \citep[e.g.][]{schaerer02}, 
dust \citep[e.g.][]{schneider04}, metals, supernova energy injection, and stellar 
black holes \citep[e.g.][]{heger02}. 
Pop~III stars with ${\rm 140 \Msun < m_{PopIII} < 260\Msun}$ end their lives as 
energetic Pair Instability Supernovae (PISN), leaving no remnants and thus 
releasing back to the surrounding gas {\it all} their mass, with $\approx 50\%$ 
in the form of heavy elements. Larger Pop~III stars, \MpopIII$\rm >260\Msun$, 
do not inject metals as they are predicted to directly collapse into stellar 
black holes \citep[e.g.][]{heger02}. Probing the existence and frequency of very 
massive Pop~III stars is thus fundamental to understand the early phases of
reionization \citep[e.g.][]{salvadori14}, metal enrichment \citep[e.g.][]{pallottini14}, 
and super massive black hole formation \citep[e.g.][]{volonteri10}.

One of the most promising ways to test the predicted existence of massive 
Pop~III stars is by identifying their direct descendants among the oldest stars, 
which can be individually resolved in the Local Group \citep[e.g.][]{beers05,
tolstoy09}. If massive Pop~III stars rapidly ended their lives as energetic 
supernovae (SN) explosions, thus preventing their present-day detection, 
their nucleosynthetic products were efficiently dispersed into the surrounding 
gas \citep[e.g.][]{greif09,smith15}. Such precious elements from the remote 
times are safegarded until the present-day in the photospheres of long-lived 
low-mass stars, $m_*<1\Msun$, which form from the ashes of Pop~III stars. 
Different groups have shown that the transition from massive Pop~III stars to 
normal Pop~II stars can efficiently take place when the metallicity of the gas 
clouds reaches a critical value, $Z_{cr} \approx [10^{-6} - 10^{-3.5}] \Zsun$ 
\citep[e.g.][]{bromm01,schneider02}. 

According to cosmological models for the Local Group formation, these 
``descendants'' of Pop~III stars can be found among the oldest stellar
populations, in particular in the Galactic halo of the Milky Way 
\citep[e.g.][]{tumlinson06,salvadori07}, in the Galactic bulge \citep{white00,
brook07,tumlinson10,salvadori10}, and in nearby dwarf galaxy satellites 
\citep[e.g.][]{salvadori09,bovill09,frebel12,salvadori15}. 
Local observations have reported the existence of many Carbon enhanced 
metal-poor (CEMP) stars, [C/Fe]$>1.0$, in both the Galactic halo \citep[e.g.
][and references therein]{spite13,bonifacio15} and ultra-faint dwarf galaxies 
\citep[e.g.][]{frebel14,spite18}. These stars predominantly appear at [Fe/H]$<-2.5$.
The most iron-poor among them \citep[e.g.][]{keller14} have chemical abundance 
patterns that are consistent with the heavy elements produced by moderately 
massive Pop~III stars, \MpopIII $< 100\Msun$, which experience mixing and 
fallback exploding as ``{\it faint SN}'' \citep[e.g.][]{bonifacio03,iwamoto05}, 
possibly because of their high rotation speed \citep[e.g.][]{meynet02}.

The CEMP stars have a bimodal distribution of carbon abundance 
\citep{rossi05,spite13,bonifacio15}, and the two groups of stars are referred to as 
``high-carbon band'' and ``low-carbon band'' \citep[see][for a precise definition]{bonifacio18}.
The available evidence points towards the conclusion that the high-carbon band stars
are the result of mass-transfer in a binary system, while the low-carbon band
stars are a reliable record of the chemical composition of the interstellar 
medium from which they were formed \citep[see][and references therein]{caffau18}.
We shall adopt  this hypothesis in what follows.

The link between low-carbon band CEMP stars and the chemical products of primordial 
\MpopIII${\rm < 100\Msun}$ stars has been confirmed by cosmological chemical 
evolution models that follow the overall formation of the Local Group 
\citep[e.g.][]{salvadori15} or that focus on the build-up of the MW halo \citep[e.g.][]{DB17,Hartwig18}. 
These models are able to interpret the declining frequency of low-carbon band CEMP 
stars at increasing [Fe/H] in a cosmological context (see also \citealt{cooke14,
bonifacio15}). Their findings show that at [Fe/H]$\leq -5$ roughly $100\%$ 
of the stars are expected to form in gaseous environments {\it uniquely} 
polluted by faint primordial SN with \MpopIII$< 100\Msun$, which produce large 
amounts of C and small amounts of Fe. At higher [Fe/H], such a key chemical 
signature is gradually washed-out because of the increasing contribution of 
normal SN type II to the chemical enrichment \citep[e.g.][]{salvadori15,DB17,Hartwig18}. 
This makes the {\it direct} descendants of moderately massive Pop~III stars
relatively straightforward to find ``just'' by searching for the most iron-poor stars.

On the other hand the descendants of very massive PISN are predicted 
to appear at higher [Fe/H], where they only represent a very low fraction of the 
total number of stars \citep{salvadori07,karlsson08}. \cite{DB17} showed 
that stars which are formed in environments polluted by PISN at $>50\%$ 
level cover a broad metallicity range, $-4<$[Fe/H]$<-1$, with a peak at 
[Fe/H]$\approx -1.8$. According to their findings these precious fossils 
only represent the $\approx 0.25\%$ of the total number of Galactic stars 
at [Fe/H]$<-2$. Such a percentage becomes even lower, $\approx 0.1\%$, 
for stars enriched by PISN at $>80\%$ level. Current observations fully 
agree with these results. Among the 500 Galactic halo stars analyzed at 
[Fe/H]$<-2$ {\it only one}, i.e. $\leq 0.2\%$ of the total, shows a chemical 
abundance pattern that might be consistent with an environment of formation 
imprinted {\it also} by massive PISN \citep{aoki14}. 
This peculiar star has [Fe/H]$\approx -2.5$, which is the typical iron-abundance 
of the {\it direct} descendant of PISN, i.e. stars enriched by PISN at $100\%$ level
\citep{DB17}. 
Yet, \citet{Chiaki18} consider it unlikely that the star observed by \cite{aoki14} 
really is a PISN descendant. In fact, their simulations predict that the direct 
descendants of PISN are much more rare: only one out of $10^5-10^6$ stars.

How can we catch these extremely rare PISN descendants, which are fundamental 
to characterize the primordial IMF and to understand early galaxy formation processes? 
In the ongoing era of wide and deep spectroscopic surveys the total number of stars 
is largely increasing. We can thus face the possibility to find many of these rare stellar 
relics if we are able to efficiently {\it identify} them. The aim of this paper is thus to 
address the following questions: 
\begin{itemize}
\item What are the key chemical elements we should look for in order to identify 
the rare descendants of massive Pop~III stars, i.e. PISN? 
\item In gaseous environments enriched both by PISN and normal Pop~II stars exploding as core-collapse 
SN, down to which fraction of PISN contribution can we still detect its peculiar chemical imprint?
\item What are the quantitative prospectives of catching the rare PISN descendants 
in ongoing and future surveys?
\end{itemize}

The main problem in addressing these issues is that many of the physical 
mechanisms regulating the early cosmic star-formation are largely uncertain. 
These include the star-formation efficiency of  ``mini-haloes", the capability of
these systems to retain the metals injected by SN, the metal mixing efficiency, 
and the typical timescale for metal diffusion and subsequent star-formation. 
In this paper we present a new, simple, and extremely general parametric 
study for the chemical evolution of the gas imprinted by the first stellar
generations. Our novel approach overcomes the current uncertainties by 
condensing the unknown physical processes into a few free parameters, 
which are then varied to explore the full parameter space. Our conclusions 
are hence general and essentially model independent. Furthermore, they do 
not rely on the usually assumed (cold) dark matter model. 
\section{A simple parametric study}    
We can start by computing the chemical properties of the gaseous environments,
or interstellar media (ISM), enriched by the nucleosynthetic products of a
{\it single} Pair Instability SN. The heavy elements yielded by PISN with 
different initial masses, $\MIII=[140,260]\Msun$, have been computed
by \cite{heger02}. Because of the peculiar physical conditions that PISN
experience during their evolution and the complete disruption of their
initial stellar mass, these results are very robust (e.g. Takahashi et al. 2018).

On the other hand, the rate at which the first stars are formed is unknown. 
State-of-the-art numerical simulations studying Pop~III star formation typically 
focus on low-mass mini-haloes, $M\approx (10^6-10^7)\Msun$, which are 
predicted to be the first star-forming sites at $z>10$ \citep[e.g.][]{abel02,wise12,bromm09}.
Molecular hydrogen, H$_2$, is the only coolant in these systems, making their 
capability to form stars limited and variable. 
The amount of H$_2$ strongly depends on the total mass, gas density, 
and formation redshift of the system, along with the presence of Lyman 
Werner flux from close-by galaxies, which can easily photo-dissociate H$_2$ 
\citep[e.g.][]{MFR01,omukai12}. On top of this variability, cosmological 
simulations of metal-enrichment show that pristine regions of gas can survive 
down to $z\leq 3$ \citep[e.g.][]{pallottini14}, implying that Pop~III stars can also
be hosted by more massive galaxies, which have higher star-formation rates.

The ability of the first star-forming systems to convert gas into stars is hence 
largely uncertain, as it strongly depends on both their intrinsic properties and environmental effects. 
To make our calculations general we encapsulate these unknowns in a single free 
parameter, the {\it star-formation efficiency}, ${\rm f_*=M_*/M_g}$, which quantifies 
the mass of Pop~III stars, ${\rm M_{*}}$, formed in a dynamical time within a galaxy 
of initial gas mass ${\rm M_{g}}$. Given the total mass of heavy elements produced
per stellar mass formed, i.e. the metal yield ${\rm Y_Z}$, we can easily compute the 
total mass of heavy elements produced by PISN as:
\be 
{\rm M_Z=Y_Z M_* = Y_Z f_* M_g}.
\label{eq:MZ_pisn}
\ee
Similarly, the net mass of each chemical element ${\rm X}$ produced by PISN can 
be computed as ${\rm M_X=Y_X f_* M_g}$ once the yield ${\rm Y_X}$ is known. 
A rough estimate of the metallicity of the ISM after a PISN
event can be obtained by assuming that all produced metals are instantaneously 
mixed within the whole galaxy, i.e. ${\rm Z = M_Z/M_g}$. In reality, however, not 
all these fresh metals will necessarily be retained by the galaxy, and/or not all the 
gas will be used to dilute metals. Indeed, depending upon the kinetic energy released 
by SN explosions and the mass of the (Pop III) galaxy, metals can fill a different volume 
of the galaxy, eventually escaping its gravitational potential. 

To determine the metallicity of the ISM out of which subsequent star-formation can 
occur, we should thus quantify: (i) the fraction of newly produced metals effectively 
retained by the galaxy, ${\rm f_Z\equiv (M^{eff}_Z/M_Z)}$; (ii) the fraction of the 
initial gas mass into which metals are effectively diluted, ${\rm f_g = (M_{dil}/M_g)}$. 
We can incapsulate these two free parameters into a single one, the diluting factor, 
${\rm f_{dil}= f_g/f_Z}$, which completely defines the ISM metallicity for subsequent 
star-formation, ${\rm Z_{ISM}\equiv (M^{eff}_Z/M_{dil})=(f_Z M_Z/f_g M_g)=M_Z/f_{dil}M_g\equiv Z/f_{dil}}$. 
A qualitative flavour of the possible values of ${\rm f_{dil}}$ and its dependance upon 
the unknown metal mixing processes is provided by the sketch in Fig.~1.
\begin{figure}
  \includegraphics[angle=-90,width=0.950\linewidth,clip=,trim={1.5cm 2.5cm 1.0cm 2.5cm}]{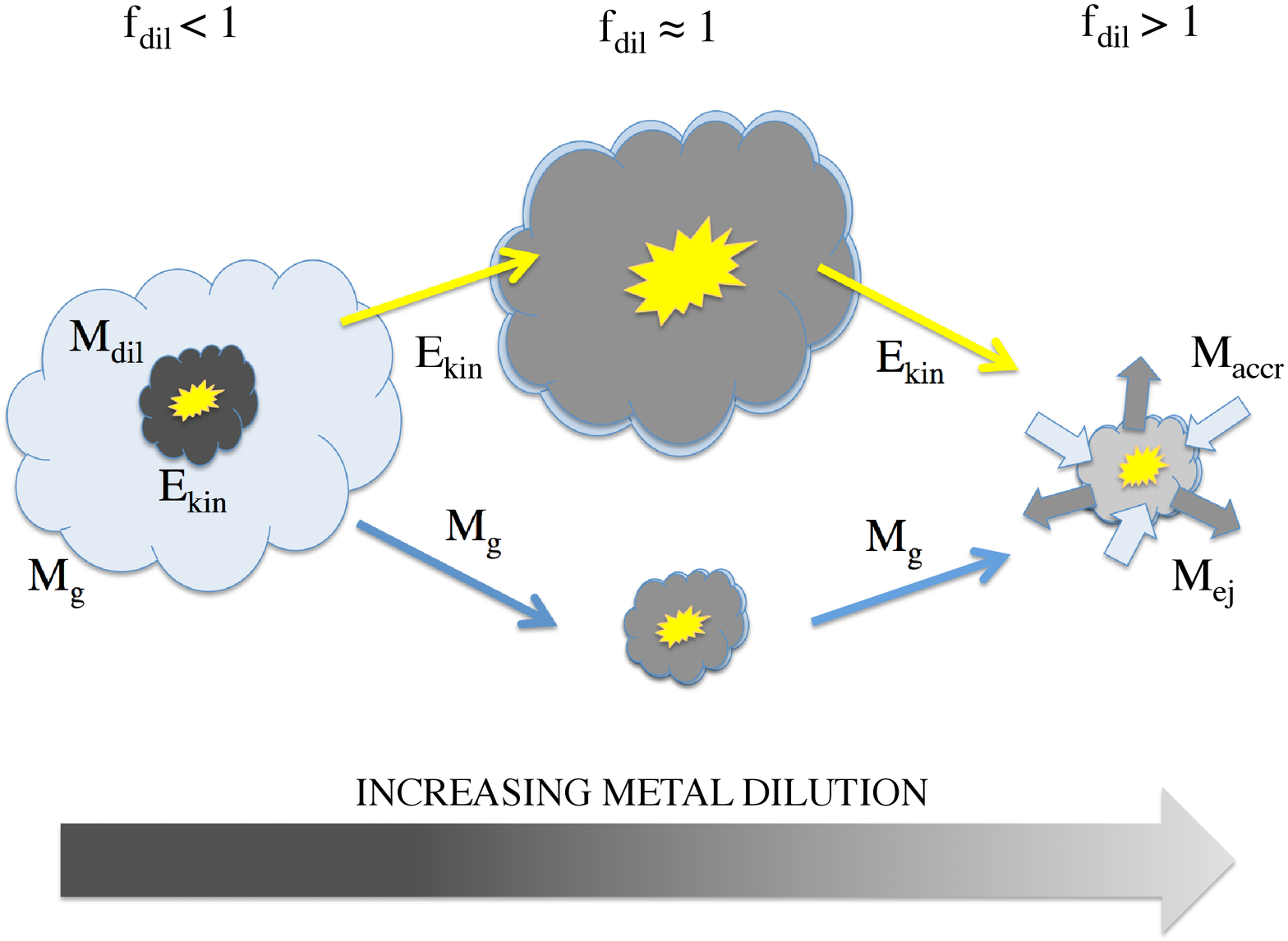}
    \caption{A simple sketch illustrating the mixing of PISN products with the environment for different
    dilution parameter, ${\rm f_{dil}}$, and how it depends on: 
    (i) the kinetic energy released by supernovae, ${\rm E_{kin}}$; 
    (ii) the gas mass of the star-forming galaxy, ${\rm M_{g}}$ (cyan cloudlike shape); 
    (iii) the mass into which the PISN products are diluted into, ${\rm M_{dil}}$ (grey cloudlike shape); 
    (iv) the mass of ejected gas, ${\rm M_{ej}}$ (grey outgoing arrows); 
    (v) the mass of pristine gas eventually accreted from the IGM, ${\rm M_{accr}}$ (cyan ingoing arrows). 
    See Sec.~2.1 for details.}   
    \label{Fig:sketch}
\end{figure}
\subsection{The cosmological context}
\label{sec:cosmo_context}
To understand what are the different physical conditions in which the enriching 
and mixing events accounted by our simple model might occur, we will discuss 
Fig.~1 in a cosmological context.

First, if the kinetic energy released by the SN explosions is sufficiently low 
(e.g. low-mass PISNe) and/or if the star-forming galaxy is massive enough 
(e.g. mini-halos formed at $z < 10$ or even more massive Ly$\alpha$-cooling 
halos), then the SN blast wave will only fill a small portion of the initial gas mass, 
${\rm M_{dil} < M_g}$. In such a region, which has been polluted with {\it all} the 
newly produced heavy elements (i.e. ${\rm f_Z = 1}$), 
the star-formation can be rapidly re-activated thanks to efficient metal-cooling 
\citep[e.g.][]{pallottini14}. 
We thus have ${\rm f_{dil} < 1}$.

Second, if the kinetic energy released by SN explosions is higher than in the previous 
case (e.g. more massive PISN) and/or if the galaxy is less massive, then the SN blast 
wave will reach larger radii, possibly mixing all newly produced metals (${\rm f_Z=1}$) 
within the {\it whole} galaxy, i.e. ${\rm M_{dil}\approx M_{g}}$. Cosmological simulations 
have shown that the metal-enriched SN-driven gas confined at the halo virial radius can 
rapidly cool-down and fall-back into the centre triggering new star-formation 
\citep[e.g.][]{onorbe15,benitez-llambay15}. Hence ${\rm f_{dil} \approx 1}$.

Third, for even higher SN explosion energy (e.g. high-mass PISNe) and/or less 
massive galaxies (e.g. the first star-forming mini-halos) the blast wave produced 
by the explosion can evacuate part (or most) of the metal-enriched gas from the hosting 
galaxy \citep[e.g.][]{greif07,Chiaki18}. Both observations and simulations show that the 
SN-driven outflows are able to eject part of the metals outside of galaxies
(e.g. most recently Jones et al. 2018). This implies ${\rm f_Z < f_g}$ and therefore 
${\rm f_{dil}=f_g/f_Z > 1}$. In addition, ${\rm f_{dil}}$ gets even higher if primordial 
gas accretes onto the galaxy, ${\rm M_{accr}}$ (e.g. via merging), since ${\rm f_Z}$ 
remains unchanged while ${\rm f_g}$ increases further. We thus have ${\rm f_{dil} > 1}$.

To be more quantitative and appreciate how much different scenarios depend upon 
the unknown mass of Pop III stars and on the properties of their hosting halos we should 
recall some recent findings from cosmological simulations. Zoomed-in simulations of the 
first star-forming mini-haloes at ${\rm z > 15}$ show that a single energetic PISN explosion 
(${\rm E_{SN}\approx 10^{53}}$~erg for $\MIII{\rm \approx 260 \Msun}$) can evacuate 
almost all the gas from mini-halos with ${\rm M\approx 10^6\Msun}$ \citep[e.g.][]{greif07,bromm03}, 
leading to ${\rm f_{dil} > 1}$. Yet, the same mini-halo retains most of the gas/metals after 
the explosion of the least energetic PISN (e.g. ${\rm E_{SN}\approx 10^{51} erg}$ for 
$\MIII{\rm \approx 150 \Msun}$ from \cite{bromm03}). In this case, therefore, 
${\rm f_{dil} \approx 1}$. As we will extensively discuss in Sec.~7, in situ enrichment of 
mini-halos by PISN can be effectively captures by assuming ${\rm f_{dil} \geq 1}$.
However, recent cosmological simulations have also shown that a part of the ejected material 
might pollute a nearby primordial mini-halo, eventually triggering new star-formation there 
\citep[e.g.][Chiaki et al. 2018]{smith15}. This is equivalent to the case ${\rm f_{dil} < 1}$. 
Although this event it is much more unlikely than the previous two \citep[see e.g.][]{Chiaki18}, 
we should also account for this scenario in our general model, which aims at catching the signature 
of PISNe formed in all the possible physical conditions. 
\subsection{Chemical imprint}
From our previous section we conclude that all the different cosmological scenarios, 
including both in situ and external pollution along with dilution due to accretion, can be efficiently 
described by varying our free parameter ${\rm f_{dil}}$ (Fig.~\ref{Fig:sketch}).
In this context, we are assume that a single PISN forms and explodes and that PISN 
with different mass have the same probability to form. In fact, a flat primordial IMF and a single 
first star per mini-halo is exactly what simulations of the first star-forming systems have found (e.g. 
Hirano et al. 2014). To account for the (possible) PISN formation in more massive and efficiently 
star-forming Ly$\alpha$ halos (Sec.~\ref{sec:cosmo_context}), we extend these hypotheses by 
assuming that in each star-forming galaxy all formed PISN have the same mass, $\MIII$. We can 
thus proceed further and compute the chemical properties of the ISM polluted by a PISN (or more) 
with mass $\MIII$. By using the previously defined relations we can compute the ISM metallicity: 
\be
{\rm Z_{\tt ISM}=M_Z/(f_{dil}M_{g})=\frac{Y^{PISN}_Z f_* M_{g}}{f_{dil}M_{g}}=\frac{f_* }{f_{dil}}Y^{PISN}_Z}
\label{eq:metals}
\ee 
along with the abundance of any $X$ element into the ISM:
\be
{\rm [X/H]_{\tt ISM}\equiv log\Big[\frac{N_{\tt X}}{N_{\tt H}}\Big] - log\Big[\frac{N_{\tt X}}{N_{\tt H}}\Big]_\odot = log\Big[\frac{M_{\tt X}}{m_{\tt X}}\frac {m_{\tt H}}{M_{\tt H}}\Big] - log\Big[\frac{N_{\tt X}}{N_{\tt H}}\Big]_\odot }
\nonumber
\ee
\be
{\rm \;\;\;\approx log\Big[\frac{M_{\tt X}}{M_{dil}}\Big] - log\Big[\frac{M_{\tt X}}{M_{\tt H}}\Big]_\odot = log\Big[\frac{f_*}{f_{dil}}Y^{PISN}_{\tt X}\Big] - log\Big[\frac{M_{\tt X}}{M_{\tt H}}\Big]_\odot}, 
\label{eq:FeH}
\ee
where both ${\rm Y^{PISN}_Z}$ and ${\rm Y^{PISN}_{\tt X}}$ depend on $\MIII$, $N_{\tt X}$ 
($N_{\tt H}$) is the number of $X$ (hydrogen) atoms, and $m_{\tt X}$ ($m_H$) the mean 
molecular weight. Using eq.~\ref{eq:FeH}, we can write the abundance ratio of any chemical 
element with respect to iron (or other key species): 
\be
{\rm [X/Fe]_{\tt ISM} = log\Big[\frac{Y^{PISN}_{\tt X}}{Y^{PISN}_{\tt Fe}}\Big] - log[\frac{M_{\tt X}}{M_{\tt Fe}}]_\odot} .  
\label{eq:ratios}
\ee
These simple calculations already provide interesting results. The total metallicity of the ISM, 
${\rm Z_{ISM}}$, along with the [X/H] abundance ratio, {\it solely} depends on the ratio between 
the two free parameters of the model, ${\rm f_*/f_{dil}}$ and on the total amount of metals 
(or X elements) produced per stellar mass formed, i.e. the yields. Furthermore, the chemical 
abundance pattern of the gas uniquely polluted by a PISN (or several PISNe with the same 
mass) is {\it solely} determined by the ratios in the chemical yields. In other words it is {\it universal}. 
We can thus study the predicted chemical properties of these environments as a function of the 
PISN mass and [Fe/H] range after defining the most plausible values for the two free parameters, 
${\rm f_*}$ and ${\rm f_{dil}}$.
\subsection{Setting the free parameter space}
The parameter space of the problem, i.e. the possible values of ${\rm f_*}$ and ${\rm f_*/f_{dil}}$, 
can be estimated analytically. First, local observations of molecular clouds show that the fraction 
of molecular hydrogen gas (${\rm f_{H2}}$) that can be turned into stars in a dynamical time
never exceed the $10\%$ of the total, i.e. ${\rm \dot{M}_*\times t_{ff} = \xi_{sf}f_{H2}M_g}$ with 
${\rm \xi_{sf}\leq 0.1}$ \citep[e.g. see][and reference therein]{pallottini17}. Second, the gas needs 
to be converted in molecular hydrogen form, whose mass fraction is necessarily lower or equal
than the total (${\rm f_{H2} \leq 1}$). By defining ${\rm f_*=\xi_{sf}f_{H2}}$ we can thus safely 
set ${\rm f_*\leq 0.1}$, i.e. we have an upper limit for the star-formation efficiency. 

In high-redshift primordial mini-haloes (${\rm T_{vir}\leq 10^4}$~K) which cool-down their gas  
via molecular hydrogen, $f_*$ is much lower, as it is limited by the fragile nature of ${\rm {H_2}}$ 
and the lack of additional cooling channels (e.g. Ciardi \& Ferrara 2001). 
Tegmark et al. (1997) showed that only mini-halos with ${\rm f_{H2} > 10^{-3}}$ can 
efficiently cool-down their gas and trigger star-formation. Assuming ${\rm\xi_{sf}=0.1}$
we infer ${\rm f_*>10^{-4}}$. A similar value is obtained by requiring the formation of at least 
one PISN per free-fall time in the most massive mini-halos: ${\rm f_*\geq  M_{\tt PISN}/M_{g}
= M_{\tt PISN}/[{{\ObOm} M]} \approx 2\times 10^{-4}}$, where $\ObOm$ is the cosmological 
baryon-to-dark matter mass ratio\footnote{We assume cosmological parameters consistent 
with the Planck 2016 results: $h=0.65$, $\Ob/h^2 = 0.022$, and $\Om= 0.32$} 
and ${\rm M(T{vir}\approx 10^4 K, z=10)=5\times10^7\Msun}$ is the mass of the most massive 
mini-halos, i.e. those with highest ${\rm T_{vir}}$ and lowest formation redshift (e.g. Barkana \& Loeb 2001). 

To evaluate the possible ${\rm f_{dil}}$ values we should distinguish among the different scenarios 
illustrated in Fig.~1. When ${\rm f_{dil} < 1}$ we can estimate the {\it minimum} gas mass into which 
all the newly produced metals are diluted as the mass enclosed by the SN-driven bubble in the 
Sedov-Taylor approximation \citep[e.g.][]{MFR01,bromm03}: 
\be
{\rm M_{dil} > \frac{4\pi}{3} \rho_{ISM} \Big(\frac{E_{kin}}{\rho_{ISM}}\Big)^{3/5}t^{6/5}},
\ee 
where ${\rm \rho_{ISM}(z) \approx 200 \Ob \rho_{cr,0}(1+z)^3}$ is the ISM gas density, ${\rm t}$ the 
time passed since the SN explosions, and 
\be
{\rm E_{kin}={\langle E_{SN} \rangle}f_{kin} N_{SN} ={\langle E_{SN} \rangle}f_{kin} \nu_{SN} f_* M_g} 
\label{eq:Ekin}
\ee
is the kinetic energy released by SN, which depends upon the average SN explosion energy, 
${\rm \langle E_{SN} \rangle}$, the fraction of energy released in kinetic form which is typically 
${\rm f_{kin}\approx 0.1}$\citep[e.g.][]{pallottini14}, and the number of SNe, $\rm N_{SN}$.
Since ${\rm N_{SN}}$ can be written as the number of SN explosion per stellar mass formed, 
${\rm N_{SN}=\nu_{SN} M_*=\nu_{SN} f_* M_g}$, where  ${\rm \nu_{SN}\approx 1/200\Msun}$ 
for PISN, it is clear from eq.~\ref{eq:Ekin} that ${\rm E_{kin}}$ depends upon the (unknown) 
star-formation efficiency. 
Since we want to know the minimum mass of the metal-enriched gas out of which a new star-formation 
event can eventually occur, we can reasonably evaluate ${\rm M_{dil}}$ at the free-fall time, 
${\rm t_{ff}=(3\pi/32G\rho)^{1/2}}$. Recalling that ${\rm f_{dil} = {f_g/f_Z}}$ and that in this case 
${\rm f_Z = 1}$, we can write ${\rm f_{dil} = f_g > {M^{min}_{dil}}/{M_g}}$ and thus:
\be
{\rm f_{dil} >2.4\times10^5\Big[\frac{\langle E_{51} \rangle f_{kin}f_*}{(1+z)}\Big]^{3/5}M_g^{-2/5}},
\label{eq:f_dil1}
\ee 
where ${\rm \langle E_{51} \rangle=\langle E_{SN}\rangle/10^{51}erg}$ is the average explosion energy 
and ${\rm M_g}$ is expressed in solar masses. 
From eq.~\ref{eq:f_dil1} we infer that the minimum 
${\rm f_{dil}}$ corresponds to the largest ${\rm M_g}$, highest formation redshift, ${\rm z\approx 30}$, 
and minimum ${\rm f_* = 10^{-4}}$ and ${\rm \langle E_{51} \rangle}$, which for the least massive PISN 
is ${\rm \langle E_{51} \rangle \approx 1}$. 
By inserting these numbers in eq.~\ref{eq:f_dil1} and assuming ${\rm M_g\approx 10^8 \Msun}$ and
${\rm f_{kin}=0.1}$, we obtain ${\rm f_{dil} > 0.02}$. Note that ${\rm f_{dil}}$ becomes one order of 
magnitude higher for ${\rm \langle E_{51} \rangle\approx 20}$, which corresponds to the average PISN
mass. Furthermore, as we will see and discuss in Sec.~\ref{sec:location}, ${\rm M_g}$, ${\rm z}$, 
and ${\rm f_*}$ are {\it not} independent. In particular, the highest ${\rm z}$ and lowest 
${\rm f_*}$ correspond to the least massive dark matter halos that can trigger star-formation (e.g. Tegmark 
et al. 1997), from which we get lower ${\rm M_g}$ and thus higher ${\rm f_{dil}}$.
\begin{figure}
 \includegraphics[width=0.990\linewidth]{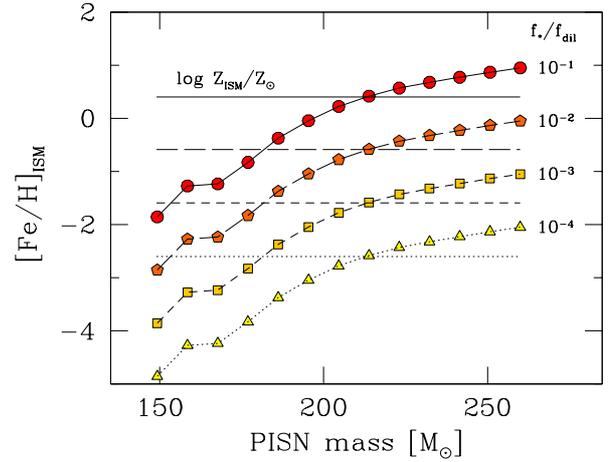}
    \caption{Iron abundance of the ISM imprinted by a PISN as a function of its mass, 
    $\MIII$ (points). Different symbols show the results obtained by assuming variuos \fratio 
    ratio (see labels). Horizontal lines show the corresponding ISM metallicity, 
    $\rm \log (Z_{ISM}/Z_\odot)$, which is independent of $\MIII$.}
    \label{Fig:FeH_PISN}
\end{figure}

To get the upper limit we should consider the case ${\rm f_{dil}>1}$, and account for the mass of gas/metals 
ejected outside of the galaxy (${\rm M_{ej}}$ and ${\rm M^Z_{ej}}$) along with the mass of pristine gas 
accreted (${\rm M^{accr}}$). We can write 
${\rm f_{g} = [M_g(1-f_*+f_*R) - M_{ej} + M_{accr}]/M_g = [1-f_{ej}-f_{accr}]}$, where ${\rm R}$ 
is the gas fraction returned from stars, which is equal to unity for the case of PISN and ${\rm f_{ej}}$ 
(${\rm f_{accr}}$) the fraction of ejected (accreted) gas mass with respect to the total. 
Assuming that the accreted gas is of primordial composition we can write 
${\rm f_{Z} = [M_Z - M^{ej}_Z]/M_Z= [1-M^{ej}_Z/M_Z]=[1-f_{ej}\eta_{ej}]}$, where in the last 
passage we have multiplied and divided for ${\rm M^{ej}_g}$ and ${\rm M_g}$, and defined 
${\rm  \eta_{ej} = Z_{ej}/(M_Z/M_g)}$. If we assume that the metallicity of the ejected gas is equal
to the mean abundance of heavy elements in the galaxy we have ${\rm  \eta_{ej} \approx 1}$, and thus 
we can write:
\be
{\rm f_{dil} = f_g/f_Z \approx \Big(\frac{1-f_{ej}+f_{accr}}{1-f_{ej}}\Big)},
\label{eq:f_dil2}
\ee 
from which we infer that the maximum ${\rm f_{dil}}$ corresponds to the highest ${\rm f_{accr}}$ and 
highest ${\rm f_{ej}}$. A merging event provides the maximum ${\rm f_{accr}}$. To be of primordial 
composition the merging system should not have experienced previous star-formation, and hence 
should have a total mass lower than our Pop~III star-forming galaxy \citep[e.g.][]{salvadori09}.
We infer ${\rm M_{accr}\leq M_g}$ and thus ${\rm f_{accr}=[0,1]}$. 
To set the maximum ${\rm f_{ej}}$ we should recall that we are interested
in probing the descendants of very massive first stars. Thus, we need to retain some metals and gas 
to trigger further star-formation even if ${\rm f_{accr}=0}$. We can fairly assume ${\rm f_{ej}=[0,0.9]}$, 
from which we obtain ${\rm f_{dil} \leq 10}$. As we will discuss in Sec.~\ref{sec:discussion}, where we 
find similar upper limits by developing more detailed calculations, ${\rm f_{ej}}$ is not independent 
on ${\rm f_{*}}$. In particular the higher ${\rm f_{ej}}$ the larger is ${\rm f_{*}}$, which means that the
maximum ${\rm f_{dil}}$ corresponds to the maximum ${\rm f_{*}}$.

In conclusion, we can reasonably state that the dilution factor varies within the range 
${\rm f_{dil} \approx(0.02,10)}$ and strongly depends on ${\rm f_*}$. Since the lowest (highest) 
${\rm f_{dil}}$ corresponds to the lowest (highest) ${\rm f_*}$ we obtain \fratio$\in [10^{-4},10^{-1}]$ 
(see also Sec.~\ref{sec:discussion} and Fig.~\ref{Fig:free_par}). We can now proceed further 
since we have established that all the unknowns of the problem are effectively captured by 
our simple parametrisation.  
\begin{figure*}
  \includegraphics[width=0.495\linewidth]{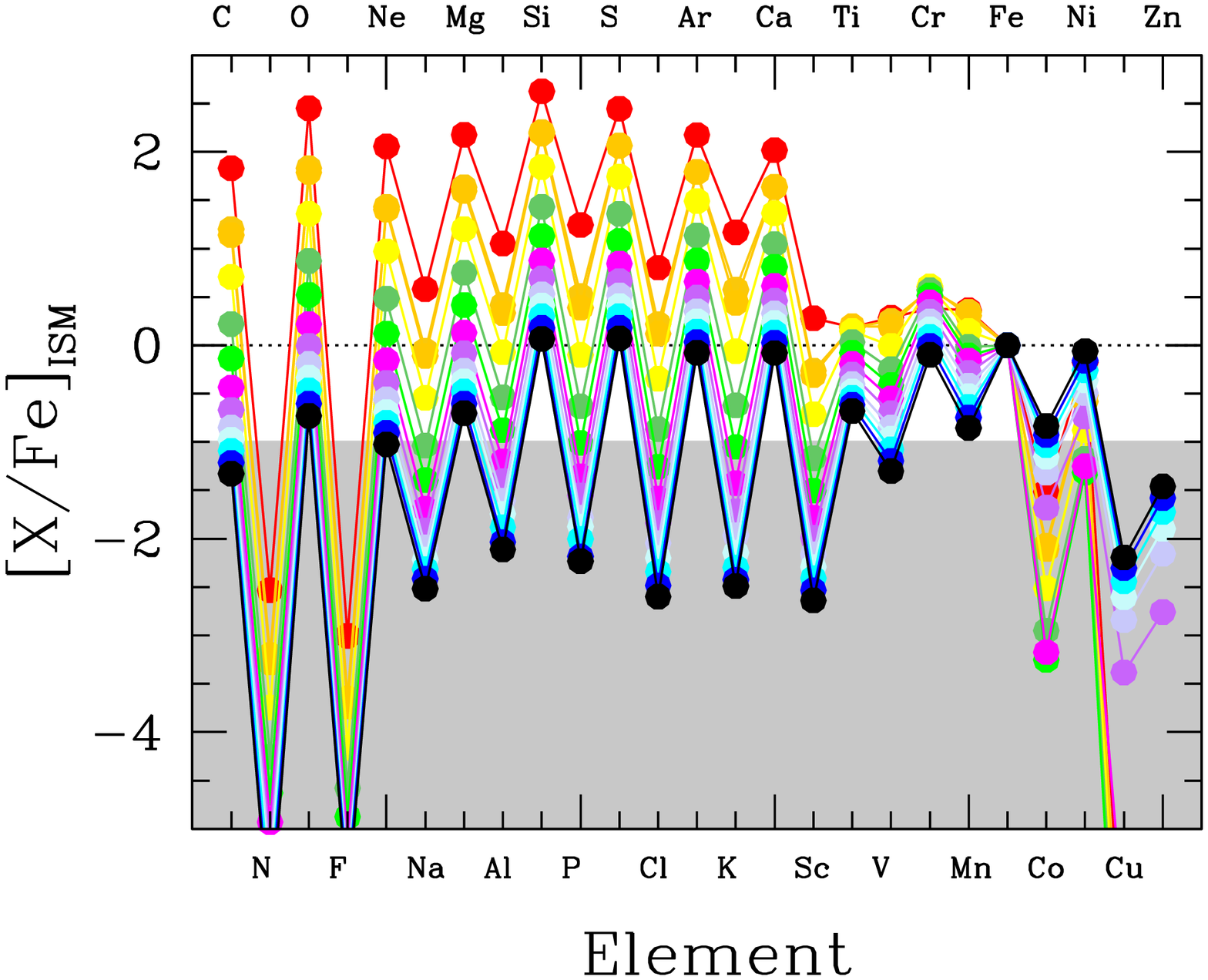}\includegraphics[width=0.495\linewidth]{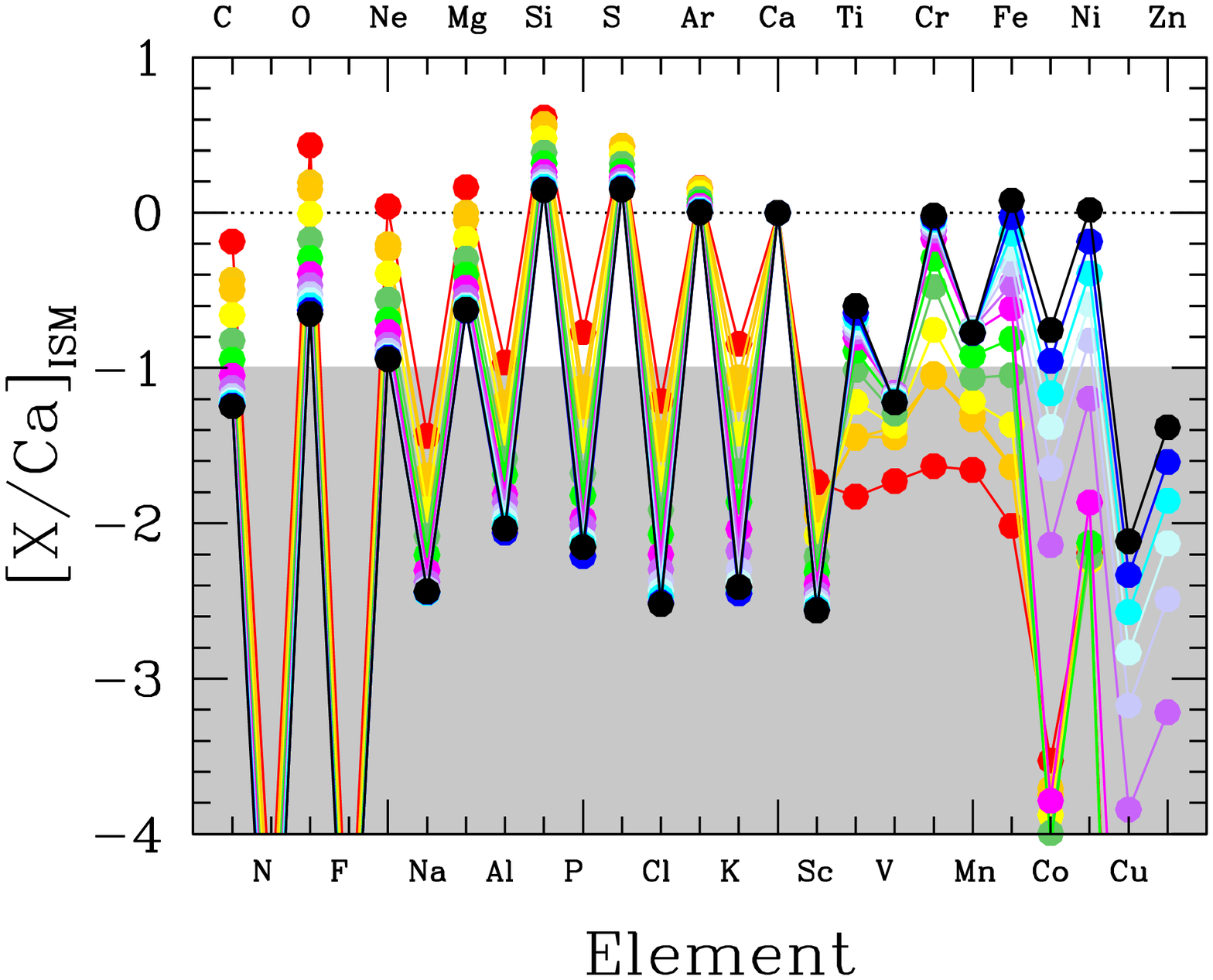}
    \caption{Chemical abundance patterns of the ISM imprinted by a single 
    PISN with increasing mass: from $m_{\rm pop~III}=150\Msun$ (red) to 
    $m_{\rm pop~III}=260\Msun$ (black). The injected chemical elements 
    are shown with respect to iron (left) and calcium (right) abundances.}
    \label{Fig:XFe_PISN}
\end{figure*}
\section{RESULTS}
In the following we present the results of our simple and general model. We will start 
by analysing the chemical properties of an ISM solely imprinted by PISN and then extend the 
calculations to the case in which the chemical products of these massive first stars represent a 
significant fraction ($\geq 50\%$) of the total mass of heavy elements in the ISM.
\subsection{ISM enrichment of a PISN event}
The predicted metallicity and iron-abundance values of an ISM imprinted by 
a single PISN\footnote{Or by several PISNe with the same mass, $\MIII$.}
are shown in Fig.~\ref{Fig:FeH_PISN}, as a function of $\MIII$, and for different 
\fratio values (see eq.~1, Sec.~2.1). We see that for any \fratio, the ISM metallicity 
is larger than any critical metallicity value to trigger the formation of low-mass 
Pop~II stars: $\rm Z_{ISM} > 10^{-3.5}\Zsun$ (horizontal lines).
Thus, long-lived second-generation stars can immediately start to form in a gaseous 
environment polluted by a single PISN event. These direct 
descendants of PISN can survive until the present day and hence be observed. 

We also notice that ${\rm [Fe/H]_{ISM}}$ strongly varies with the PISN mass, while 
${\rm Z_{ISM}}$ is basically independent of $\MIII$. In fact, the lower the iron
produced by PISN the higher the oxygen \citep{heger02}, so that each PISN always injects 
the same amount of heavy elements into the ISM ($\approx 50\%$ of its mass).
Furthermore, at a fixed \fratio, we have that ${\rm [Fe/H]_{ISM}}$ spans $\leq 1$~dex 
if $\MIII\geq 200\Msun$, while it might vary by $\geq 2$~dex when $\MIII< 200\Msun$. 
As a result the ``direct descendants of PISN'', i.e. second-generation stars that have been 
enriched {\it only} by the chemical product of these massive first stars, span a large [Fe/H] 
range and they are biased towards [Fe/H]$>-3$ values. This implies that a very low Fe 
abundance is {\it never} a good tracer of these rare stars. What about the other chemical elements? 

In Fig.~\ref{Fig:XFe_PISN} we show the predicted chemical abundance pattern 
of an ISM polluted by PISN with different masses normalized to iron, [X/Fe]$_{\tt ISM}$, 
and calcium, [X/Ca]$_{\tt ISM}$. {\it We recall that these 
abundance ratios are independent on the assumed free parameters} (eq.~4).
By focusing on the chemical imprint of a {\it single} PISN (each color) we can 
see the so-called ``odd-even'' effect, which has been historically looked for in 
ancient metal-poor stars but which has never been unambiguously observed. 
By inspecting the results for different $\MIII$ more closely, we note that [X/Fe]$_{\tt ISM}$ 
can vary by more than 2 orders of magnitudes for each chemical species. 
This implies that it is extremely difficult to identify a ``typical" abundance pattern of an 
ISM imprinted by a population of PISN, i.e. by more than one PISN with different mass. 
Yet, we can see that there are some key elements, i.e. N, F, Cu, and Zn, which 
are systematically under-produced with respect to iron, i.e. [X/Fe]$_{\tt ISM}< -1$. 
{\it Searching for the lack of such key chemical species in large stellar surveys 
can be an effective way to pre-select PISN candidates for high-resolution follow-up}.   
\begin{figure*}
 \includegraphics[width=0.950\linewidth]{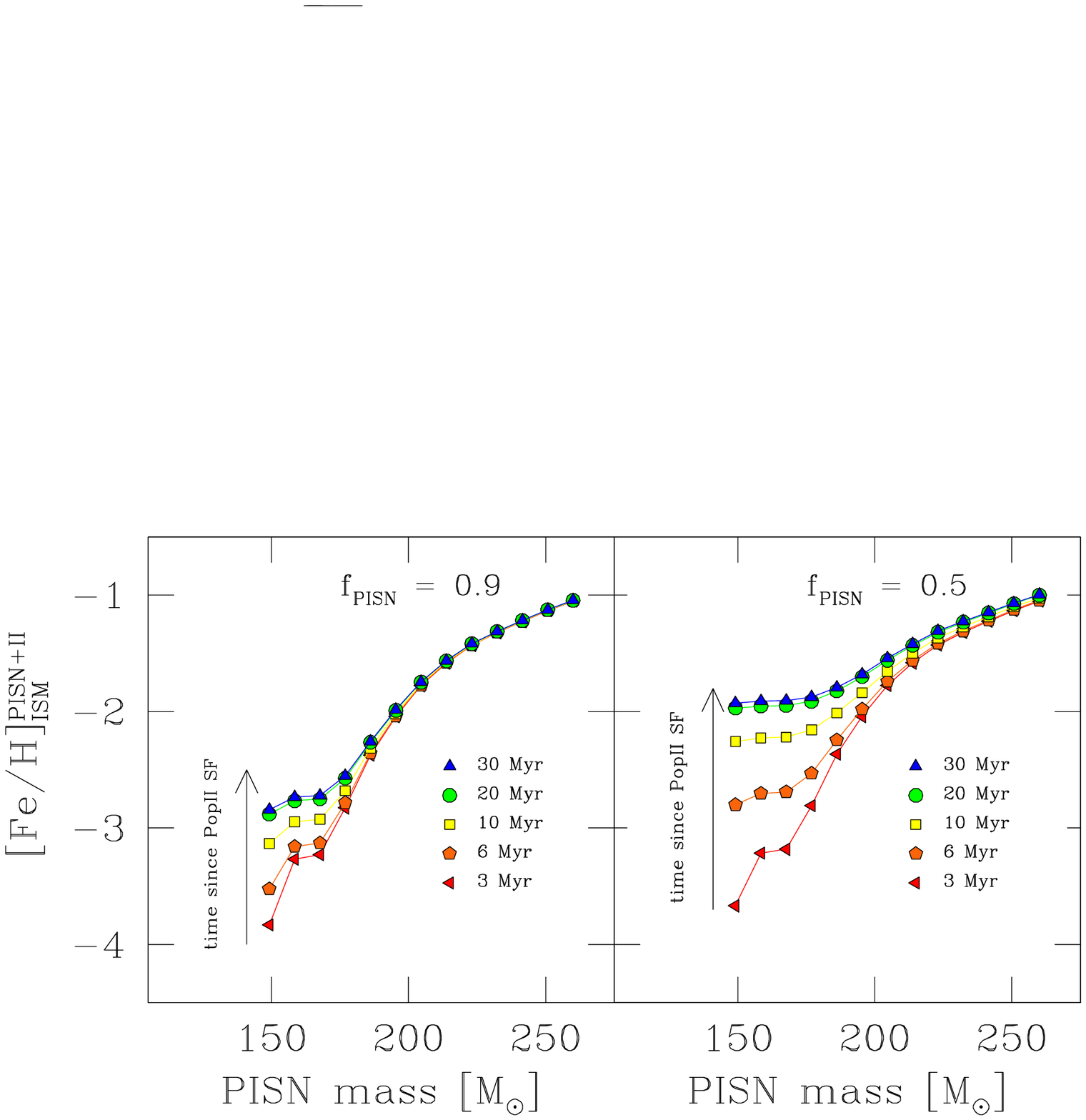}
    \caption{Iron abundances of an ISM imprinted by both PISN of mass 
    ${\rm m_{\tt PISN}}$ and the subsequent generation of Pop II stars. The results refer 
    to the model \fratio$=10^{-3}$ (Fig. \ref{Fig:FeH_PISN} and text), and assume 
    ${\rm f_{PISN}=0.9}$ ({\it left}) and ${\rm f_{PISN}=0.5}$ ({\it right}). Different 
    symbols show the expected values at different times since the onset of Pop II star-formation
    (see labels).} 
    \label{Fig:FeH_popII}
\end{figure*}

We should note here that in large stellar surveys  
often [Fe/H] is inferred either from the Ca{\sc ii} K line only  \citep[e.g][]{Pristine1}, 
or this line has a very  high weight in the determination of [Fe/H] \citep[e.g][]{Aguado17,Carlos14}.
In other words Ca is implicitly assumed to be a good tracer of Fe and the well known
evolution of [Ca/Fe] in the Milky Way, is folded into the analysis method \citep[e.g.][]{Bonifacio11}. 
Our results, however, show that this is not the case for an ISM uniquely imprinted by PISN, 
as [Ca/Fe] spans more than two orders of magnitudes depending on $\MIII$. 
In Fig.~3b we show the predicted chemical abundance pattern of an ISM polluted by a 
single PISN normalized to Ca. We see that several elements are under-produced
with respect to Ca: N, F, Na, Sc, V, Cu, and Zn. Among them, we can identify our 
{\it four key elements}, which can then be used to guide observational searches of 
gas/stars imprinted by PISN. 
\subsection{Subsequent Pop~II star-formation}
In our previous Section we showed that an ISM imprinted by PISN at a $100\%$ level, 
i.e. the birth environment of the direct descendants of very massive first stars, shows a 
deficiency of N, F, Cu, and Zn. We can thus use these key chemical elements to identify 
their missing descendants. However, we know that second-generation stars only imprinted
by PISN are extremely rare \citep[e.g.][]{salvadori07,DB17,Chiaki18}.
We can thus try to extend our search to stars imprinted by both PISN and a subsequent 
generation of Pop~II stars. We then face two questions. How does the chemical abundance 
pattern of an ISM pre-enriched by PISN evolve when we account for the contribution of  
``normal" Pop~II stars? Is the deficiency of the four key chemical elements preserved?

\begin{figure*}
 \includegraphics[width=0.495\linewidth]{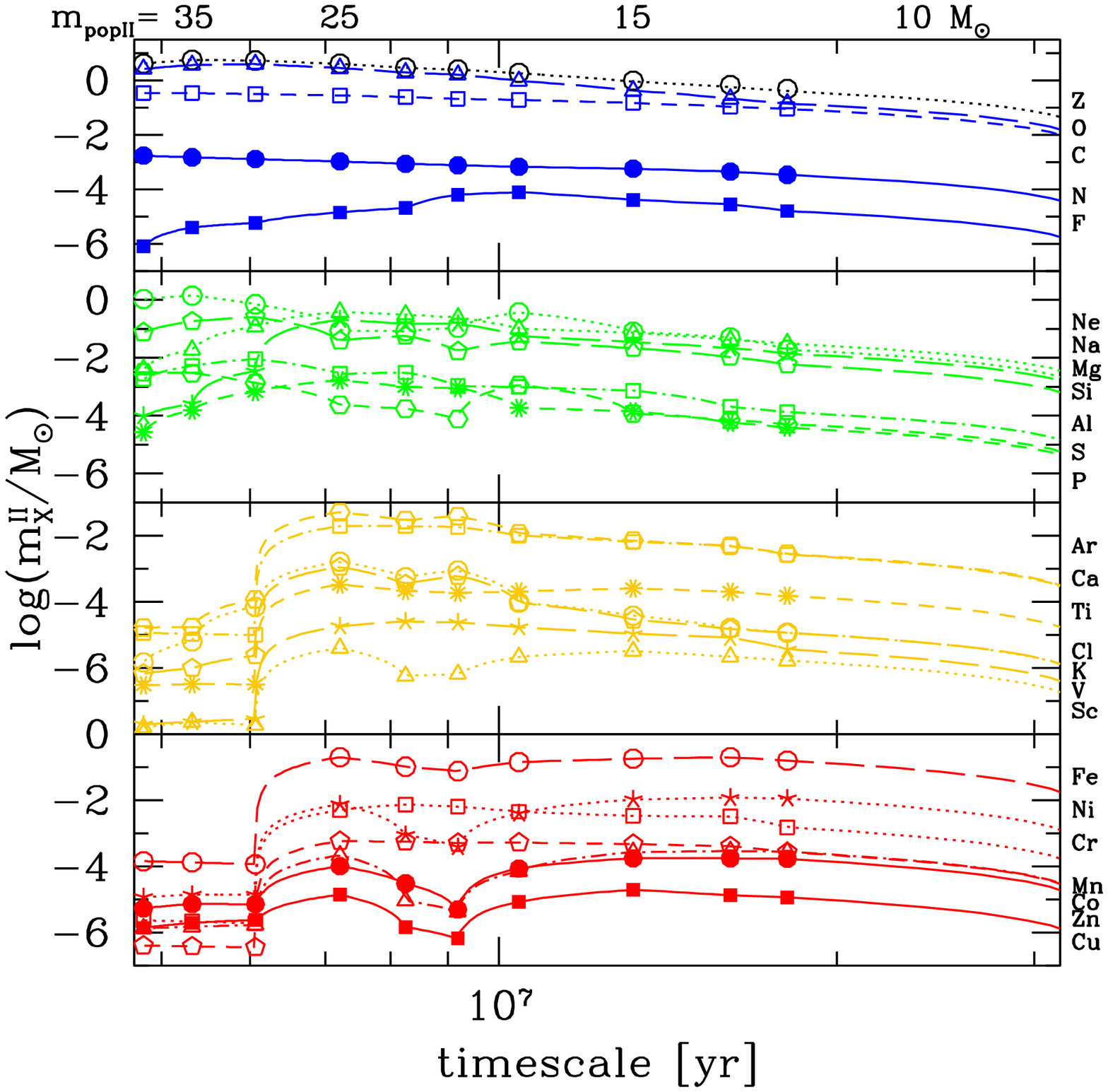} \includegraphics[width=0.495\linewidth]{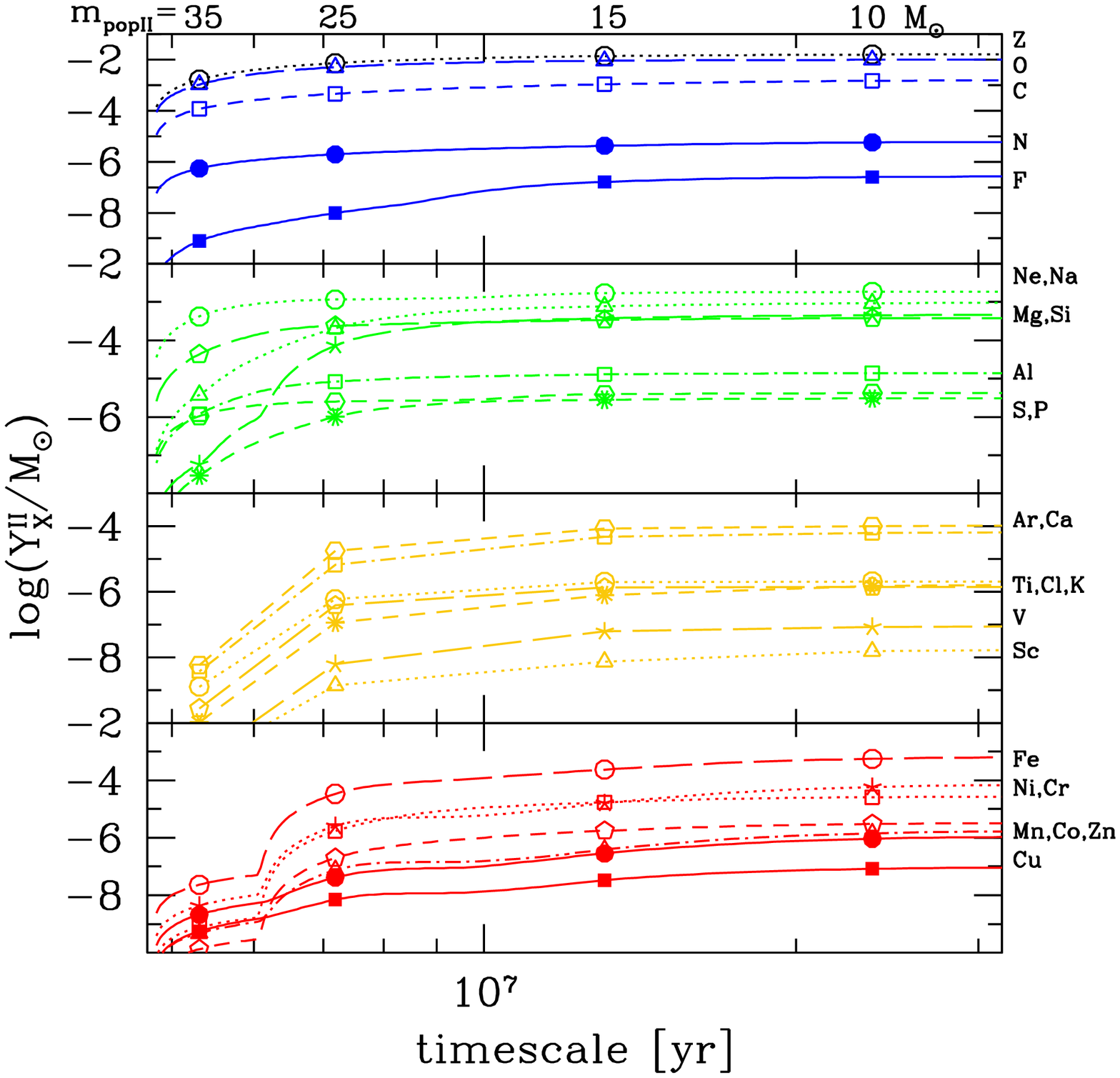}
    \caption{Chemical products of Pop~II stars with  ${\rm Z_*=0.01} \Zsun$, i.e. typical of an environment
    enriched by a single PISN. Symbols/lines refer to different chemical species (see labels). 
    {\it Left:} Mass of chemical elements produced by individual Pop II stars with various 
    ${\rm m_{popII}}$, which therefore evolve with different (life)times. 
    {\it Right:} Yields of Pop~II stars, i.e. cumulative mass of chemical elements produced per stellar 
    mass formed, as a function of time since the Pop II formation.
    In both panels the points show the values obtained by using the data from Woosley \& Weaver (1995).
    The lines are the interpolated values over the SNII mass range}.
    \label{Fig:Yx_popII}
\end{figure*}

To address these questions we can re-write the equations presented in Sec.~2 as a function 
of the mass of heavy elements contributed by PISN, $\rm{M^{eff}_Z= f_Z M_Z}$, with respect 
to the total mass of metals effectively injected into the ISM, i.e. 
$\rm{f_{\rm PISN} = (f_Z M_Z)/M^{tot}_{Z}}$, where ${\rm M^{tot}_Z = f_Z M_Z + M_Z^{II}}$ 
is the total mass of metals into the ISM and ${\rm M_Z^{II}}$ is the mass dispersed by Pop~II 
stars. To ensure that PISN dominate the metal enrichment we will vary this new free parameter 
in the range $f_{\rm PISN} = [0.5,1.0]$.  Analogously to what we have done in eq.~\ref{eq:MZ_pisn}, 
we can write ${\rm M_Z^{II}=M_*^{II}Y_Z^{II}(m_{II}(t,Z_*))}$ from which we get:
\be
{\rm M_Z^{II}= Y_Z^{II}(t,Z_*)f_*^{II} M^{II}_{g}\equiv Y_Z^{II}(t,Z_*)f_*^{II} f_g M_{g}}, 
\ee
where ${\rm Y_Z^{II}}$ is the metal yields of Pop~II stars and ${\rm f_*^{II}}$ is the efficiency 
at which Pop~II stars formed in the PISN-polluted ISM, with a gas mass ${\rm M^{II}_{g}\equiv f_g M_g}$
(see Sec.~ 2.1 for details). Note that the yields of Pop~II stars depend on both the stellar metallicity 
(i.e. ${\rm Z_*=Z_{ISM}}$, eq.~\ref{eq:metals}, Fig.~\ref{Fig:FeH_PISN}) and the time passed since 
the Pop~II star formation, ${\rm t_{popII}}$, which sets the mass of Pop~II stars contributing to the
metal enrichment, ${\rm m_{popII}}$. 
Given the broad mass range of Pop~II stars that can explode as core-collapse supernovae 
(SNII), ${\rm m_{popII}=(8-40)\Msun}$, such timescale can vary from a few Myr to tens of Myr,
as is shown in Fig.~\ref{Fig:Yx_popII}. 

To ensure the generality of our results we study the problem at various times, i.e. for 
each ${\rm f_{PISN}}$ we investigate how the ISM abundance pattern varies when SNII 
of different masses contribute to the chemical enrichment. Before doing so we determine the 
star-formation rate of Pop~II stars in the ISM pre-enriched by PISN.
By using the relation 
${\rm M_Z^{II}= M_Z^{tot}-f_Z M_Z=M_Z\times [(1-f_{\rm PISN})/f_{\rm PISN}]}$ we get: 
\be
{\rm f_*^{\tt II}=
\frac{f_*}{f_{dil}} \frac{{Y^{PISN}_Z}}{Y^{\tt II}_Z}\, \frac{1-f_{\rm PISN}}{f_{\rm PISN}}
=\frac{f_*}{f_{dil}} \frac{Y^{PISN}_Z}{Y_Z^{\tt II}}\, \beta},
\label{eq:fs_II}
\ee 
where for simplicity we have omitted the yield dependence on ${\rm Z_*}$, ${\rm t_{popII}}$ 
(or ${\rm m_{popII}}$), and ${\rm m_{PISN}}$, and so we will do hereafter. 
The relative amount of metals contributed by Pop~II with respect to PISN is 
${\rm \beta = \Big(\frac{1-f_{PISN}}{f_{PISN}}\Big)}$, where   
${\rm f_{PISN}\in [0.5,1.0]}$. Hence, eq.~\ref{eq:fs_II} is telling us that for any \fratio model, 
${\rm f_*^{\tt II}}$ is fully determined by the PISN-to-Pop~II metal yield ratio, 
${\rm Y_Z/Y^{II}_Z}$, once  $\beta$ has been fixed. Note that, analogously to what we 
discussed for ${\rm f_*}$, the PopII star-formation efficiency should be ${\rm f_*^{\tt II} \leq 0.1}$ 
(Sec.~2). 

The metallicity of an ISM polluted by both a PISN and a subsequent generation of normal SNII is then:
${\rm Z^{PISN+II}_{ISM}=(f^{II}_ZM^{tot}_Z)/(M_g f_{dil}f^{II}_{g})=M^{tot}_Z/(M_g f_{dil}f^{II}_{dil})}$, 
where ${\rm f^{II}_{dil}}$ accounts for the dilution of metals into the ISM. Similarly to what we saw for 
${\rm f_{dil}}$ (see Sec.~2 for details), ${\rm {f^{II}_{dil}}}$ resides in a relatively tight range, which is 
even narrower in the case of Pop~II stars since ${\rm\langle E_{51}\rangle \approx 1}$ independent 
of the SNII mass. Therefore we approximate ${\rm {f^{II}_{dil}\approx1}}$ and get the equivalent 
of eq.~\ref{eq:metals}:
\be
{\rm Z^{PISN+II}_{\tt ISM}=\frac{Y^{PISN}_Z}{f_{\rm PISN}}\,\frac{f_*}{f_{dil}}}\,
\label{eq:metals_popII}
\ee 
and of eq.~\ref{eq:FeH}:
\be
{\rm [X/H]^{PISN+II}_{\tt ISM}=log\big[\frac{f_*}{f_{dil}}[Y^{PISN}_X+\beta {\frac{Y^{II}_X Y^{PISN}_Z}{Y^{II}_Z}} ]\big]-log\big[\frac{M_X}{M_{Fe}}\big]_\odot}\;.
\label{eq:FeH_popII}
\ee
which are functions of the free parameters ${\rm f_*/f_{dil}}$, ${\rm{ f_{PISN}}}$, $\MIII$, and 
of the time passed since the formation of Pop II stars, ${\rm t_{popII}}$. Thus, for each $\MIII$ 
and \fratio value (see Fig.~\ref{Fig:FeH_PISN}), we can compute how ${\rm Z_{ISM}}$ and 
${\rm [Fe/H]_{ISM}}$ evolve for different $\rm{ f_{PISN}}$. 

The results of our calculations are displayed in Fig.~\ref{Fig:FeH_popII}. By construction the 
SNII contribution only alters the total ISM metallicity by a factor ${\rm{(1- f_{PISN})\leq 0.5}}$. 
On the other hand, it has a large impact on the iron abundance. This effect is particularly relevant 
for an ISM enriched by PISN with ${\rm \MIII < 200\Msun}$, which only convert $<0.001\%$ 
of their mass in iron. Furthermore, it is more pronounced 
when the PISN contribution to the ISM enrichment is lower, $\rm{ f_{PISN}}=0.5$. In conclusion, 
{\it the subsequent contribution of normal Pop II stars pushes towards more similar, and thus higher, 
${\rm [Fe/H]_{ISM}}$ values}.
\begin{figure*}
  \centering
 \includegraphics[width=0.85\linewidth,clip=,trim={1.5cm 12.70cm 5.75cm 0cm}]{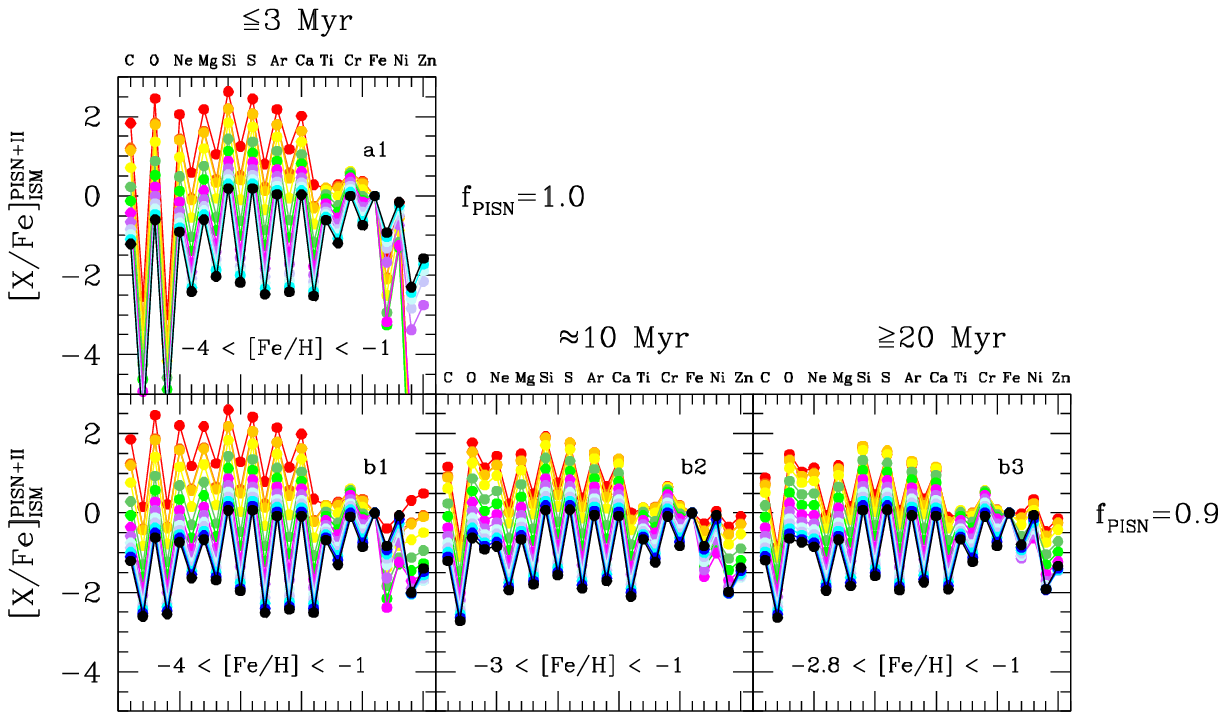}
 \includegraphics[width=0.85\linewidth,clip=,trim={1.5cm 8.25cm 5.75cm 4.35cm}]{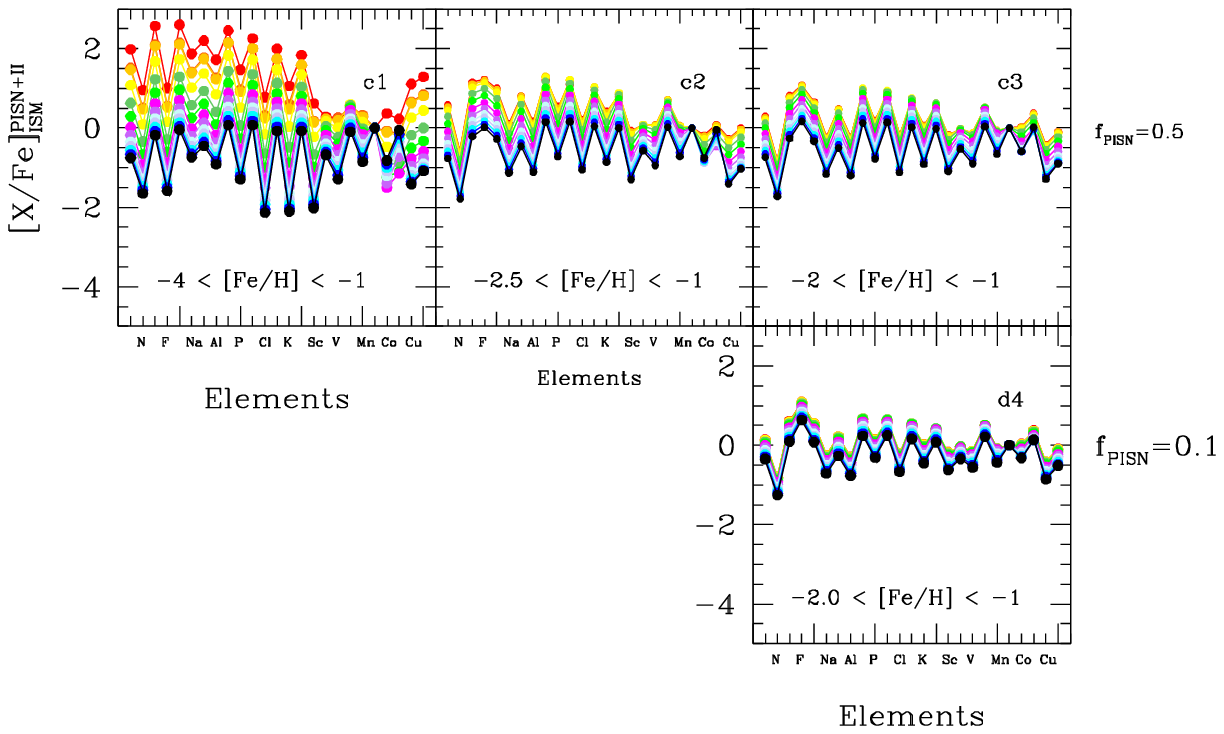}
  \caption{Possible chemical abundance patterns of an ISM imprinted by PISN with different masses 
    (colors as in Fig.~\ref{Fig:XFe_PISN}) and a subsequent generation of normal Pop II stars exploding 
    as core-collapse SN. Different rows show how the results change by varying the total amount of
    metals in the ISM contributed by PISN: $90\%$ ({\it panels b}), $50\%$ ({\it panels c}). 
    For comparison we also show two extreme cases: $100\%$ contribution from PISN 
    ({\it panel a1}), and $10\%$ contribution from PISN, which implies that $90\%$ of 
    the heavy elements are provided by Pop~II stars in a $\approx 20$~Myr time-scale ({\it panel d4}). 
    Different columns refer to different times since the Pop II formation: ${\rm t_{popII}\approx (3, 10, 20)}$~Myr 
    ({\it panels 1, 2, and 3}). In each panel the corresponding {\rm [Fe/H]} range for the ${\rm f_{PISN}}$ 
    and ${\rm t_{popII}}$ values is reported in the labels. The results refer to the case ${\rm f_*/f_{dil}=10^{-3}}$.}
    \label{Fig:XFe_popII}
 \end{figure*}

These results can be better understood by inspecting Fig.~\ref{Fig:Yx_popII} (left), which shows the 
mass of each chemical species, ${\rm m^{II}_X}$, yielded by Pop~II stars with various masses as 
predicted by the models of \cite{woosley95}. These quantities depends both on the mass of Pop~II
stars, which evolve on different timescales, and on their metallicity. As example, we only show the 
case ${\rm Z_*=Z_{ISM}=0.01\Zsun}$, which is a typical ISM metallicity after a PISN event 
(Fig.~\ref{Fig:FeH_PISN}). It is clear from the Figure that that while most of the metals (mainly 
oxygen) are provided by massive Pop~II stars, ${\rm m_{popII}>20\Msun}$, the iron produced 
by SNII is predominantly yielded by ${\rm m_{popII}<20\Msun}$ stars, i.e. on longer time-scales.

In the right panel of Fig.~\ref{Fig:Yx_popII}, we show the time dependent SNII yields, ${\rm Y^{II}_X}$, 
i.e the cumulative mass of heavy elements injected by Pop~II stars as a function of time since their 
formation epoch. Once ${\rm Z_*}$ has been fixed the time-dependent yields are obtained as: 
${\rm Y^{II}_X(t)=\int_{m_{popII}(t)}^{100\Msun} m^{II}_x \Phi(m_*) dm_*}$, where we have used 
the time and metallicity-dependent stellar lifetimes by \cite{raiteri1996}, and a standard Larson 
IMF\footnote{We adopt a Larson IMF: ${\rm \Phi(m_*) = A (m_*/0.35\Msun)^{-2.35}}$ with 
${\rm m_*=(0.1,100)}\Msun$, and normalization constant ${\rm A}$.}\citep[e.g.][]{salvadori08}.
We see that the longer the time since the Pop II star-formation, the higher the number of SNII contributing 
to the ISM enrichment, and so the larger ${\rm Y^{II}_{X}}$. Yet, while ${\rm Y^{II}_{Z}}$ reaches approximately 
the final value at ${\rm t_{popII}\approx 10}$~Myr, at the same ${\rm t_{popII}}$ we have that 
${\rm Y^{II}_{Fe}}$ is only ${\rm < 10\%}$ of the total iron mass produced after ${\rm t_{popII}\approx 20}$~Myrs.
\begin{figure*}
  \centering
 \includegraphics[width=0.45\linewidth,clip=,trim={3.25cm 1.5cm 2.5cm 0cm}]{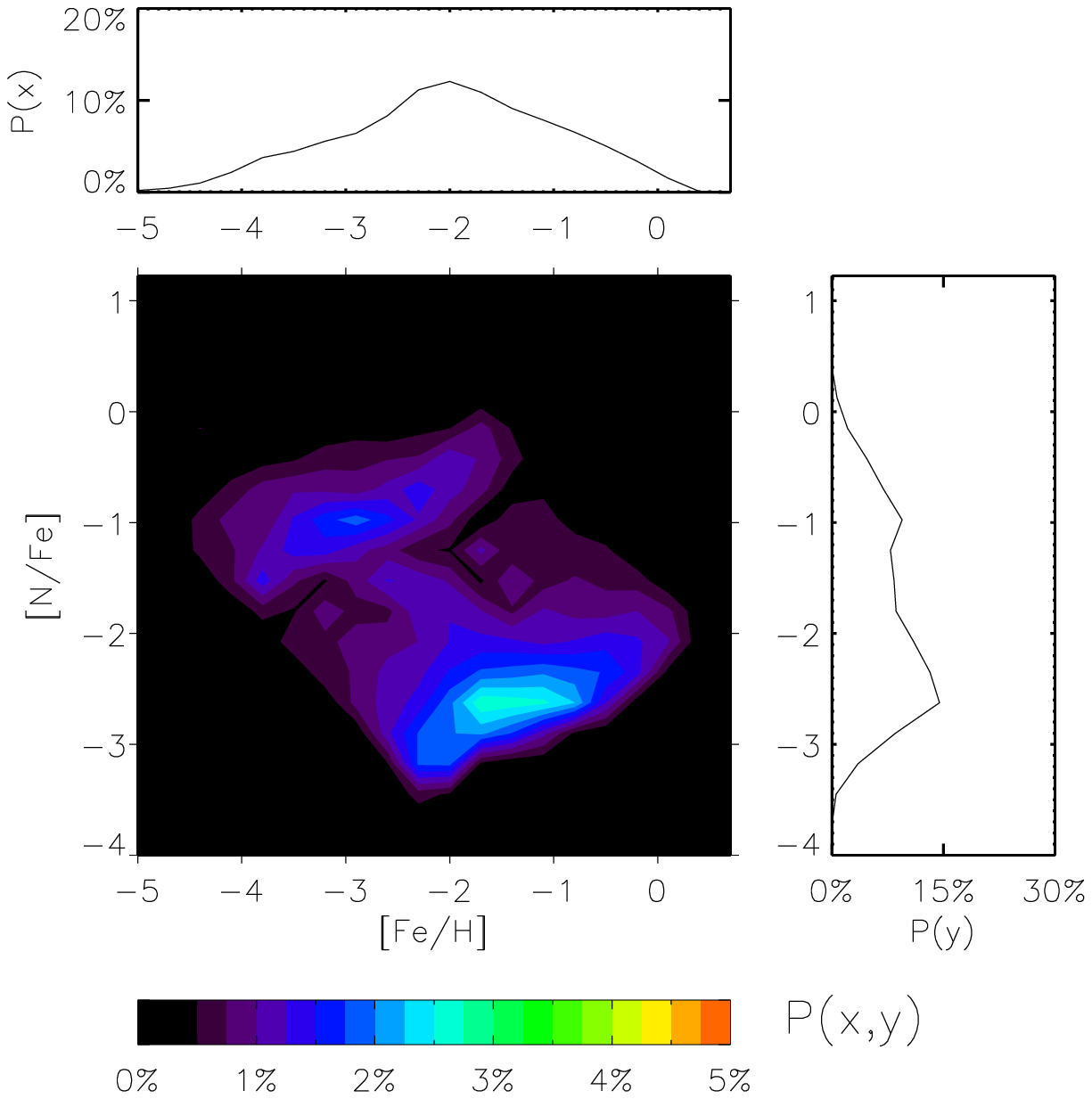}
 \includegraphics[width=0.45\linewidth,clip=,trim={3.25cm 1.5cm 2.5cm 0cm}]{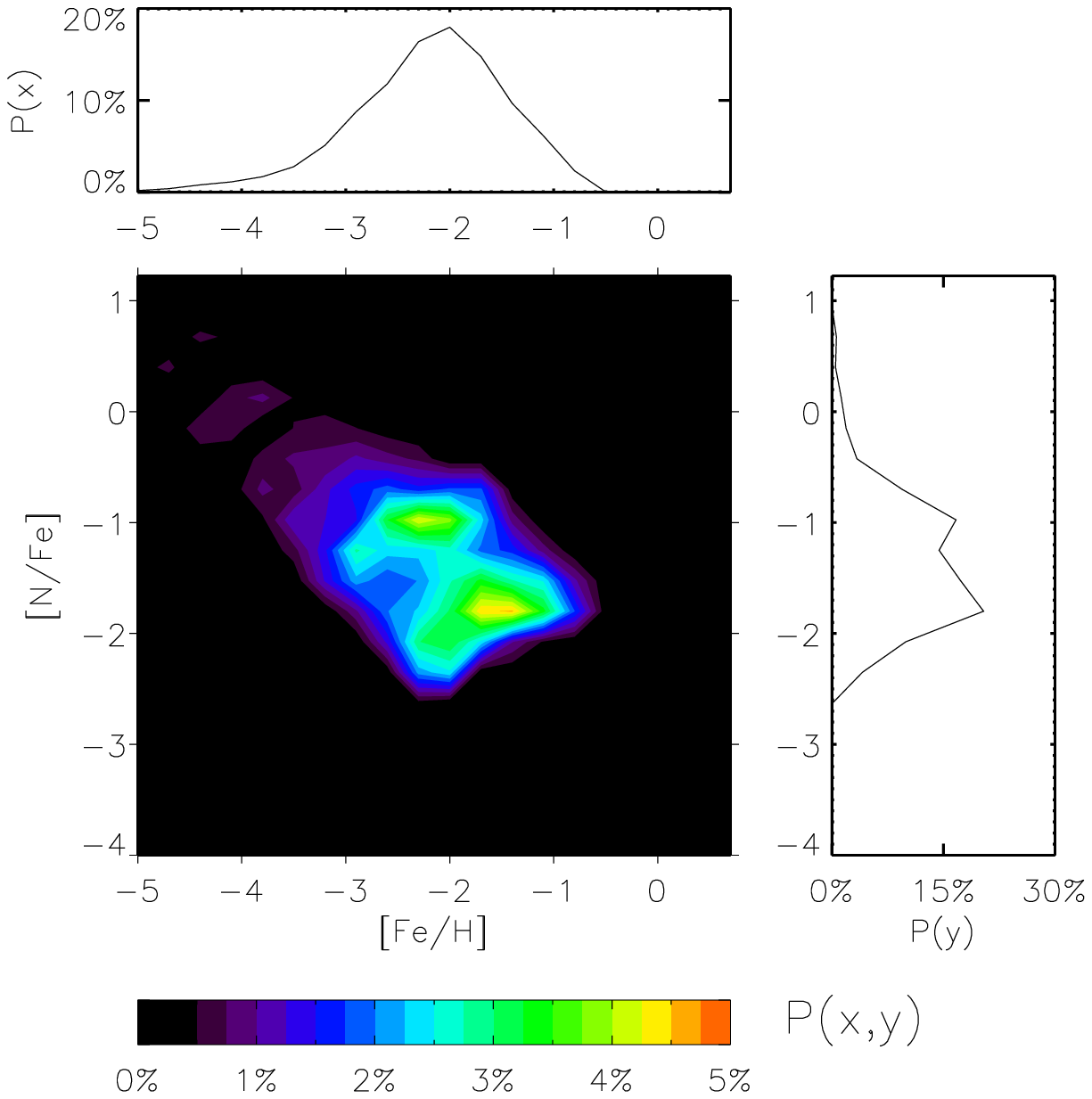}
 \includegraphics[width=0.45\linewidth,clip=,trim={3.25cm 1.5cm 2.5cm 3cm}]{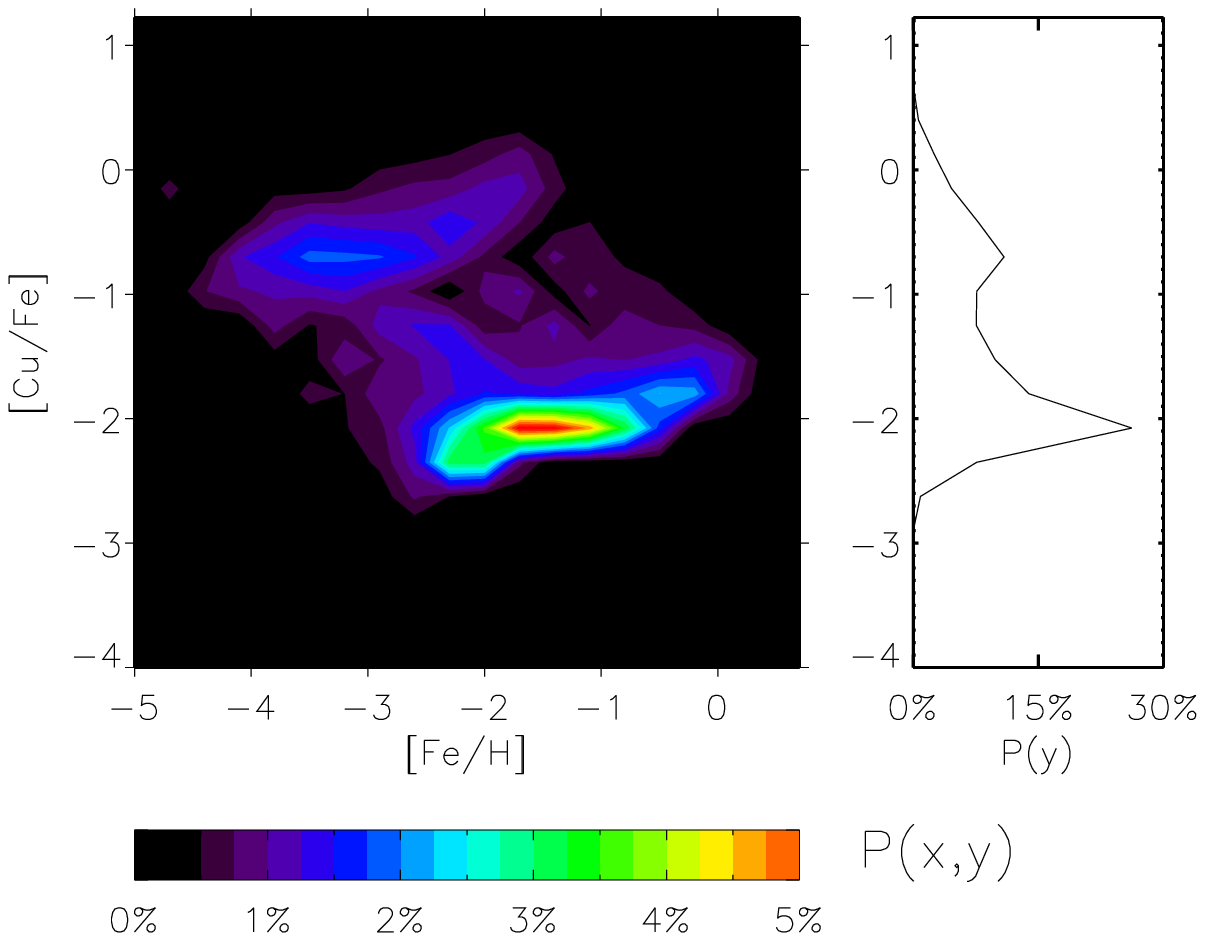}
 \includegraphics[width=0.45\linewidth,clip=,trim={3.25cm 1.5cm 2.5cm 3cm}]{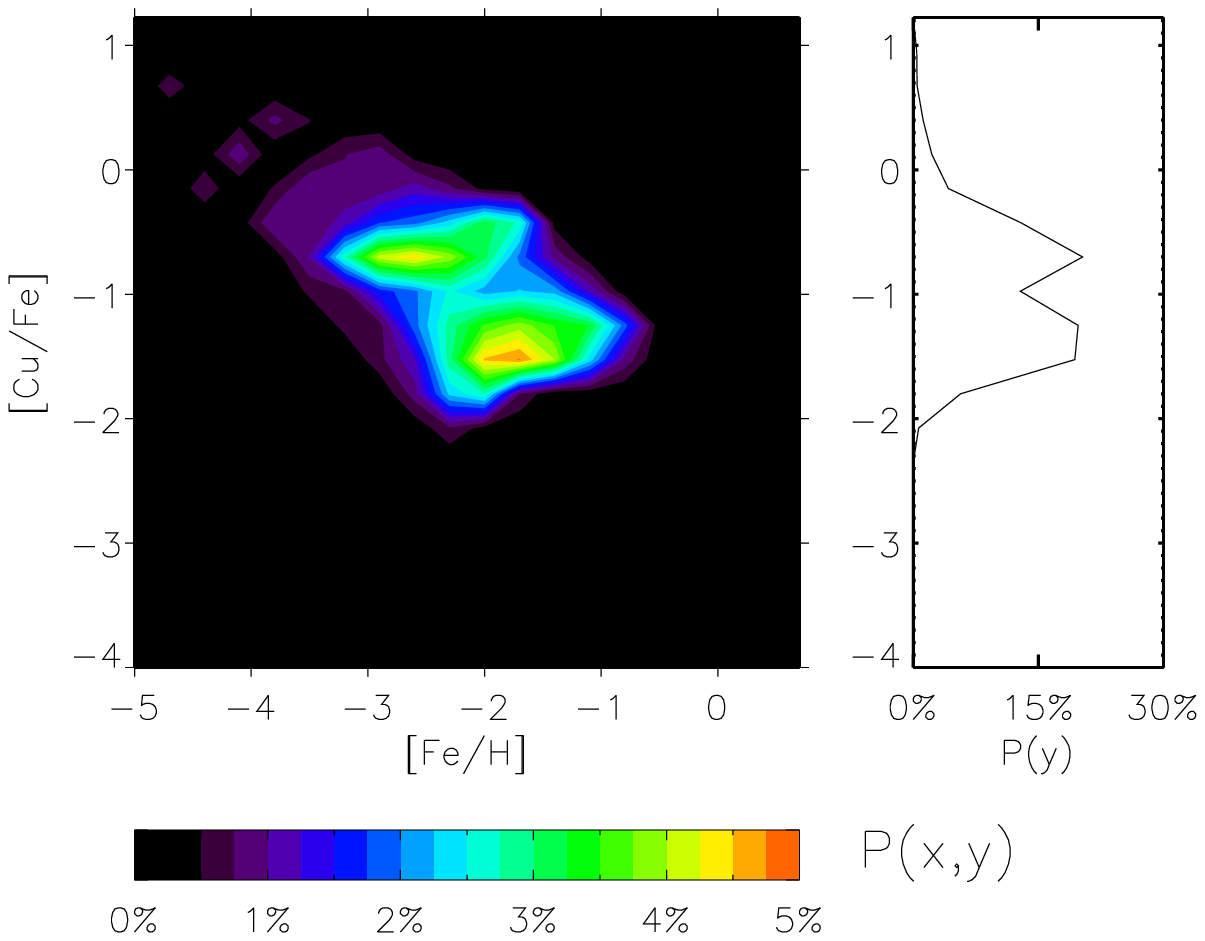}
 \includegraphics[width=0.45\linewidth,clip=,trim={3.25cm 0.cm 2.5cm 3cm}]{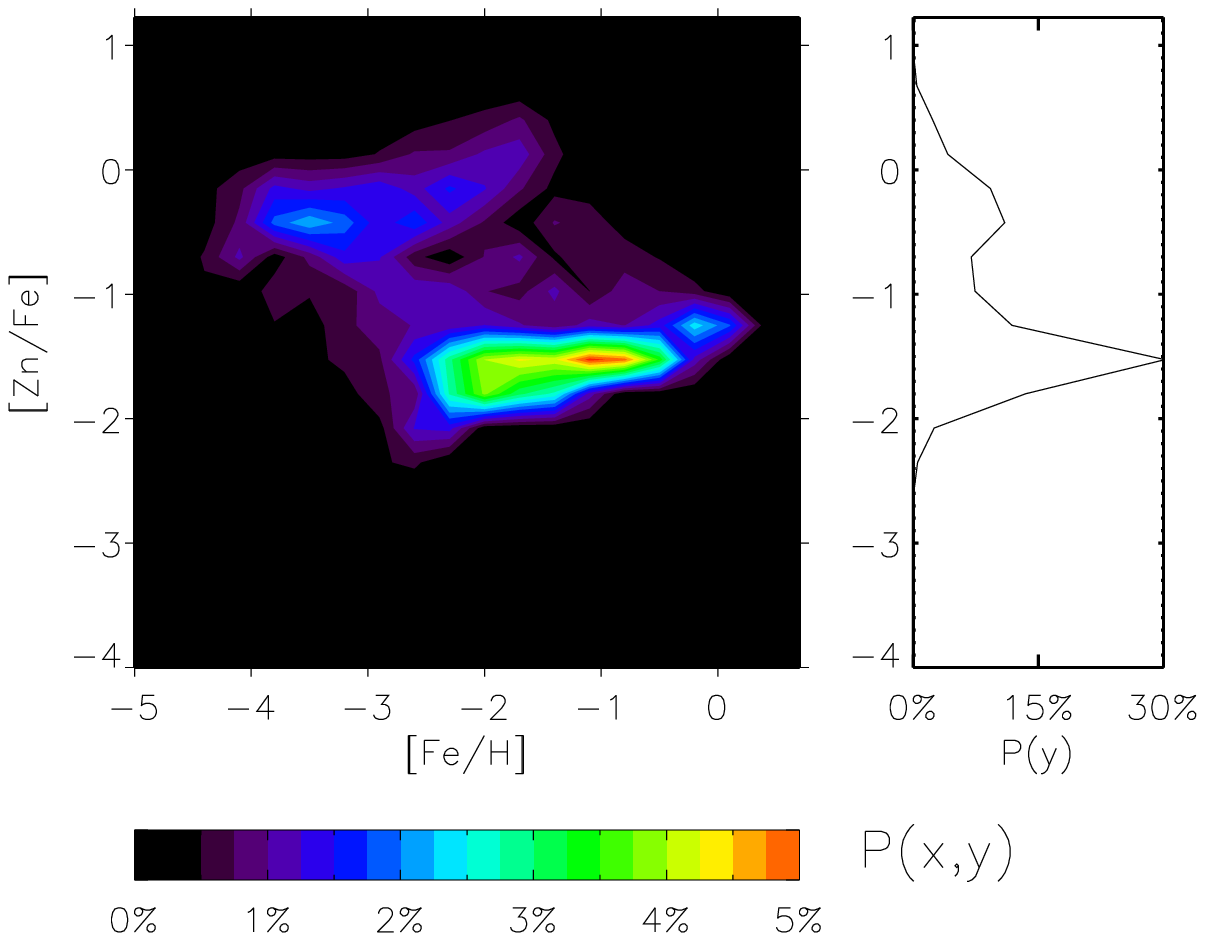}
 \includegraphics[width=0.45\linewidth,clip=,trim={3.25cm 0.cm 2.5cm 3cm}]{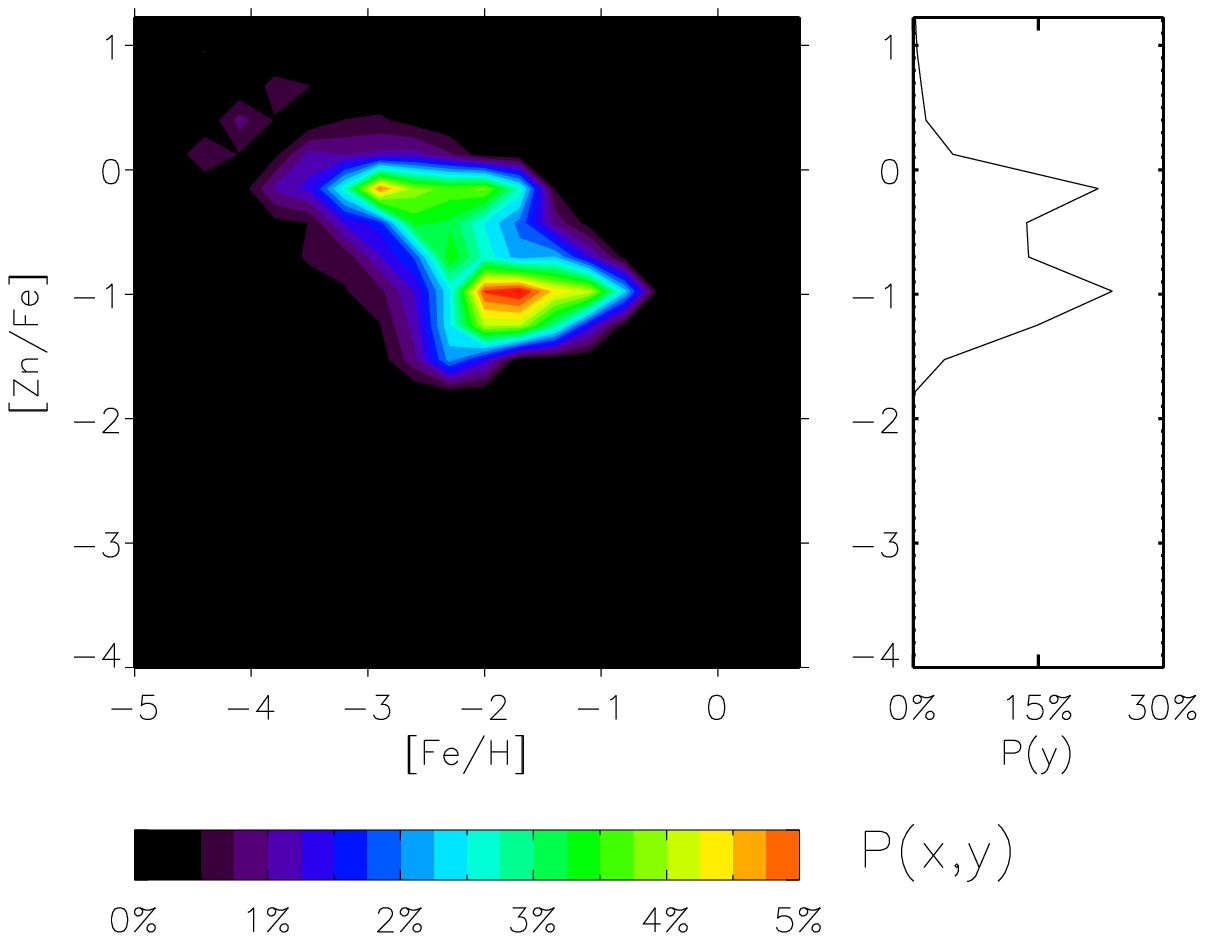}
   \caption{For the three key elements N ({\it upper panels}), Cu ({\it middle}), and Zn ({\it lower}), 
    we show the most likely [X/Fe] vs [Fe/H] values for an ISM imprinted by PISN at a ${\rm 90\%}$ 
    ({\it left}) and ${\rm 50\%}$ ({\it right}) level, i.e. for ${\rm f_{PISN}=0.9}$ and ${\rm f_{PISN}=0.5}$. 
    The histograms on the top (right) show the corresponding probability functions for [Fe/H] ([X/Fe]).}
    \label{Fig:XFe_FeH}
\end{figure*}

By using ${\rm Y_X^{II}}$ and eq.~\ref{eq:FeH_popII}, we can finally compute the 
chemical abundance pattern of an ISM enriched by a PISN and subsequent SNII:
\be
{\rm [X/Fe]^{PISN+II}_{ISM}
= log\Big[\frac{Y^{PISN}_X+ {\beta} {\frac{Y^{PISN}_Z}{Y^{II}_Z}} Y^{II}_{X} }
{Y^{PISN}_{Fe}+{\beta} {\frac{Y^{PISN}_{Z}}{Y^{II}_Z}} Y^{II}_{Fe}}\Big] -log[\frac{M_X}{M_{Fe}}]_\odot}\;.
\label{eq:ratios_popII}
\ee
Equations~\ref{eq:metals_popII}, \ref{eq:FeH_popII}, and \ref{eq:ratios_popII} respectively give 
back eq.~\ref{eq:metals}, eq.~\ref{eq:FeH}, and eq.~\ref{eq:ratios} when ${\rm \beta \to 0}$ 
or ${\rm f_{PISN}=1}$, i.e. when the overall amount of metals into the ISM is provided by PISN. 
Similar to the case of PISN (eq.~\ref{eq:ratios}), we see that eq.~\ref{eq:ratios_popII} does not depend 
on \fratio. Note however that \fratio, together with ${\rm m_{PISN}}$, sets the stellar metallicity, 
${\rm Z_*=Z_{ISM}}$, thus influencing the chemical yields (Sec. \ref{sec:key_elements}). 
We therefore conclude that ${\rm [X/Fe]^{PISN+II}_{ISM}}$ depends on all the free parameters of our model: 
(i) the PISN mass, ${\rm m_{\tt PISN}}$, which sets ${\rm Y^{PISN}_X}$ and together with \fratio 
determines ${\rm Z_{ISM}=Z_*}$; (ii) the time ${\rm t_{popII}}$ passed since the Pop~II formation, which sets 
${\rm Y^{II}_X(t,Z_*)}$; and (iii) the total fraction of metals contributed by PISN, 
${\rm f_{PISN}}$. Hence, by exploring the overall parameter space
we can fully investigate how the chemical abundance patterns of an ISM imprinted by PISNe varies 
after the formation of subsequent Pop~II stars, and under which conditions the deficiency of the key 
chemical species discussed in Sec.~3.1 is preserved.
\begin{figure}
  \centering
 \includegraphics[width=0.95\linewidth,clip=,trim={3.25cm 0.cm 2.5cm 3cm}]{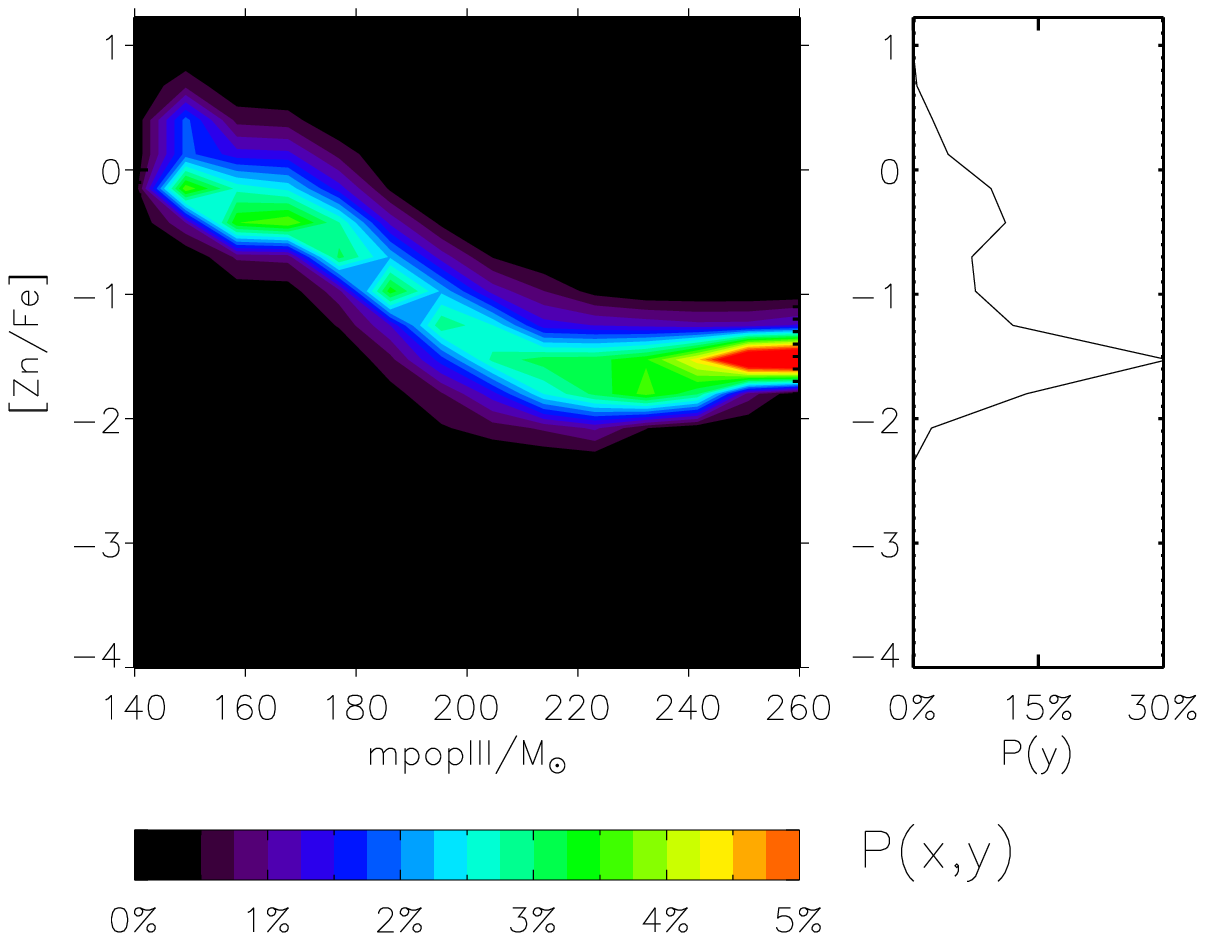}
 \includegraphics[width=0.95\linewidth,clip=,trim={3.25cm 1.5cm 2.5cm 3cm}]{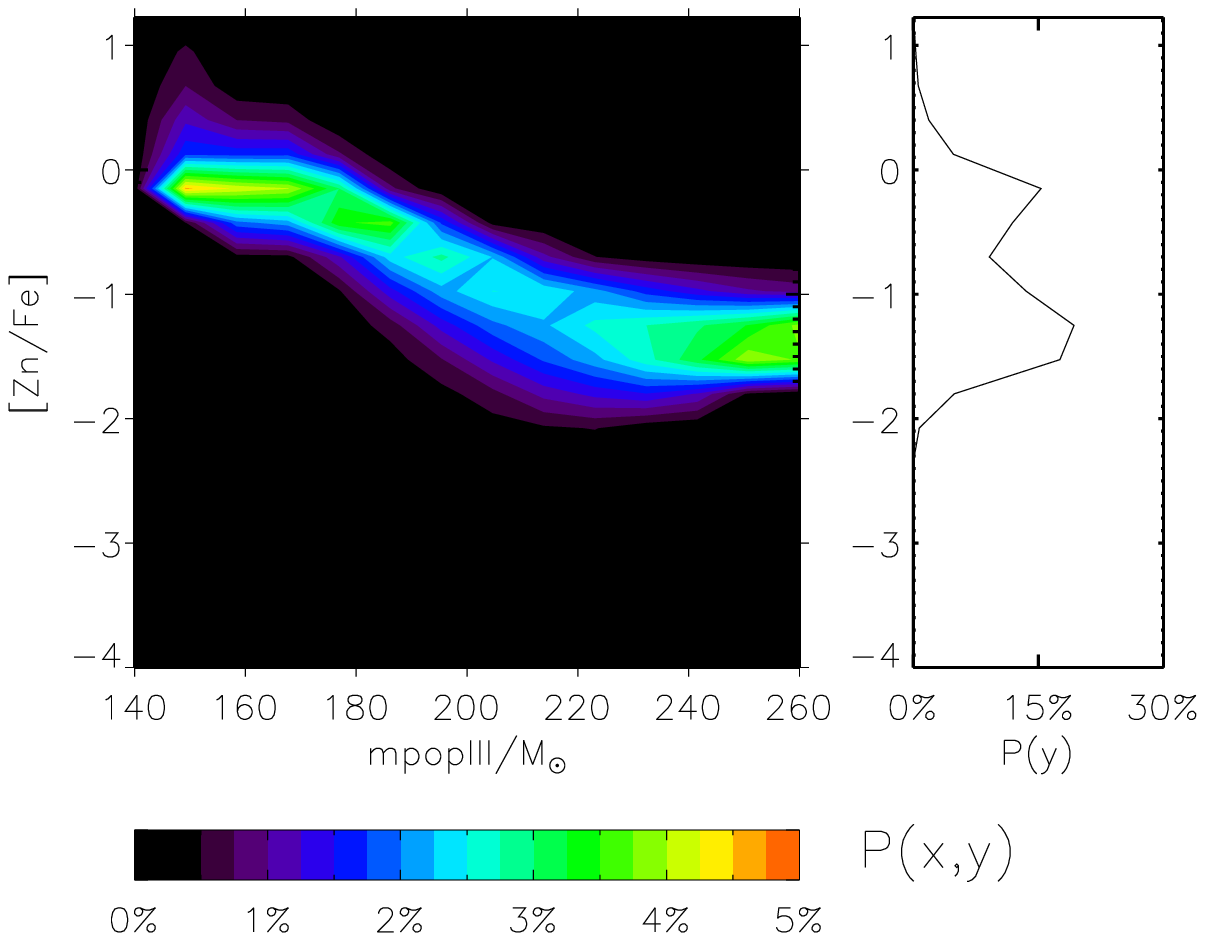}
   \caption{Most likely [Zn/Fe] vs ${\rm m_{PISN}}$ values for an ISM imprinted by PISN at a ${\rm 90\%}$ 
    ({\it top}) and ${\rm 50\%}$ ({\it bottom}) level. As in Fig.~\ref{Fig:XFe_FeH} the histograms on the right 
    show the corresponding probability distribution functions for [Zn/Fe].} 
    \label{Fig:ZnFe_mIII}
\end{figure}

The results of eq.~\ref{eq:ratios_popII} are shown in Fig.~\ref{Fig:XFe_popII} for \fratio ${\rm  = 10^{-3}}$.
We see that the peculiar chemical signatures of PISNe, i.e. the so-called odd-even effect (panel a), is 
mostly preserved when the contribution of Pop~II stars to the chemical enrichment is $\approx 10\%$ 
of the total (panels b). Yet, the effect is less prominent than before since those elements that were 
completely lacking in ejecta of PISN (e.g. N, F) are largely produced by massive SNII, 
yielding a finite, although low, [X/Fe] abundance ratios (panels b vs panel a). Furthermore, while 
the time passes and less massive Pop~II stars begin to contribute to the ISM enrichment, 
the over-abundance of even elements becomes less and less pronounced because of the larger 
amount of iron produced, which reduces the [X/Fe] ratios (panels b2-b3).
Finally, both the odd-even chemical signatures and the large variation of [X/Fe] for different $\MIII$ 
get gradually lost when the total amount of metals from low-mass SNII, ${\rm m_{popII}\leq 20}$, 
equals those from PISN (panels c2-c3). When ${\rm f_{PISN}=0.1}$, they are almost completely 
washed-out (panel d). With the only exception of F, which is largely produced by Pop~II stars 
yielding ${\rm [F/Fe]>0}$ in most cases, we can see in Fig.~\ref{Fig:XFe_popII} that when 
${\rm f_{PISN}\geq 0.5}$ the majority of our models provide ${\rm [(N, Cu, Zn)/Fe] < -1}$. 
Hence, we can effectively use the deficiency of these three key species to trace ISM/stars 
imprinted by very massive PISN at $>50\%$ level. 
\section{PISN descendants: key elements}
\label{sec:key_elements}
With the only exception of F, which is largely produced by normal Pop~II stars, the lack of N, 
Cu, and Zn with respect to Fe is preserved in environments enriched by PISN at $>50\%$ level. 
To be able to use these key elements as tracers of PISN enrichment, we should then address 
the following questions: (1) How much do our results depend upon \fratio? 
(2) What are the typical [Fe/H] values of ISM and stars imprinted by PISN?

Fig.~\ref{Fig:XFe_FeH} shows the most likely [(N,Cu,Zn)/Fe] vs [Fe/H]  ISM abundance ratios 
for ${\rm f_{PISN}=(0.9,0.5)}$ and all the possible \fratio, $\MIII$, and ${\rm t_{popII}}$ values. 
The histograms reported in upper/right panels are obtained by respectively integrating the probability 
maps over [X/Fe] and [Fe/H]. Independent of the relative contribution of PISN to the ISM enrichment, 
i.e. for different ${\rm f_{PISN}}$, we see that the [Fe/H] covers a large range of values the most
common being [Fe/H]$\approx -1.8$. These histograms represent the metallicity distribution 
function (MDF) of stars imprinted by PISN at $>90\%$ and $>50\%$ level. The peak of these 
MDFs nicely coincides with the one found by \cite{DB17}. However, our calculations are simpler 
and more generic, since they do not depend on the assumed (cold) dark matter model, nor on 
the specific environment of PISN formation (e.g. mini-halos).

For all the key elements in Fig.~\ref{Fig:XFe_FeH} the probability distribution functions of the 
most likely [X/Fe] abundance ratios exhibit two peaks, one at ${\rm [X/Fe]_{low}<-1}$, and 
the other at higher values, ${\rm [X/Fe]_{high}}$. 
When ${\rm f_{PISN}=0.9}$, ${\rm [X/Fe]_{low}}$ is more pronounced than the other one 
and we see that ${\rm P([X/Fe]_{low}) \approx 2\times P([X/Fe]_{high})}$ for both Cu and Zn. 
On the other hand, when ${\rm f_{PISN}=0.5}$ the two peaks are of similar size. Moreover, 
while ${\rm [X/Fe]_{high}}$ almost coincides with what we found for ${\rm f_{PISN}=0.9}$, we 
see that ${\rm [X/Fe]_{low}}$ is shifted towards higher abundance ratios by $\approx0.5$ dex.

In Fig.~\ref{Fig:ZnFe_mIII} we demonstrate the origin of these peaks by showing the 
case of [Zn/Fe]. We immediately see that ${\rm [X/Fe]_{high}}$ is produced 
by the chemical enrichment of low-mass PISN, ${\rm m_{PISN} \leq 180 \Msun}$, while 
${\rm [X/Fe]_{low}}$ is caused by more massive PISN. To understand why one peak is more
pronounced than the other when ${\rm f_{PISN}=0.9}$ we should inspect the b-panels of 
Fig.~\ref{Fig:XFe_popII}. By comparing the panels b1, b2, and b3, which respectively refer 
to ${\rm t_{popII} = (3, 10, 30)~Myr}$, we see that the [Zn/Fe] ratios of the ${\rm m_{PISN} 
\leq 180 \Msun}$ models (yellow, orange, and red symbols) vary by more than 1~dex. 
Conversely, models with ${\rm m_{PISN} > 180 \Msun}$ (from green to black symbols), 
show very similar [Zn/Fe] values across the whole ${\rm t_{popII}}$ range, which is mainly
due to the constant [Fe/H] values they produce (Fig.~4, left). As a result of these features
we have that $P({\rm [X/Fe]_{high}})\approx 2P({\rm [X/Fe]_{low}})$ for ${\rm f_{PISN}=0.9}$.
\begin{figure}
  \centering
 \includegraphics[width=0.99\linewidth,clip=,trim={3.5cm 0cm 3.75cm 3.75cm}]{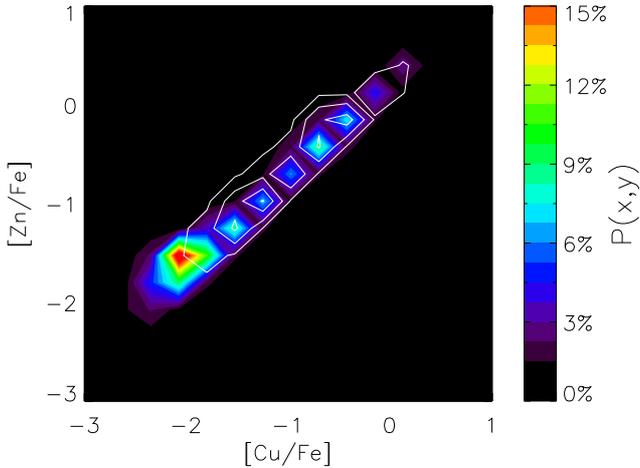}
   \caption{Most likely [Zn/Fe] vs [Cu/Fe] values for an ISM imprinted by PISN at a ${\rm 90\%}$,
    ({\it coloured contours and bar for the corresponding probability}) and ${\rm 50\%}$ level 
    ({\it over-plotted white contours corresponding to probabilities $P=(1, 5, 10, 15)\%$}).}
    \label{Fig:ZnFe_CuFe}
\end{figure}
In the right panel of Fig.~4, we see that when ${\rm f_{PISN}=0.5}$ models with 
${\rm m_{PISN} > 180 \Msun}$ provide different [Fe/H] for various ${\rm t_{popII}}$.
As a result their [Zn/Fe] distributions are less peaked then in the previous case 
(Fig.~\ref{Fig:ZnFe_mIII}).
On the other hand, in Fig.~4 (right) we see that models with ${\rm m_{PISN} \leq 180 \Msun}$
show exactly the same [Fe/H] when ${\rm t_{popII}\geq 20}$~Myr, which results in a higher
$P({\rm [X/Fe]_{high}})$. The combination of these effects for ${\rm f_{PISN}=0.5}$ causes
$P({\rm [X/Fe]_{high}})\approx P({\rm [X/Fe]_{low}})$.

In Fig.~\ref{Fig:ZnFe_CuFe} we show the results for the most likely [Zn/Fe] and [Cu/Fe] 
of an ISM enriched by PISN at a $90\%$ and $50\%$ level. When the highly varying [Fe/H] is not included 
(e.g. Fig.~4 for its dependance on ${\rm f_{PISN}}$, ${\rm m_{PISN}}$, and ${\rm t_{popII}}$), the most 
probable abundance space is greatly reduced. Furthermore it essentially lies in a line since Zn and Cu 
are both Fe-peak elements, which are mainly produced during the explosive phase of SN.
\begin{figure}
  \centering
 \includegraphics[width=0.95\linewidth,clip=,trim={1.5cm 1.5cm 2.5cm 3cm}]{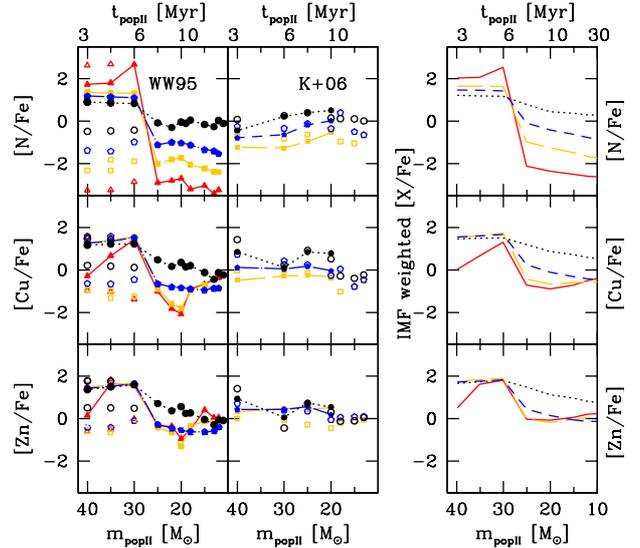}
   \caption{Abundance ratio from Pop II stars with different masses/lifetimes and metallicities. 
    Lines show the linearly interpolated values among different masses.
     {\it Left:} [X/Fe] values from Woosley \& Weaver (1995) for metallicities ${\rm Z_* = 10^{-4}\Zsun}$ 
     ({\it red triangles, solid lines}), ${\rm Z_* = 0.01\Zsun}$ ({\it yellow squares, long-dashed}), 
     ${\rm Z_* =  0.1\Zsun}$ ({\it blue pentagons, short-dashed}), and  ${\rm Z_* = \Zsun}$ 
     ({\it black circles, dotted}). Model A assumed in our calculations is shown with filled symbols. 
     Open symbols show models B and C (see text). {\it Middle:} [X/Fe] values for core-collapse 
     SN ({\it open symbols}) and hypernovae ({\it filled symbols}) from Kobayashi et al. (2006) 
     for metallicities 
     ${\rm Z_* = 0.05\Zsun}$ ({\it yellow squares, long-dashed}), ${\rm Z_* =  0.2\Zsun}$ ({\it blue pentagons, 
     short-dashed}), and ${\rm Z_* = \Zsun}$ ({\it black circles, dotted}).
     {\it Right:} IMF weighted [X/Fe] yields used in our calculations and derived using model A of 
     Woosley \& Weaver (1995) (see left panel, same colors/symbols).}
    \label{Fig:key_el}
\end{figure}

In conclusion, by identifying ISM gas or stars with ${\rm [N/Fe] < -1.5}$, ${\rm [Cu/Fe] < -1}$, 
${\rm [Zn/Fe] < -0.5}$, and that reside in the broad range ${\rm -3 < [Fe/H] <-1}$, we should 
maximize the probability to find gas/stars imprinted by PISN at ${\rm \geq 50\%}$ level.
Before comparing our results with available data from the literature to look for interesting candidates, 
we should discuss the robustness of these findings. To this end we will examine in detail the SNII 
time-dependent yields exploited in our  calculations and we will analize the possible effects on the 
ISM enrichment of supernovae Ia (SNIa), which represent important sources of metals at [Fe/H]$> -2$. 
\subsection{A deeper look at the SNII yields}
We have shown that the deficiency of the three key elements N, Cu, and Zn with respect 
to Fe is a good tracer of a chemical imprint from PISN. Indeed, such elements are so scarcely 
produced by PISN (Fig~\ref{Fig:XFe_PISN}) that even when the ISM is polluted by subsequent 
SNII at $\geq 50\%$ level their under-production with respect to Fe persist. We can then ask:
are we sure not to underestimate [(N,Cu,Zn)/Fe] yields from SNII?

In Fig.~\ref{Fig:key_el} we show the [(N,Cu,Zn)/Fe] yields for Pop II stars with different masses
(lifetimes) and metallicities as derived by various authors (left vs middle) and by using different 
models. In the left-most panel of Fig.~\ref{Fig:key_el} we report the \cite{woosley95} outputs 
for the three models published in their paper (A, B, and C) for core-collapse SN with masses
${\rm m_{popII}=(30, 35, 40) \Msun}$. These models differ for the SN explosion energy, 
the A model being the standard case ($E_{SN}\approx 10^{51}$~erg). For all these models 
we halved the iron yields as suggested by Timmes et al. (1995). 
This automatically produces higher [X/Fe] ratios, so we are not underestimating them. 
From Fig.~\ref{Fig:key_el} we can see that the standard model (A), which is assumed in our 
calculations, gives the largest [X/Fe] abundance ratios in almost all cases. Furthermore, we note 
that massive core-collapse SN with ${\rm m_{popII}\geq 30} \Msun$ provide the highest contribution 
to [(N,Cu,Zn)/Fe] abundance ratios, and that such values are even higher than those provided by 
\cite{kobayashi06} (right panel), which include the effects of both normal core-collapse SN and 
more energetic ``hypenovae".

Massive ${\rm m_{popII}\geq 30} \Msun$ stars have the shortest lifetimes, ${\rm t_{popII} < 6}$ Myr, 
and hence they can promptly start contributing to the chemical enrichment by injecting large 
amounts of [(N,Cu,Zn)/Fe]. Our calculations account in two ways for these high [(N,Cu,Zn)/Fe] 
producers. In fact, to evaluate the possible abundance patters of ISM/stars polluted by both 
PISN and Pop~II stars we use IMF integrated yields over times (masses) since the Pop~II f
ormation (Fig.~\ref{Fig:key_el}, right). So, on one hand, while evaluating the results at 
${\rm t_{popII} < 6}$~Myr we consider the possibility to have an ISM {\it only} imprinted by 
PISN and {\it massive} core-collapse SN. On the other hand, 
while computing the results at later times, we {\it always} account for the contribution of 
${\rm m_{popII}\geq 30} \Msun$ core-collapse SN that exploded before 
Note that in low-mass inefficiently star-forming systems (e.g. mini-halos), massive stars might 
form at a much lower rate than predicted by the theoretical IMF because of its poor sampling 
\citep[e.g.][]{DB17}. Hence massive Pop~II stars, i.e. the higher [(N,Cu,Zn)/Fe] producers, 
might form at a lower rate. Ultimately, we can surely assert that the lack of [(N,Cu,Zn)/Fe] 
in ISM/stars imprinted by PISN and Pop~II stars is not an artefact of our assumptions.
\subsection{Is there a degeneracy with the SNIa imprint?}
Recent measurements of zinc and iron abundances for individual red giant branch stars 
in the Milky Way \citep{duffau17,barbuy15} and in nearby dwarf galaxies \citep{skuladottir17,
skuladottir18} have reported low zinc over iron ratios, [Zn/Fe]$\approx (-0.5,0.0)$ for stars at
different [Fe/H]. To have a global view of these observations the reader can have a look at
Fig.~4 from \cite{skuladottir18}. 

The analytical calculations presented in the above papers to explain the observed trends, 
suggest that supernovae Ia (SNIa) can be responsible of such unusual low [Zn/Fe] ratio. 
Indeed, SNIa produce large amount of iron with respect to zinc. Assuming that the low 
[Zn/Fe] is just a consequence of the SNIa enrichment, the authors computed the predicted 
abundance ratios of other elements, finding a general good agreement with the measured 
values. The natural question is then: is there a degeneracy between the imprint of SNIa and 
massive primordial stars? And if so, how can we discriminate among these two different 
chemical enrichment channels?

To address these issues we estimate the possible chemical enrichment paths of an ISM 
imprinted by normal Pop~II stars exploding as both core-collapse SN (SNII) and supernovae 
Ia (SNIa), thus adopting a very similar approach of that developed by \cite{duffau17} and 
\cite{skuladottir17}. To avoid including the unknown contribution from primordial stars to the 
chemical enrichment, we assume an initial ISM composition equal to the average abundance 
pattern of "normal" Galactic halo stars at $\langle {\rm [Fe/H]} \rangle \approx -3$ 
\citep{cayrel04,yong13}. These stars, which are not enhanced in carbon, i.e. [C/Fe]$ < 1.0$, 
have very similar abundance patterns, and indeed exhibit extremely small scatters. 

Starting from these chemical abundance ratios as initial conditions, we varied the final 
iron-abundance within the range ${\rm[Fe/H]_{out} = [-2, -0.5]}$ to compute the possible 
${\rm [Zn/Fe]}$ and ${\rm [X/Fe]}$ abundance ratios of an ISM enriched by ${\rm N_{II}}$ 
core-collapse supernovae and ${\rm N_{Ia}=f_{Ia} N_{II}}$ supernovae Ia, where ${\rm f_{Ia}}$ 
is the fraction of SNIa with respect to SNII. Such a quantity can be derived from observations 
and it is found to be ${\rm f_{Ia}\approx 0.1}$ \citep[e.g.][]{mannucci06}. By adopting the 
yields of \cite{iwamoto99}, writing 
${\rm M_{Fe}^{in}=M_{g}\times 10^{[Fe/H]_{in} - log (M_{Fe}/M_{H})_{\odot}}}$, we
computed the number of core-collapse SN required to get different iron abundances:
\be
{\rm [Fe/H]^{II+Ia}_{out}=log \big[\frac{{M_{Fe}^{in}}+{Y_{Fe}^{II}}N_{II}+{Y_{Fe}^{Ia}}f_{Ia}N_{II}}{{M_{g}}}\big] -    log \big[\frac{M_{Fe}}{M_{H}} \big]_{\odot}}
\nonumber
\ee
where we assume a constant gas mass, ${\rm M_g}$, as done in \cite{duffau17} and \cite{skuladottir17}. 
Once derived ${\rm N_{II}}$, we computed the expected abundance ratio of the various chemical elements
\be
{\rm [X/Fe]^{II+Ia}_{out}=log \big[\frac{{M_{X}^{in}}+{Y_{X}^{II}}N_{II}+{Y_{X}^{Ia}}f_{Ia}N_{II}}{{M_{Fe}^{in}}+{Y_{Fe}^{II}}N_{II}+{Y_{Fe}^{Ia}}f_{Ia}N_{II}}\big] -    log \big[\frac{M_{X}}{M_{Fe}} \big]_{\odot}}\,
\nonumber
\ee
by varying ${\rm [Fe/H]_{out}}$, ${\rm f_{Ia}}$, and ${\rm M_{g}}$. As noticed by \cite{duffau17} 
and \cite{skuladottir17} these findings are independent on the assumed ${\rm M_{g}}$. 
 \begin{figure}
 \includegraphics[width=0.95\linewidth]{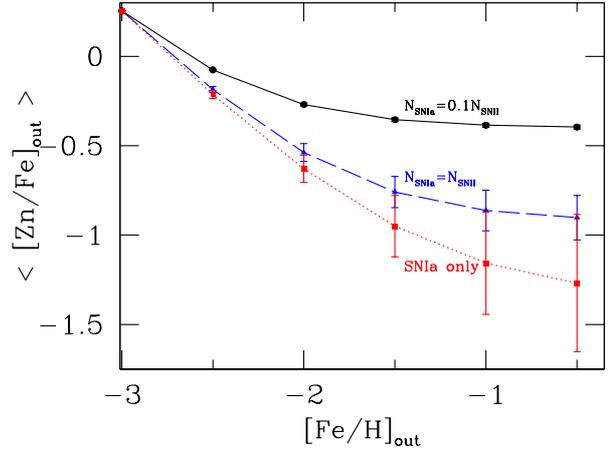}
    \caption{Average ${\rm \langle [Zn/Fe] \rangle }$ of an ISM enriched by both SNII and SNIa up to different
    iron-abundance values ${\rm [Fe/H]_{out}}$, and starting from an initial chemical composition of
    extremely metal-poor stars, ${\rm [Fe/H]_{in}\approx -3}$ and ${\rm [Zn/Fe]_{in}\approx 0.3}$ (see text and
    Fig.~\ref{Fig:XFe_SNIa}). The different curves refer to different SNIa-to-SNII ratios: 
    ${\rm f_{Ia}=N_{SNIa}/N_{SNII}= (0.1,1, >10)}$ (solid, dashed, dotted).}
     \label{Fig:ZnFe_SNIa}
\end{figure}
 \begin{figure*}
  \centering
    \includegraphics[width=0.99\linewidth]{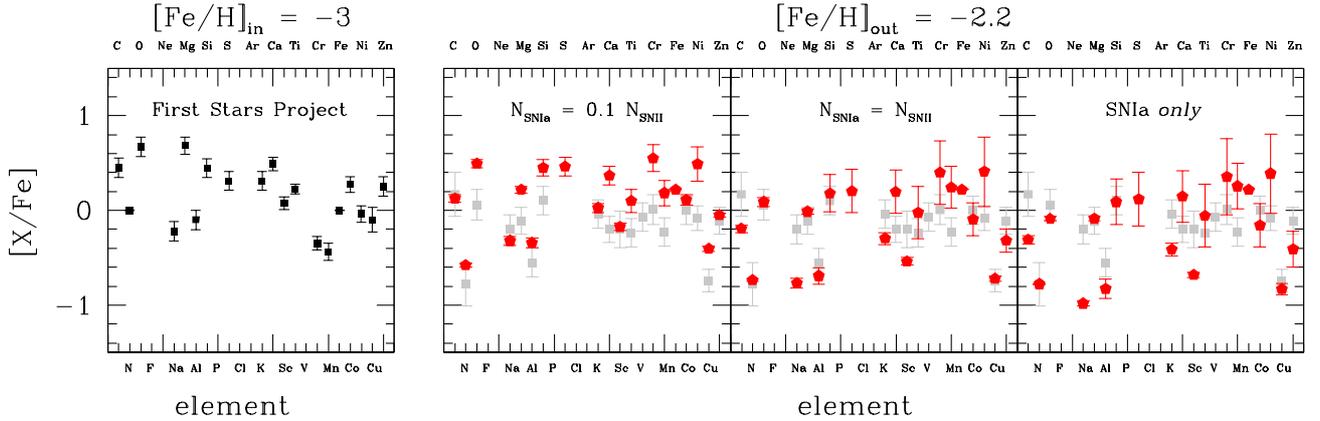}        
       \caption{{\it Left}: chemical abundances of Galactic stars at [Fe/H] $\approx -3$, which are assumed 
       as the initial conditions for the ISM in our calculations (see text). To derive [X/Fe] we have averaged the 
       measurements of the First Stars project sample, using NLTE results when available 
       (Cayrel et al. 2004; Andrievsky et al. 2007, 2008; Bonifacio et al. 2009; Spite
       et al. 2011, 2012). For Cu we have used the results of Andrievsky et al. (2008)
       and for Cr the average of Cr{\sc I} abundances from (Cayrel et al. 2004).        
       Errorbars show the star-to-star scatter.
       {\it Right}:  chemical abundance pattern of the ISM enriched up to ${\rm [Fe/H]_{out}\approx -2.2}$ 
       with the contribution of both SNII and SNIa (red penthagones). Errorbars represent the $1\sigma$ 
       deviation among different SNIa models. The panels show results for different SNIa-to-SNII ratios 
       (see labels). Grey squares show the chemical abundance pattern of a possible PISN descendant, 
       the star BD +80$^\circ$ 245, for which we measured $\mathrm{[Fe/H]}= -2.2$ (Sec.~5).}
        \label{Fig:XFe_SNIa}
\end{figure*}
In Fig.~\ref{Fig:ZnFe_SNIa} we show ${\rm \langle [Zn/Fe]_{out} \rangle}$ with respect to 
${\rm [Fe/H]_{out}}$ for different choices of ${\rm f_{Ia}}$, where the points and errorbars have 
been obtained by averaging among the different SNIa yields (models) by Iwamoto et al. 1999. 
For this reason, the larger the contribution of SNIa, the larger the errorbars.
Fig.~\ref{Fig:ZnFe_SNIa} shows two interesting features. First, ${\rm \langle [Zn/Fe]_{out} \rangle}$ 
declines with increasing ${\rm [Fe/H]}$, implying an overall trend that is different with respect 
to the ${\rm [Zn/Fe]}$ plateau predicted for an ISM enriched by PISN and SNII (Fig.~\ref{Fig:XFe_FeH}). 
Second, none of the models reach values ${\rm [Zn/Fe] <-1.5}$, which characterise an 
ISM imprinted by PISN at a $\geq 90\%$ level. This is also true for the case ${\rm f_{Ia}>10}$,
which is extreme and unlikely as it is equivalent to assume that {\it only} SNIa pollute the ISM
while observations suggest ${\rm f_{Ia}=0.1}$. In conclusion, our simple estimate show that 
there is not apparent degeneracy between the imprint of SNIa and massive first stars. 

However, around ${\rm [Fe/H] \approx -2}$, i.e. where the probability to find the descendants of PISN
is highest (Fig.~\ref{Fig:XFe_FeH}), we can see that ${\rm \langle [Zn/Fe]_{out} \rangle \approx -0.5\pm 0.3}$ 
for all ${\rm f_{Ia}}$, which is similar to the high ${\rm [Zn/Fe]}$ abundance ratio of an ISM imprinted by 
PISN at a $50\%$ level (Fig.~\ref{Fig:XFe_FeH}, right bottom panel). 
How can we discriminate these different chemical enrichment channels?
Fig.~\ref{Fig:XFe_SNIa} shows the complete chemical abundance patterns we do expect when 
${\rm [Fe/H]_{out}\approx -2}$ for different ${\rm f_{Ia}}$. We note that independent on ${\rm f_{Ia}}$ 
we have ${\rm \langle [(N, Cu, Zn)/Fe] \rangle \leq 0}$, i.e. all the values are similar to the highest 
[X/Fe] peaks of an ISM enriched by PISN at a $50\%$ level. Yet, there are some key differences. 
For all ${\rm f_{Ia}}$, ${\rm \langle [(Cr, Ni)/Fe] \rangle >0}$, while in environments enriched by PISN 
at a ${\geq 50\%}$ level we have different valued depending upon the ${\rm m_{PISN}}$. Furthermore, 
for the most likely and observationally constrained SNIa-to-SNII ratio, i.e. ${\rm f_{Ia}\approx 0.1}$, the 
light elements O, S, and Si, show similar super-solar values. In conclusion, there are several elements 
that can allow us to discriminate among an ISM polluted by SNII and a PISN or by SNII and SNIa.

\section{BD +80$^\circ$ 245: the smoking gun of a PISN explosion ? }
\label{Sec:bd}
In the following section, we suggest that the peculiar abundance pattern 
of the star BD +80$^\circ$ 245 may be explained if it has been formed 
from gas polluted by a primordial PISN, with ${\rm f_{PISN}\ge 0.5}$. 
This star, with [Fe/H]$= -2.09$, was discovered as an $\alpha-$poor 
star by \citet{carney97}.

The star has been subsequently analysed by \citet{fulbright00}, based on 
spectra taken at Keck telescope with the HIRES spectrograph \citep{hires}, 
by \citet{ivans03} and \citet{roederer14AJ}. The latter two investigations, as 
well as the original discovery paper, are all based on spectra obtained at the 
McDonald Observatory with the ``2d-coud\'e'' echelle spectrograph at the 
2.7m Harlan J. Smith telescope. Each of the three investigation acquired an 
independent spectrum. All four investigations agree, within errors, on atmospheric 
parameters and detailed abundances of the different elements. The two most 
interesting  studies for our purposes are those of \citet{ivans03} and \citet{roederer14AJ}.

Both studies report a significant under-abundance of Zn with 
respect to iron, which is unusual. Here and in the following we
transform the published abundances to our preferred solar abundace
scale (see Table~1), namely A(N) = 7.86, 
A(Fe$_{\odot}$) = 7.52 \citep{CaffauSolar} and
A(Cu$_{\odot}$) = 4.25, A(Zn$_{\odot}$) = 4.47 \citep{lodders09}.
On this scale \citet{ivans03} reports  [Zn/Fe]=--0.29,
and \citet{roederer14AJ} [Zn/Fe]=--0.27. 
\citet{ivans03} reports also an upper limit on Cu, 
namely [Cu/Fe]$< -1.07$ and \citet{roederer14AJ} an upper limit
on N, namely [N/Fe]$<+1.0$. 
\subsection{BD +80$^\circ$ 245: new spectra and analysis}
\label{Sec:obs_bd}
We observed two spectra of  BD +80$^\circ$ 245 with the SOPHIE spectrograph
\citep{SOPHIE} on the 1.93\,m telescope at Observatoire de Haute Provence, 
on March 16th with an integration of 466\,s and March 17th 2018 with an 
integration of 1800\,s. In both nights the meteorological
observations were far from ideal and we observed through thin to thick 
clouds conditions. We used the high efficiency mode (HE)
that provides a resolving power of R=39000. 
The radial velocities measured by the SOPHIE pipeline
with a G2 mask where 4.96 \kms\ and 4.98 \kms\ respectively, 
in excellent agreement with the mean radial velocity
measured by \citet{latham2002}. The S/N ratio of the coadded spectrum 
is 53 at 510\,nm. The goal of our observations was to try to improve on
the upper limits on the abundances of N and Cu, but the quality
of the spectra does not allow that. 

Nevertheless the SOPHIE spectrum allowed a robust determination of 
12 elements, of which three are in two different ionisation stages.
To derive the abundances we use the \mygi\ code \citep{mygi}
and a grid of synthetic spectra computed with the SYNTHE code
\citep{kurucz05} in its Linux version \citep{sbordone04}, from
a grid of one-dimensional LTE, plane-parallel model atmospheres computed
with version 12 of the ATLAS code \citep{kurucz05}.
The adopted atomic line list is described by \citet{Heiter}
and the molecular lines were those compiled by 
R. Kurucz\footnote{{http://kurucz.harvard.edu/linelists/linesmol}}. 
The results are reported in Table\,\ref{abundanceBD} and they are, by and large, 
in good agreement with those of \citet{ivans03}. The small offsets
of the order of 0.15\,dex, for some elements, can be ascribed to the
different lines and atomic parameters adopted in the two studies.

\begin{figure}
\centering
  \resizebox{!}{5.5cm}{\includegraphics[clip=true]{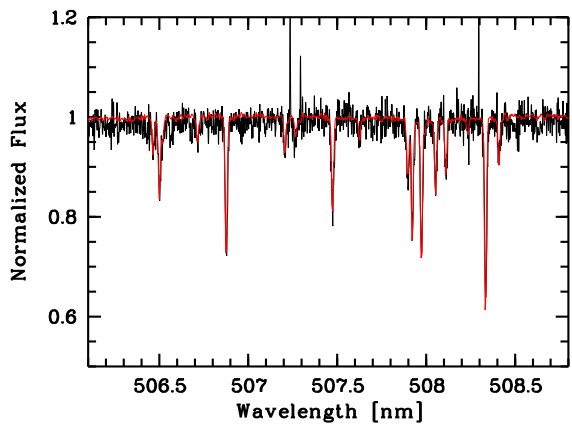}}
    \caption{A portion of our our SOPHIE spectrum of BD +80$^\circ$ 245 \relax  compared
to the HIRES spectrum. The HIRES spectrum is at much higher S/N (350 to be compared 
to 53 of the SOPHIE spectrum in this region) and slightly higher resolution. However, the 
two wavelength scales are in good agreement.}
    \label{Fig:SOPHIE_HIRES}
\end{figure}

We retrieved from  the Keck Observatory Archive (KOA\footnote{\url{https://koa.ipac.caltech.edu/}}) 
one HIRES \citep{hires} spectrum of BD +80$^\circ$ 245 which was available already
reduced. This spectrum was observed on June 12th 2006 with a slit
of 0\farcs{861}, which provides a resolving power of R=45000. The spectral
coverage is from 290\,nm to 585\,nm, the exposure time 1200\,s and the S/N ratio
at 500\,nm is about 350. 
We are aware that the KOA extracted spectra are intended for quick look and
are not guaranteed to be ``science ready'', however in this case we could verify
the HIRES spectrum against the SOPHIE spectrum (see Fig.\,\ref{Fig:SOPHIE_HIRES} for an example)
and convinced ourselves that both the wavelength scale and the intensity scale are
free from significant systematic errors, therefore suitable for a science analysis. 

There is a good agreement in the literature on the atmospheric parameters of 
BD +80$^\circ$ 245 we decided to adopt those of \citet{ivans03} to simplify the 
comparison with their results: \teff\ = 5225\, K log g = 3.00 and a metallicity of --2.0. 
We computed an ATLAS 9 model \citep{kurucz05}, using
its linux version \citep{sbordone04} using the updated Opacity Distribution Function
of \citet{CK03} with a 1\kms\ microturbulence.
\subsubsection{BD +80$^\circ$ 245: the key element Cu}
\begin{figure}
  \centering
  \resizebox{!}{5.5cm}{\includegraphics[clip=true]{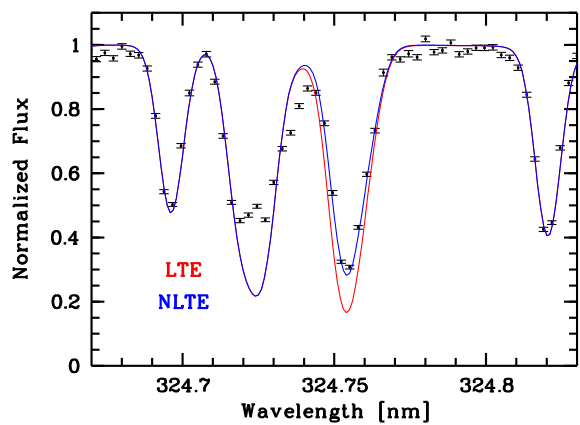}}
    \caption{The HIRES spectrum of BD +80$^\circ$ 245 \relax 
    in the region of the Cu{\sc i} resonance line at 324.7\,nm (dots). Two synthetic spectra 
    are shown for an ATLAS 9 model with \teff = 5225\,K log g = 3.00 and [M/H]=--2.0.
    The red line corresponds to the LTE computation and the blue line to the NLTE computation.  
    In both cases  A(Cu)=1.32.}
    \label{Fig:Cu_3247}
\end{figure}
\begin{figure}
  \centering
  \resizebox{!}{5.5cm}{\includegraphics[clip=true]{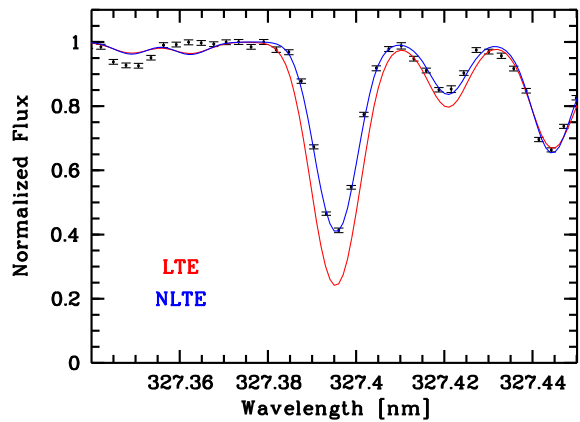}}
    \caption{The same as in Fig.~14 for the CuI resonance line at 327.3nm. Synthetic spectra assume A(Cu)=1.26.} 
    \label{Fig:Cu_3273}
\end{figure}

\label{Sec:Cu_bd}
The line of Cu{\sc i} at 510.5\,nm cannot be detected in the HIRES spectrum, 
however both the UV resonance lines at 324.7\,nm and 327.3\,nm can be readily 
detected as shown in Fig.\,\ref{Fig:Cu_3247} and Fig.\,\ref{Fig:Cu_3273}.
These lines are known to suffer from strong NLTE effects \citep{andrievsky18,korotin18}, 
we therefore computed NLTE profiles to determine the Cu abundances and these profiles 
are shown in Fig.\,\ref{Fig:Cu_3247} and Fig.\,\ref{Fig:Cu_3273}. The two lines provide a 
consistent Cu abundance and the average is A(Cu)$=1.29\pm 0.03$\,dex, or [Cu/Fe]$=-0.75$.
We verified that this Cu abundance is consistent with the non-detection
of the  Cu{\sc i}  510.5\,nm line. 
\citet{bonifacio10} have also shown that the UV resonance lines are strongly
affected by granulation effects that go in the opposite direction: 3D corrections
to the Cu abundance are always negative, while NLTE corrections are always positive. 
At the present time we are not able to perform a full 3D--NLTE computation, 
but we can gain some insight from the work of \citet{roederer18}. Since 
Cu{\sc ii} is the majority species in the atmospeheres of these stars, we can assume
that its lines form in conditions that are close to LTE, thus the difference 
between the LTE abundance of Cu{\sc ii} and that of Cu{\sc i} can be taken as a proxy
of the NLTE correction for Cu{\sc i}. In their figure 3 they show how for their sample of stars
this proxy follows the general trend of the NLTE corrections of \citet{andrievsky18}, being
somewhat lower. From this comparison we can expect that the full 3D--NLTE correction
is smaller than the 1D NLTE. The consequence for 
BD +80$^\circ$ 245 is that our NLTE Cu abundance is probably an upper limit
to the true 3D--NLTE abundance. Thus our conclusion that Cu is strongly underabundant
with respect to Fe in this star is robust against NLTE and 3D effects. 
\begin{figure}
  \centering                                      
  \resizebox{!}{5.5cm}{\includegraphics[clip=true]{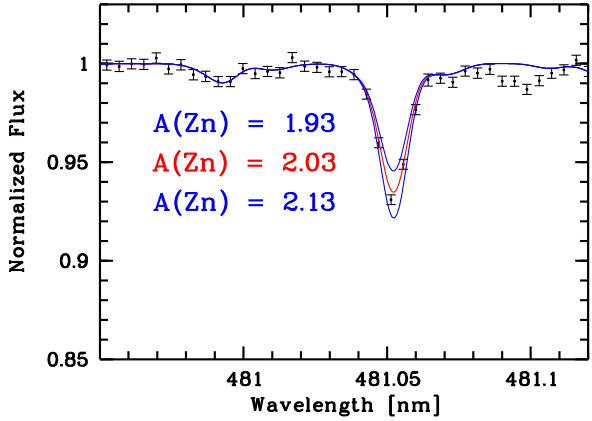}}  
\caption{The same as in Fig.~14 for the Zn{\sc i} line at 481.0\,nm (dots). 
    The red line corresponds to the abundance determined by us with a $\chi^2$ fit,
    and the two blue lines correspond to the $\pm 0.1$\,dex variation in the Zn abundance.}
    \label{Fig:Zn_HIRES}
\end{figure}
\begin{figure}
  \centering                                      
  \resizebox{!}{5.5cm}{\includegraphics[clip=true]{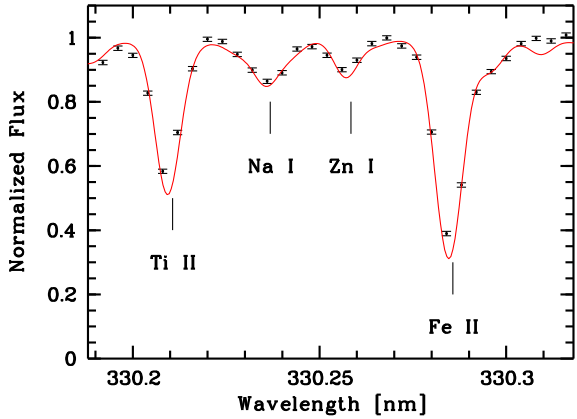}}  
\caption{The same as in Fig.~14 for the Zn{\sc i} line at 330.2\,nm. Synthetic spectra assume A(Zn) = 2.21.} 
   \label{Fig:Zn3302}
\end{figure}
\subsubsection{BD +80$^\circ$ 245: the key element Zn}
\label{Sec:Zn_bd}
The line of Zn{\sc i} at 481.0\,nm can be readily detected in the HIRES spectrum, 
as shown in Fig.\,\ref{Fig:Zn_HIRES}. To derive the Zn abundance we use a 
$\chi^2$ fitting of synthetic line profiles and derive A(Zn)=2.03, this is in excellent 
agreement with the values derived by both \citet{ivans03} and \citet{roederer14AJ}. 
The difference of 0.06 dex in our Zn abundance and that of \citet{ivans03} is due 
to both the different log gf assumed for this line (-0.137 assumed by us and -0.170 
assumed by Ivans et al. (2003)) and to the slightly different equivalent width.
Our fitted equivalent width is 0.67\,pm in excellent agreement with 
\citet[][0.65\,pm]{roederer14AJ}. In addition we could detect the Zn{\sc i} doublet lines 
at 330.2\,nm in the HIRES spectrum. Our best fit is shown in Fig.~\ref{Fig:Zn3302} 
and it provides A(Zn)=2.21. 
The region is rather crowded and the continuum is difficult to define, and it is necessary 
to model also the neighbouring lines. For this reason this 
line is probably less reliable than the 481.0\,nm line.  Nevertheless the two lines provide 
consistent abundance the straight mean of the two is A(Zn) = 2.12\, dex with a $\sigma$ 
of 0.09\, dex. The Zn and Fe abundances should be fairly immune both to departures from 
LTE ad from 3D effects at the metallicity of BD +80$^\circ$ 245. The departures from LTE 
for Zn have been studied by  \citet{takeda05} and any correction is expected to be less than 
0.1\,dex. This is also supported by the recent study of \citet{roederer18}, who measured both 
{Zn}{\sc i} and {Zn}{\sc ii} lines in a set of stars. As for Cu, on the assumption that the lines 
of the majority species in the stellar atmosphere, {Zn}{\sc ii}, are formed under condition that 
are close to LTE, then the difference of abundance between {Zn}{\sc ii} and  {Zn}{\sc i} can be 
taken as a proxy of the NLTE correction that has to be applied to the abundances derived from 
the Zn{\sc i}. For all the sample of of \citet{roederer18} the abundances derived from the two
ions are very similar, implying small NLTE corrections, in agreement with the computations of 
\citet{takeda05}. For Fe the NLTE corrections at [Fe/H]$\approx -2.0$ are expected to be 
less than 0.1\, dex for giants with log g $\ge 2.0$ \citep[][]{mashonkina16}. \citet{duffau17} studied 
the effects of granulation (3D effects) on Zn, for solar metallicity and for --1.0, concluding that 
these effects are small and the [Zn/Fe] ratio should  not be strongly affected.
\begin{figure}
  \centering                                      
  \resizebox{!}{5.5cm}{\includegraphics[clip=true]{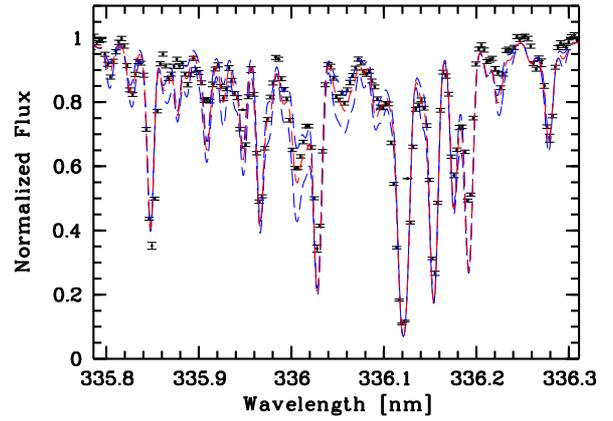}}  
\caption{The HIRES spectrum of BD +80$^\circ$ 245 \relax 
    in the region of the NH band at 336\,nm (dots) and the best
fit synthetic spectrum corresponding to A(N) = 4.87 (solid red line).
We also show synthetic spectra compute with $\pm 0.2$\,dex of our best
fit N abundance (dashed blue lines) and take this as error estimate. }
    \label{Fig:NH}
\end{figure}
\subsubsection{BD +80$^\circ$ 245: the key element N}
\label{Sec:N_bd}
The HIRES spectrum also allowed us to detect the NH band at 336\,nm, as shown in 
Fig.~\ref{Fig:NH}. Our best fit synthetic spectrum provides A(N)=4.86, or [N/Fe]$=-0.79$.
To fit the whole region we assumed the oxygen abundance derived from the OH lines
(see next Section) and the abundance of Ti and Sc provided by \citet{ivans03}.
The uncertainty in this fitting is due to the continuum placement and we estimate it to be 0.2\,dex.
The NH band is subject to granulation corrections, which always decrease the abundance 
with respect to that derived from a 1D analysis \citep[see e.g.][and references therein]{Bonifacio13,Collet18}. 
We have not yet performed this computation, since it is computationally demanding nor could we
find computations in the literature for stars of similar atmospheric parameters. 
For the purpose of the present investigation we simply note that since the 1D analysis implies
that N is under-abundant with respect to iron, a 3D analysis will result in an even
lower N abundance.
\subsubsection{BD +80$^\circ$ 245: oxygen determination}
\label{Sec:O_bd}
We could detect many OH lines of the $A^2\Sigma^+ - X^2\Pi$ system, that allowed us to 
determine the oxygen abundance in BD +80$^\circ$ 245. We used wavelengths from 
\citet{Stark94} and log\,$gf$ values calculated from the theoretical lifetimes of \citet{GG81}. 
Note that these are almost identical to the more recent theoretical calculations of 
\citet{gillis01}. 
The only exception was the line at 311.0780nm for which we adopted 
log\,$gf=-3.274$ from the updated lists of 
R. Kurucz\footnote{{http://kurucz.harvard.edu/molecules/oh/ohaxupdate.asc}}.
Using a synthetic spectrum as a guideline we selected only lines that were either clean 
or mildly blended. In some cases two or more OH lines were blended together and we 
fit the whole feature. When a line appeared in two adjacent echelle orders we took two 
independent measures. In total we have 46 independent measures of oxygen that provide 
a mean abundance A(O) = 6.85 with a line-to-line scatter of 0.11. This corresponds 
to [O/Fe]=+0.30, and it is lower than the mean ratio in Galactic stars at these [Fe/H].
The OH lines are also known to be affected by granulation effects 
\citep[see e.g.][and references therein]{Gallagher17,Collet18}.
We apply the correction computed by \citet{JGH10}, which amounts to $-0.25$\,dex, 
implying A(O)=6.60, or [O/Fe]$=+0.06$. 
\subsubsection{BD +80$^\circ$ 245: age and orbit determination}
\label{Sec:age_bd}
\begin{figure}
  \centering                                      
  \resizebox{!}{7.5cm}{\includegraphics[clip=true]{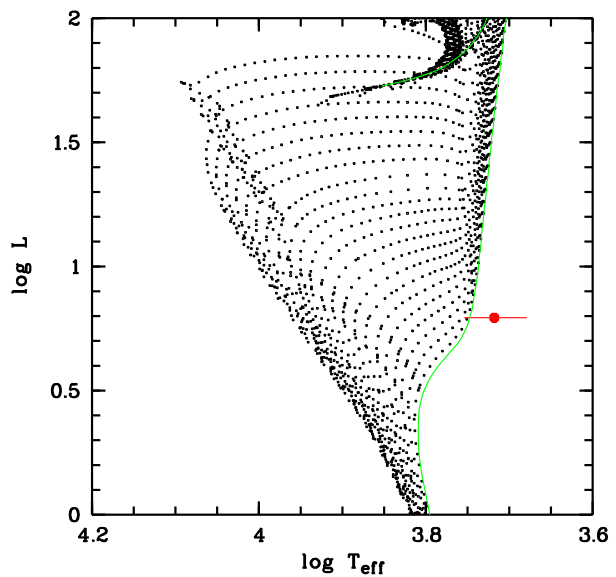}}  
\caption{PARSEC isochrones \citep{bressan12} with $Z=0.000152$ and ages 
between 1 and 13.5 Gyr. The green isochrone corresponds to 13.5 Gyr. 
The red point shows the position of BD +80$^\circ$ 245 in this diagram, and error
bars correspond to 3 $\sigma$ errors in \teff\ and parallax. }
    \label{Fig:HR}
\end{figure}

The second data release of the Gaia mission \citep{Gaia,GaiaDR2,GaiaDR2_valid}
has provided a very accurate parallax for this star $4.296 \pm 0.0270$ mas,
which allows us to have an accurate measurement of its luminosity. 
In Fig.\,\ref{Fig:HR} we show the Hertzsprung-Russel diagram with the
position of the star compared to a set of PARSEC isochrones \citep{bressan12}
with $Z=0.000152$ and ages between 1 Gyr and 13.5 Gyr. 
We assumed an error of 150\,K on the effective temperature of the star
and draw $3\sigma$ error bars in the diagram.
The position of the star is anomalous, it appears too cool for its luminosity. 
An increase in \teff\ would also implies an increase in metallicity, 
which also would help to make the position of this star 
more compatible with the isochrones. An increase in metallicity moves the isochrones
towards cooler temperatures in this diagram.
It may also be that the position of this star is due to its
peculiar chemical composition. The fact that the star does not show any radial
velocity variations does not support the hypothesis that the star is a binary. 
In spite of its anomalous position we can say that it certainly points towards 
a very old age. BD +80$^\circ$ 245 was probably born only a few $10^8$ 
years after the Big Bang.  

Finally, we used the Gaia parallax and proper motions, together with our measured
radial velocity, to compute a Galactic orbit for BD+80$^\circ$ 245. To this end we used 
{\tt GravPot16} \citep{GravPot16} and its default gravitational potential, based on the 
Besan{\c c}on model\footnote{\url{https://fernandez-trincado.github.io/GravPot16/index.html}}.
The orbit of BD +80$^\circ$ 245 is a typical ``halo'' orbit: highly eccentric ($e=0.46$), 
maximum height above the Galactic plane of 13\,kpc, apocentric distance of 14.6\,kpc 
and a slow retrograde rotation. This is compatible both for a star accreted from a satellite 
galaxy, and with an ``in situ'' halo star.
\subsection{BD +80$^\circ$ 245: model comparison}
\label{Sec:comparison_bd}
In conclusion, BD +80$^\circ$ 245 can be a descendant of very massive first stars since: (i) it is likely
a very ancient star and (ii) it exhibits under-abundances of the three key elements, N, Cu and Zn, which
are consistent with the abundance ratios shown in Fig.~\ref{Fig:XFe_FeH} for $\rm f_{PISN}=0.5$. 
We can thus directly compare the chemical abundance pattern of BD +80$^\circ$ 245 with our model 
predictions for an ISM imprinted by very massive first stars and determine the best fitting model. 
To this end we computed the ${\chi}^2$ distribution normalized to the degrees of freedom, 
${\chi^2}_{\nu}={\chi}^2/({\rm N-M})$ (e.g. see Ishigaki et al. 2018), where ${\rm N}$ is the number 
of data points to be fitted, ${\rm M}$ the number of model free-parameters (\fratio, ${\rm f_{PISN}}$, 
${\rm t_{popII}}$, ${\rm m_{PISN}}$) and ${\chi^2}$ has been computed as:
\be
{\rm \chi^2 = \Sigma^N_{i=1} \frac{\big([X_i/Fe]_{obs} - [X_i/Fe]_{th}\big)^2}{\big(\sigma^2_{{[X_i/Fe]}_{obs}}+\sigma^2_{{[X_i/Fe]}_{th}}\big)}}.
\label{eq:chi_square}
\ee
In the above expression ${\rm [X_i/Fe]_{obs}}$ is the observed abundance ratio of the element $i$ with 
respect to iron and ${\rm \sigma^2_{{[X_i/Fe]}_{obs}}={\sigma^2}_{Xi, obs}+{\sigma^2}_{Fe, obs}}$ 
the associated observational error. These values are reported in Table 1 for all measured elements. 
For Na, Al, Si, K, V, Mn, Co, we used the values measured by Ivans et al. 2004 and we estimated 
${\rm \sigma^2_{{[X_i/Fe]}_{obs}}={\sigma^2}_{Al, obs}+{\sigma^2}_{Fe, obs}\approx 0.15}$.
The terms ${\rm [X_i/Fe]_{th}}$ in eq.~\ref{eq:chi_square} represent the theoretical chemical abundance 
ratios of an ISM imprinted by both a PISN and normal Pop~II stars. Given the large yield uncertainties 
among different models for Pop~II stars (Fig.~10, left panels) we assume ${\rm \sigma^2_{{[X_i/Fe]}_{th}}=0.5}$ 
for all elements.
\begin{figure}
  \includegraphics[width=0.99\linewidth]{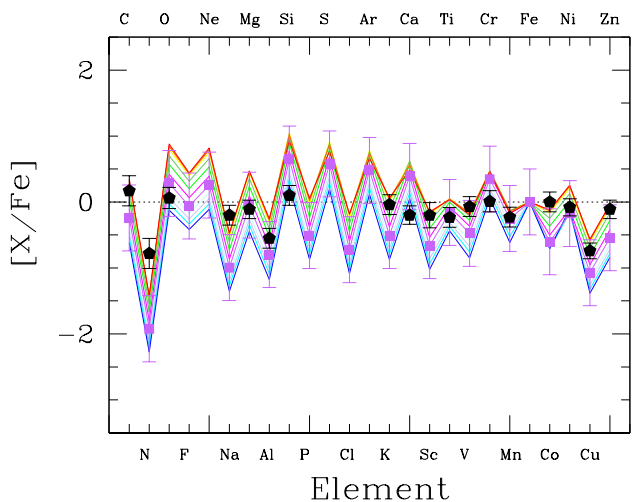}        
    \caption{Comparison between the chemical abundance pattern of BD +80$^\circ$ 245 inferred
    from spectroscopic observations (black circles with errorbars, see text and Table 1) and the 
    best fitting model parameters: \fratio$=10^{-4}$, ${\rm f_{PISN}=0.5}$, and ${\rm t_{popII} = 30}$~Myr.
    The different colours/curves show the chemical abundance pattern predicted for different PISN
    masses (see Fig.~\ref{Fig:XFe_PISN}). {\it Violet squares with errorbars} show the best-fitting
    model, which corresponds to ${\rm m_{PISN}=223 \Msun}$.}
    \label{Fig:data_predictions}
\end{figure}
The results of the ${\chi^2}_{\nu}$ distribution for all our model parameters is reported in Appendix~A.
Our analysis show that models with ${\rm f_{PISN} = 0.5}$, ${\rm t_{popII} = 30}$~Myr, and \fratio$=10^{-4}$,
are those that best-fit the abundance pattern of BD +80$^\circ$ 245, providing ${\chi^2}_{\nu} < 2$ in the 
overall PISN mass range. In Fig.~\ref{Fig:data_predictions} the chemical abundance pattern of all these 
models are compared with BD +80$^\circ$ 245. Our best fitting-model (violet points with errobars) 
corresponds to a PISN mass ${\rm m_{PISN}=223 \Msun}$, which gives ${\chi^2}_{\nu} = 0.99$. 
As expected, the agreement between this model and BD +80$^\circ$ 245 is quite good, thus suggesting 
that this rare star is really a PISN descendant. It is worth noticing that although \fratio$=10^{-4}$ is typical 
of mini-halos (see Sec.~7.1) we cannot exclude that BD +80$^\circ$ 245 has been polluted by more than 
one PISN, a possibility that we did not explore with our current modelling.
On the contrary, it is unlikely that BD +80$^\circ$ 245 formed out of an ISM enriched by both SNII 
and SNIa as the star is very ancient and the models provide a partial agreement to the data (see 
Fig.~\ref{Fig:XFe_SNIa}). Only by assuming an equal contribution of SNIa and SNII, ${\rm f_{Ia}=1}$, 
we obtain a partial agreement with the data (middle panel in Fig.~\ref{Fig:XFe_SNIa}). 
Yet, even this model largely under-estimates the [(C, Na, K, Sc)/Fe] abundance ratios and it 
overestimates [(Ni, Cr)/Fe]. In conclusion, all available data suggest that BD +80$^\circ$ 245 
might really be the first evidence of a PISN descendants.
\section{Observational prospects}
In Sect.\,\ref{Sec:bd} we argued that  BD +80$^\circ$ 245  has all the chemical signatures 
that allow to single out a true descendant of a PISN. As suggestive as this may be, a single 
star cannot strongly constrain the PISN pollution models. 
We therefore believe it is very important to search for other stars with the PISN chemical signature, 
both in order to be sure that this is not an exception, and, if such stars are found, to know what is 
their frequency. The latter would in fact constrain the number of PISN exploding in the early Galaxy 
and thus the IMF of Pop III stars \citep[e.g.][]{DB17}.
\citet{aoki14} claimed that SDSS\,J001820.5--093939.2 is another such PISN descendant. Their 
claim is based mainly on the low $\alpha$/Fe ratios in this star, however this signature could also 
indicate a prompt enrichment of the ISM by SNIa as suggested by \citet{ivans03} for BD +80$^\circ$ 245.
Furthermore it is now becoming clear that there is a population of low-$\alpha$ stars at low metallicity 
\citep{caffau13,bonifacio18}. Are all of these PISN descendants? Some fraction of them?  
SDSS\,J001820.5--093939.2 could well be a PISN descendant, however unfortunately, it is too faint 
to allow a measurement of the key elements, N, Cu and Zn. 
\begin{table*}
\centering
\caption{Chemical abundance ratios of BD$+80^\circ 245$ derived with our analysis. The lines used 
to determine the different chemical elements are reported in Appendix B (Tables 2-10).}
\label{abundanceBD}
\begin{tabular}{lccccccrrl} 
\hline
Ion & A(X$_\odot$) & A(X) & uncertainty & [X/H] & [X/Fe] & $\Delta$[X/Fe] & N lines & spectrum & Comment \\ 
\hline
C          & 8.50 & 6.46 & 0.20 & $-2.045$& $+0.17$& $0.23$&  & HIRES  & G-band\\
N          & 7.86 & 4.86 & 0.20 & $-3.00$& $-0.78$& $0.23$&  & HIRES  & NH band 336\,nm \\
O          & 8.76 & 6.60 & 0.11 & $-2.16$& $+0.06$& $0.16$& 43 & HIRES  & OH, 3D corrected\\
Mg{\sc i}  & 7.54 & 5.20 & 0.08 & $-2.33$& $-0.11$& $0.01$& 3 & SOPHIE & \\
Ca{\sc i}  & 6.33 & 3.91 & 0.08 & $-2.42$& $-0.2$& $0.14$& 18 & SOPHIE &\\
Sc{\sc ii} & 3.10 & 0.68 & 0.15 & $-2.42$& $-0.2$& $0.19$& 5 & SOPHIE &\\
Ti{\sc i}  & 4.90 & 2.39 & 0.08 & $-2.51$&  $-0.29$& $0.14$& 9 & SOPHIE & \\
Ti{\sc ii} & 4.90 & 2.50 & 0.10 & $-2.40$& $-0.18$& $0.15$& 25 & SOPHIE & \\
Cr{\sc i}  & 5.64 & 3.27 & 0.07 & $-2.37$& $-0.15$& $0.16$& 11 & SOPHIE & \\
Cr{\sc ii} & 5.64 & 3.60 & 0.12 & $-2.04$& $+0.17$&  $0.16$& 5 & SOPHIE & \\
Fe{\sc i}  & 7.52 & 5.31 & 0.11 & $-2.21$&    &     & 150 & SOPHIE & \\
Fe{\sc ii} & 7.52 & 5.30 & 0.06 & $-2.22$&    &     &  23 & SOPHIE & \\
Ni{\sc i}  & 6.23 & 3.94 & 0.06 & $-2.30$& $-0.08$& $0.13$& 14 & SOPHIE & \\
Cu{\sc i}  & 4.25 & 1.29 & 0.03 & $-2.96$& $-0.74$& $0.12$& 2 & HIRES  & NLTE \\
Zn{\sc i}  & 4.47 & 2.14 & 0.09 & $-2.33$& $-0.11$& $0.14$& 2 & HIRES  & \\
Sr{\sc ii} & 2.92 & $-0.12$ & 0.17 & $-3.04$& $-0.82$& $0.2$& 2 & SOPHIE & \\
Y{\sc ii} & 2.21 & $-1.25$ & 0.20 & $-3.46$& $-1.24$& $0.23$& 1 & SOPHIE & \\
Ba{\sc ii} & 2.17 & $-1.93$ & 0.25 & $-4.1$& $-1.89$& $0.27$& 3 & SOPHIE & \\
\hline
\end{tabular}
\end{table*}
\subsection{Looking for the killing elements}
Our simple model, and the study of BD +80$^\circ$ 245 can give us some insight into what stars and 
lines we should look for. The strongest spectral features to look for are the Cu{\sc i} UV resonance 
lines and the NH 336\,nm band as shown in Fig.\,\ref{Fig:Cu_3247}, \ref{Fig:Cu_3273} and \ref{Fig:NH}. 
The penalty is that observing in the UV is difficult from the ground and one needs a fairly high resolution
($R\approx 20000$). For all three features a resolution that is one half of that of the spectra analysed 
here is sufficient and the S/N ratio can be as low as 30. 

However the real problem is that as of today, to our knowledge, there are only 5 spectrographs that can 
observe at these wavelengths with a sufficient, or barely sufficient resolution and they are all installed on 
6 to 10\,m class telescopes. At the VLT there are UVES \citep{UVES}  and X-Shooter \citep{X-Shooter}. 
The latter only if one could observe with a 0\farcs{3} slit or an image slicer of the same width, currently 
neither is available. There is HIRES \citep{hires} at Keck, as used in this paper and there is HDS \citep{HDS} 
at Subaru. It is unrealistic to conduct a survey of a large number of stars with any of these instruments,
because of the high pressure on the telescopes and because they all operate in single object mode. 
Probably the most promising instrument is the fifth one: IMACS+MOE \citep{IMACS,MOE} on the Magellan 
telescope. This provides a resolving power R$\sim 20000$ covering the spectral regions relevant for N and 
Cu and has a multi-object capability, over a field-of-view of $15'\times 15'$. This field of view is interesting, 
but, as we shall see below, probably not sufficient to conduct a survey searching for PISN descendants.

None of the wide-field multi-object spectrographs that either exist or are being built has access to the far 
UV region necessary for this study. This is due to a combination of the difficulty of having a fibre-fed 
spectrograph that is efficient in the UV and of the difficulty of having a wide-field high resolution multi-slit 
spectrograph. IMACS+MOE here is the exception and it would be interesting to see if the concept could 
be transferred to a really wide field, of the order of one degree diameter.  

So in spite of the fact that the features of N and Cu are stronger, it turns out that the most promising candidate 
to search for true PISN descendants is Zn. One needs to have a S/N ratio and resolving power sufficient to 
measure the Zn{\sc i} line at 481.0\,nm when it has an equivalent width of 0.5 to 1.0\,pm. Clearly one would gain 
by observing stars cooler than BD +80$^\circ$ 245 since for a given Zn abundance the line becomes stronger. 
However the isochrones in Fig.\,\ref{Fig:HR} show how the Red Giant Branch climbs almost vertical and the cool 
giants of \teff\ $\sim 4000$\,K are found only at the RGB tip and thus are a very rare population. 
\subsection{Zn surveys to catch the PISN descendants}
There are two instruments that can conduct a survey of Zn abundances in metal-poor stars to catch 
the descendants of very massive first stars: HERMES at the AAT \citep{HERMES_AAT} and WEAVE 
at the WHT \citep{WEAVE}. Unfortunately 4MOST on Vista \citep{4MOST} will not cover the Zn{\sc i} 
481.0\,nm line at high resolution. Yet, we should note that 4MOST might also help in finding more 
PISN descendants since it will easily measure the [Sc/Ca] ratio, which is predicted to be extremely 
low in PISN-enriched environments (e.g. Fig.~2).
\subsubsection{The HERMES and WEAVE spectrographs}
The HERMES spectrograph uses the 2dF corrector and positioner on the AAT \citep{2dF} to deploy 400 fibers 
over a field of view with 2 degrees diameter. It was commissioned at the end of 2013 \citep{HERMES_firstlight} 
and has been operating for the last 5 years. One of the main surveys being conducted with this instrument is the 
Galactic Archaelogy with HERMES (GALAH) survey \citep{GALAH}. The survey has recently published its second 
data release \citep{GALAH_DR2} that contains 342\,682 stars and derived atmospheric parameters and chemical 
abundances. It normally observes stars in the magnitude range $12 < V < 14$ with no colour selection. This data 
set contains Zn abundances for 248\,993 stars all of which with [Fe/H]$> -2.0$. There are no stars with [Fe/H]$<-1.5$
that have a [Zn/Fe] that is significantly below the Solar abundance.

WEAVE should arrive on the sky at the end of 2019 and will deploy about 1000 fibres over a field of view of two 
degrees of diameter. The fibers feed  a two arms spectrograph \citep{WEAVE_SPEC} that can be operated either in 
a low resolution mode (R$\sim 5000$) or in a high resolution mode (R$\sim 20000$). The high resolution mode covers 
two non contiguous intervals of about 60\,nm in the blue and 90\,nm in the red. The red interval is fixed and is 595\,nm 
to 685\,nm. In the blue one can chose between two gratings which cover either the 404\,nm to 465\,nm (blue grating)
range or the 473\,nm to 545\,nm range (green grating). Only the observations with the green grating are relevant to determine the Zn abundance, since it covers the 481.0\,nm line.  
\subsubsection{The observational strategy}
In order to understand what could be observed with HERMES and WEAVE we used the {\tt galaxia} \citep{galaxia} 
implementation of the Galaxy Besan\c{c}on model \citep{robin03}. We selected as potential targets giants stars with 
\teff\ $< 5300$\,K and log g $\le 3.0$. If we consider a GALAH-like magnitude selection $12 < V < 14.5$, with a 
metallicity in the interval $-1.5 \le$ [Fe/H] $\le -2.5$ the number of candidate stars per square degree is on average 
one for all Galactic latitudes $|b| > 40^\circ$. As we go to lower Galactic latitudes the number increases and is of the 
order of two, with about 1 star out of four coming from the thick disc.  Up to now most of GALAH observations have 
been taken at $|b|< 40^\circ$, it is thus not surprising that there is a lack of metal-poor stars. 

The WEAVE Galactic Archaelogy survey aims at high resolution observations in the magnitude range $13 < V < 16$. 
With this magnitude range and the above conditions on \teff , log g and metallicity the number of targets per square 
degree is of the order of three for $|b|> 40^\circ$ and increases to about six in the range $20 < |b| \le 40$, due to 
the contribution of the metal-weak thick disc.

Considering that both HERMES and WEAVE spectrographs have a field of view larger than three square degrees, 
if we push observations down to $V<16$ we can expect about nine giants in the metallicity range 
$-1.5 \le$[Fe/H]$\le -2.5$ for each pointing. On the assumption that a 1h integration allows to reach a high enough 
S/N ratio to measure Zn and that we can do 7 such exposures per night we would need of the order of 160 nights 
of observations on either or both telescopes. If we can pre-select the metal-poor giants without need of taking 
spectra, such a survey could be executed as a parasite survey, since it needs very few fibres (with respect to the 
total number of fibres available), while the majority of the fibres is dedicated to other targets. Such a pre-selection 
is indeed possible. The Gaia parallaxes are already available for all stars with $V< 16$ and can be effectively used 
to select only giant stars. This allows to filter out nearby K dwarfs that have the same colours of K giants and are 
metal-rich. The temperature range can be easily selected from a colour, and now Gaia provides homogeneous 
$G_{BP} - G_{RP}$ photometry that is well adapted for the purpose. 

The most difficult part is the iron-abundance selection. However in the northern hemisphere the Pristine Survey \citep{Pristine1} 
should be able to select stars with [Fe/H]$<-1.5$ and in the southern hemisphere the same can 
be accomplished with the SkyMapper Survey \citep{SkyMapper}. Both photometric surveys can be usefully 
complemented by the Gaia spectrophotometry. Furthermore, when the Gaia RVS photometry becomes available 
it has been shown that it can provide a very powerful metallicity diagnostic for F,G and K stars \citep{bonifacio_colours}. 
It thus seems that a pre-selection of metal-poor giants is indeed possible. An investment of 160 nights, spread 
over several years and two telescopes seems achievable. Especially considering that the strategy we are 
proposing is little intrusive in the sense that it requires only few fibres to be allocated, while the instrument is 
conducting other surveys. It is more intrusive for WEAVE than for HERMES, for two reasons: in the first place
it requires that the high resolution mode be used, in the second place it requires the use of the green grating. 
The green grating is good for Zn and for metal-rich stars, while the blue grating is optimized for metal-poor stars. 
Independently of the measure of Zn there is an interest for WEAVE to switch to the green grating when going at 
lower galactic latitudes. 

In all those cases targeting 10 to 20 candidate metal-poor giants selected by photometry and parallax would allow 
to assemble a significative sample of Zn measurements in metal-poor giants. If stars with low [Zn/Fe] are found, 
the strategy would be to go to an 8-10m class telescope and observe the UV to derive N and Cu abundances and 
have the confirmation that the star is a PISN descendant.
\section{Summary and discussion}
\label{sec:discussion}
The lack of a convincing chemical signature from massive first stars exploding as Pair Instability 
SN (PISN) is at odds with predictions from cosmological simulations, which find that Pop~III stars 
with masses ${\rm m_{popIII} > 150 \Msun}$ should have formed. All ideas proposed to explain 
these negative results underline the {\it challenges} in finding the rare PISN descendants: it is 
very likely that these stars are less frequent than the descendants of lower mass Pop III stars 
\citep[e.g. Cooke \& Madau 2014][Chiaki et al. 2018]{DB17}. Furthermore, they likely appear at 
higher metallicities, [Fe/H]$\approx -2$, where they only represent $< 0.01\%$ of the total stellar 
population \citep{salvadori07,karlsson08,DB17}. New observational strategies are thus required 
to search for these elusive stellar fossils. 

With this aim, we developed a novel, simple, and generic parametric study to define what are 
the key chemical abundance ratios we should look for to quickly identify the PISN signature. 
We first investigated what are the chemical properties of an inter stellar medium (ISM) 
imprinted by a single PISN. Indeed, state-of-the-art hydro dynamical simulations of the first 
star-forming systems find that typically there is only one star per mini-halo, and that the probability 
to form Pop III stars with different masses is roughly constant (e.g. Hirano et al. 2014, 2015).
To make a general study and account for the different cosmological scenarios (e.g. star-forming
mini-halos vs more massive Ly$\alpha$-cooling systems, in situ vs external metal enrichment)
we modelled the chemical enrichment as a function of the unknown star-formation efficiency, 
${\rm f_*}$, and dilution factor, ${\rm f_{dil}}$, which quantifies the effectively fraction of metals 
injected into the ISM and of the gas used to dilute them (Fig.~1, Sec. 2.1 and 2.2). If several
PISN are formed, e.g. in more efficiently star-forming Ly$\alpha$-cooling halos, we simply
assume that they all have the same mass.

We found that the ISM Fe-abundance depends upon \fratio (eq.~\ref{eq:FeH}), while its 
abundance pattern is independent from both free parameters (eq. \ref{eq:ratios}). 
The key results for an ISM only imprinted by PISN with a single mass, ${\rm m_{PISN}}$, 
can be summarised as follow:
\begin{itemize}
\item for {\it all free parameters} ${\rm Z_{ISM} > 10^{-3}\Zsun > Z_{cr}}$, which implies 
that normal Pop~II stars can promptly form,
\item yet, depending upon \fratio and ${\rm m_{PISN}}$, we have an ISM enriched in the 
broad range, ${\rm -4 < [Fe/H] < 1}$;
\item similarly, an ISM imprinted by PISN with different ${\rm m_{PISN}}$ show a very large 
variety of [X/Fe];
\item {\it independent on all free parameters}, there is an under-abundance 
of the key elements ${\rm [(N, F, Cu, Zn)/Fe]< -1}$.
\end{itemize}

Searching for the lack of N, F, Cu, and Zn with respect to Fe might then be an effective way 
to {\it pre-select} stellar candidates $100\%$ enriched by the chemical product of PISNe. 
However, stars polluted by PISN only, i.e. truly second-generation stars, are 
predicted to be extremely rare \citep[e.g.][Chiaki et al. 2018]{salvadori07,karlsson08,DB17}. 
We then investigated if the under-abundance of these four key elements is preserved in an 
ISM {\it also} polluted by subsequent generation of Pop~II stars, which rapidly evolve as 
as normal core-collapse SN (SNII). By limiting our investigation to ISM with a metal
mass fraction injected by PISN ${\geq 50\%}$ of the total mass, i.e. ${\rm f_{PISN}\geq 0.5}$, 
we were able to derive simple expressions for both the final ISM metallicity and chemical 
abundance ratios. In particular, we find that ${\rm [X/Fe]}$ {\it solely} depends upon the 
fraction of metals from PISN, ${\rm f_{PISN}}$, the PISN mass, ${\rm m_{PISN}}$, and 
the time passed since the Pop~II formation, ${\rm t_{popII}}$, which sets the mass of SNII 
contributing to chemical enrichment. Our results show that after the formation and evolution 
of normal Pop~II stars, an ISM imprinted by a single PISN evolves as follow:
\begin{itemize}
\item the longer ${\rm t_{popII}}$ the more uniform ${\rm [Fe/H]}$ among different 
${\rm m_{PISN}}$, i.e. ${\rm [Fe/H]}$ shrinks towards higher values; 
\item the ${\rm [Fe/H]}$ homogeneity increases for larger Pop~II contribution (lower ${\rm f_{PISN}}$) 
reducing the ${\rm [X/Fe]}$ scatter and hence the signature of PISN with different masses; 
\item independent of ${\rm f_{PISN}}$ the most likely ${\rm [Fe/H]}$ of an ISM imprinted 
by PISN at a $>50\%$ level is ${\rm [Fe/H]}\approx -2$;
\item {\it independent on all free parameters} the under-abundance of the key elements
is preserved: ${\rm [(N, Cu, Zn)/Fe]< 0}$.
\end{itemize}

After checking that the choice of different Pop~II yields (including ``hypernovae") does not 
affect our findings we firmly concluded that ${\rm [(N, Cu, Zn)/Fe]}$ are ``killing element" ratios 
whose under-abundance reveals the unique signature from very massive first stars. 
Armed with these knowledge we explored literature data to pinpoint the the PISN descendants. 
By searching for stars at [Fe/H]$\approx -2$ with [(N, Cu, Zn)/Fe]$< 0.0$ we identified as a 
possible candidate the star BD +80$^\circ$ 245, which was first discovered by \cite{carney97} 
and subsequently observed at high resolution by several authors \citep{fulbright00,ivans03,roederer14AJ}. 
We acquired two new spectra of BD +80$^\circ$ 245 with the SOPHIE spectrograph, which 
allowed us to determine 12 different chemical elements, and re-analysed the high-resolution 
spectra from the Keck archive. Our analysis and model comparison allowed us to conclude 
that BD +80$^\circ$ 245 is likely a smoking gun of a PISN explosion. In fact:
\begin{itemize}
\item {[N/Fe]$=-0.79$, [Cu/Fe]$=-0.75$, and [Zn/Fe]$=-0.12$, consistent with an ISM 
of formation imprinted by PISN at $50\%$ level;}
\item the overall chemical abundance pattern is in good agreement ($\chi^2_\nu = 0.99$) 
with our \fratio$=10^{-4}$ model for an ISM equally polluted by Pop~II stars with masses 
${\rm \geq 8 \Msun}$ (i.e. evolving on a time-scale  ${\rm t_{popII}\approx 30}$~Myr) and 
a single PISN with ${\rm m_{PISN}=233\Msun}$;
\item conversely, it cannot be explained with an enrichment driven by Pop~II stars and SNIa. 
\end{itemize}

Our working hypotheses, driven by state-of-the-art numerical simulations of the first star-forming 
systems (e.g. Hirano et al. 2014, 2015), are the following: 1) the probability to form first stars with 
different masses is flat; 2) in each mini-halo only one PISN forms\footnote{In more massive Ly$\alpha$ 
halos, we assume that PISN formed in each system have all the same mass, ${\rm m_{PISN}}$.}
By relaxing the first hypothesis, e.g. by assuming a decreasing probability to form more massive first 
stars, we might obtain that models that predict an enrichment driven by one PISN and several {\it primordial} 
core-collapse SN with masses ${\rm >8 \Msun}$ still provide a good match of BD +80$^\circ$ 245. 
By relaxing the second one, we might realise that the abundance pattern of BD +80$^\circ$ 245 is 
also well-fitted by models that account for an imprint by more PISNe, which have different masses. 
In spite of these possibilities, which we plan to 
investigate in future studies, we can ultimately conclude that BD +80$^\circ$ 245 seems to be the 
first convincing example of a star imprinted {\it also} by very massive first stars. Ultimately, 
BD +80$^\circ$ 245 probes that primordial stars exploding as Pair Instability SN should have existed.
With current facilities the best strategy to identify more of these very rare PISN descendants is 
to perform Zn surveys, which are achievable with both WEAVE and HERMES spectrographs. 
But where should we look for in order to maximise the probability to find the rare PISN descendants?
\subsection{Were should we look for PISN descendants?}
\label{sec:location}
Our parametric study provide general (and hence very robust) results for the under-abundance 
of the three killing elements and predicts that the most likely iron abundance over the whole 
parameter space is [Fe/H]$\approx -2$. Yet, the typical [Fe/H] values are predicted to vary in 
a broad range depending upon \fratio (Fig.~\ref{Fig:FeH_PISN}), which should reflect different 
hosting halo properties (Fig.~\ref{Fig:sketch}).
\begin{figure*}
 \includegraphics[width=0.99\linewidth]{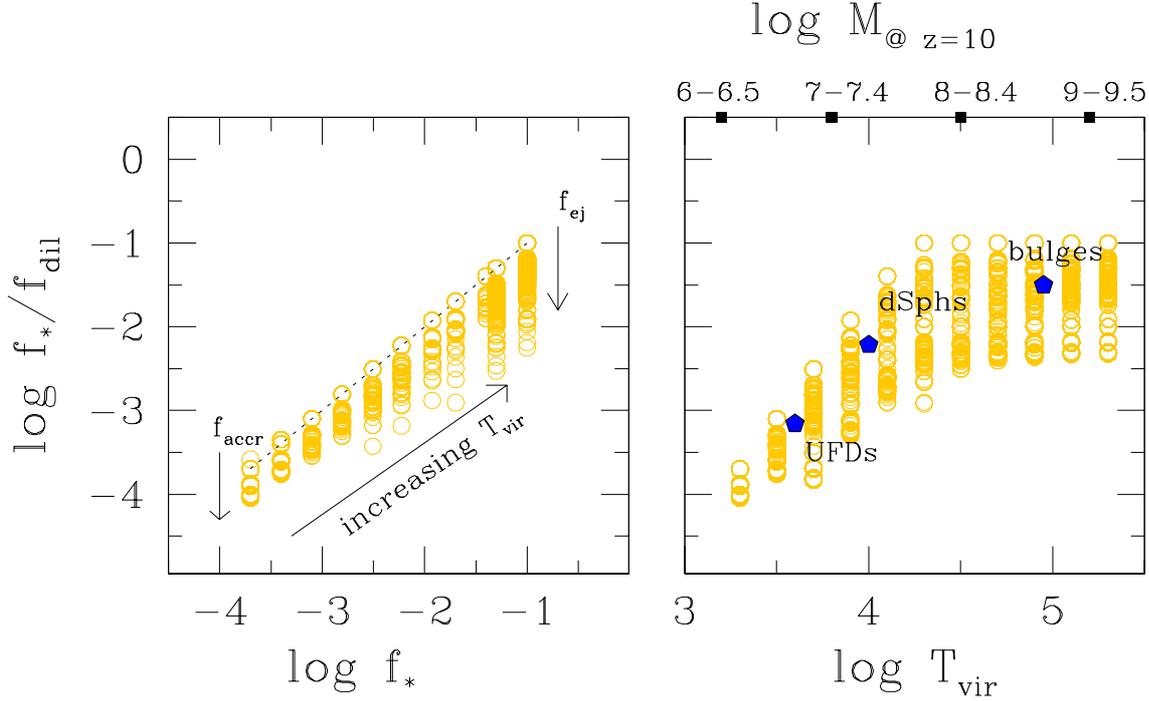}        
    \caption{{\it Left:} Possible values of \fratio as a function of ${\rm f_*}$ (or ${\rm T_{vir}}$, see text) 
    and the other unknowns of the problem, ${\rm f_{accr}}$ and ${\rm f_{ej}}$ (open yellow circles). 
    The black dotted line corresponds to the case ${\rm f_{accr} = 0}$ and minimum ${\rm f_{ej}}$. 
    {\it Right:} \fratio as a function of ${\rm T_{vir}}$ and the corresponding halo virial mass evaluated 
    at ${\rm z=10}$ (open yellow circles). The filled blue penthagones show the expected mean \fratio 
    in environments with different ${\rm T_{vir}}$ or different dark matter halo masses.}
    \label{Fig:free_par}
\end{figure*}

To gain insights on the environments of formation of very massive first stars and hence on the 
actual location of their descendants and typical [Fe/H], we need to link our free parameters 
to the properties of star-forming halos. To this end we should assume a cosmological model
and we will hereafter refer to the $\Lambda$CMD cosmology. In this scenario the first stars are 
predicted to form in low-mass primordial mini-haloes, with total mass ${M\approx 10^{6-7}}\Msun$ 
and gas virial temperature ${T_{vir}\leq 10^4}$~K (e.g. Bromm \& Yoshida 2011; Bromm 2013). 
These low-mass objects cool-down their gas via H$_2$ molecules, which are easily photo-dissociated 
by Lyman Werner photons. This makes mini-halos inefficiently star-forming systems meaning 
that their star-formation efficiency is much lower than the one of more massive (Ly$\alpha$-cooling) 
halos. Recently, \cite{DB17} demonstrated that the semi-empirical formula proposed by \cite{salvadori09} 
provides an excellent tool to evaluate the reduced star-formation efficiency of mini-haloes 
with respect to Ly$\alpha$-cooling systems. The formula reads:
\be
{\rm f_* =  \frac{2\times f^{\rm Ly\alpha-cool}_* }{[1+ (2\times 10^4 {\rm K} /{\rm T_{vir}})^3]}}
\label{eq:f_mini}
\ee
and can be used to link ${\rm f_*}$ of mini-halos to their gas virial temperature, which is a 
function of the hosting halo mass and virialization redshift \citep[e.g.][]{barkana01}:
\be
{\rm T_{vir} = 1.98\times 10^4 \; (\frac{\mu}{0.6}) \; (\frac{1+z}{10})\Big[\frac{M}{10^8h^{-1}M_{\odot}}\Big]^{2/3}},
\label{eq:M_Tvir}
\ee
where $\mu$ is the mean molecular weight respectively equal to $\mu = 1.22$ and $0.59$ for 
neutral and fully ionized gas. Hence, according to the $\Lambda$CDM scenario, the smaller 
$f_*$ the lower ${\rm T_{vir}}$ and the smaller the dark matter mass of the hosting halo. For
example, by setting ${\rm f^{\rm Ly\alpha-cool}_*}= 0.1$ and a minimum virial temperature to 
form stars ${\rm T_{vir}} = 2\times 10^3$~K (e.g. Tegmark et al. 1997, Dijkstra et a.l 2004), we 
obtain ${\rm f_*}=(10^{-4},10^{-1})$, which is what we assumed in our model. 

Fig.~\ref{Fig:free_par} (left) illustrates the tight relation between ${\rm f_*}$ and ${\rm T_{vir}}$ while 
showing the expected \fratio values for different ${\rm f_*}$ obtained by using eqs.~\ref{eq:f_dil1}-\ref{eq:f_dil2}. 
According to eq.~\ref{eq:f_dil1}, we have that ${\rm f_{dil}}$ is a function of ${\rm f_*}$ and ${\rm M_g = \ObOm M}$, 
which can be also expressed in terms of ${\rm f_*}$ (and ${\rm z}$) using eqs.~\ref{eq:f_mini}-\ref{eq:M_Tvir}. 
By varying ${\rm f_*}$ in our reference range we get that in most cases ${\rm f_{dil} \geq 1}$. This means 
that, when ${\rm f_*}$ and ${\rm M_g}$ are tightly linked to each others, the amount of metals/gas ejected 
by PISN (or subsequent stellar generations) will always fill the entire galaxy and/or part of them will actually 
escape out of it. We thus need to use eq.~\ref{eq:f_dil2}. 

To evaluate the ejected gas fraction in eq.~\ref{eq:f_dil2}, ${\rm f_{ej}=M_{ej}/M_g}$, we can write ${\rm M_{ej}}$
as a function of the escape velocity of the galaxy, ${\rm v_{e}}$, and of the SN explosion energy transformed 
in kinetic form, ${\rm 1/2 M_{ej}v^2_{e}=E_{kin}}$ (e.g. Larson 1974), where 
${\rm E_{kin}=\nu_{SN}f_*f_{kin}<E_{51}>M_g}$ as discussed in Sec.~2.1. 
By assuming a Navarro-Frenk-White (NFW) prescription for the halo density profile, we can write the circular 
velocity at a position ${\rm x=r/r_{vir}}$ with respect to the halo virial radius \citep{NFW97}. We get 
${\rm v^2_{e}(x)=2v_c^2 [F(c_hx)+\frac{c_hx}{1+x}]/xF(c_h)}$ where ${\rm F(c_h) = ln(1+c_h) - (c_h/1+c_h)}$ 
and ${\rm c_h}$ is the halo concentration parameter. The escape velocity from the centre (${\rm x=0}$) of a 
NFW halo is ${\rm v_e^2=2[c_h/F(c_h)]v_c^2}$, where ${\rm v_c}$ is the circular velocity of the halo, which 
is again a function of ${\rm T_{vir}}$ and ${\rm z}$ (Barkana \& Loeb 2001). In other words we have that 
${\rm f_{ej} =f_{ej}(f_*,f_{kin},z,c_h)}$, i.e. it also depends upon ${\rm f_*}$ and ${\rm z}$.

By varying ${\rm f_* = (10^{-4}-10^{-1})}$, ${\rm z = (30-5)}$, the concentration parameter ${\rm c_h=(1-30)}$, 
${\rm f_{kin} = (0.01-0.1)}$, and ${\rm f_{accr} = (0.0-1.0)}$ we get the results shown in Fig.~\ref{Fig:free_par} 
(left) for the mean PISN explosion energy (and mean stellar mass), ${\rm <E_{51}> = 20}$. It is clear from these 
calculations that \fratio$\in[10^{-4}-10^{-1}]$ as assumed in our study. Furthermore, when ${\rm f_*}$ and 
${\rm f_{dil}}$ are linked to each other via the $\Lambda$CDM cosmological model, we see that \fratio is mostly 
driven by ${\rm f_*}$, i.e. the lower (higher) ${\rm f_*}$, the lower (higher) \fratio.

We can use these findings to study \fratio as a function of ${\rm T_{vir}}$ and gain physical insights about the 
environment of formation of massive first stars. This is illustrated in Fig.~\ref{Fig:free_par} (right), which also 
shows the correspondence\footnote{The two values of $M$ refer to different choices of the mean molecular 
weight $\mu$, se eq.~{\ref{eq:M_Tvir}}.} between ${\rm T_{vir}}$ and ${\rm M}$ at $z=10$. From this Figure 
we infer that ${\rm <f_*/f_{dil}>\approx 0.0007}$ in the smallest mini-haloes, ${\rm T_{vir}\approx 6\times 10^{3}}$, 
which are likely associated to ultra-faint dwarf galaxies 
\citep{salvadori09, bovill09, frebel14, wise14, blandhawthorn15,salvadori15}. More massive systems, 
which possibly become more luminous ``classical'' dwarf galaxies \citep[e.g.][]{salvadori15} have larger 
${\rm <f_*/f_{dil}>\approx 0.007}$. Finally, we obtain that ${\rm <f_*/f_{dil}>\approx 0.03}$ in the most massive 
and rare haloes, ${\rm T_{vir}\approx 6\times 10^{4}}$, which likely associated to the bulges of galaxies that form 
via merging of high-sigma density peaks \citep[e.g.][]{salvadori10}. 

We recall that \fratio only determines the most likely [Fe/H] range of PISN descendants (eq.~\ref{eq:FeH}) 
while [X/Fe] is {\it not} a function of \fratio (eq.~\ref{eq:ratios}). 
To have an estimate of the typical [Fe/H] for different \fratio we can use Fig.~\ref{Fig:FeH_PISN}, 
since we know that the subsequent Pop~II contribution does only alter the lowest [Fe/H]-tail of 
the metallicity distribution function of PISN descendants Fig.~\ref{Fig:FeH_popII}. 
Hence, we infer that in classical dwarf spheroidal galaxies the PISN descendants can be most likely found 
at ${\rm [Fe/H]\approx -1.5}$ while in the bulges of more massive systems at ${\rm [Fe/H]\approx
 -0.5}$. Interestingly, these values are similar to the ${\rm [Fe/H]}$ range of sub-solar ${\rm [Zn/Fe]}$ 
measurements respectively reported in the classical dwarf galaxy Sculptor and in the Milky Way 
bulge (see references in Skuladottir et al. 2018 and their Fig.~4 for a complete view of ${\rm [Zn/Fe]}$
across different environments). 

If the stellar halos of galaxies are the result of accreting satellites with different masses, then the 
corresponding \fratio might span a very wide range of values (Fig.2). As a consequence, to catch 
the PISN descendants in these environments we should look for stars around the most likely [Fe/H] 
range among {\it all} different models, i.e. ${\rm -2.5 < [Fe/H] < -1.5}$ (Fig.7), as discussed in our 
observational strategy (Sec. 6.2.2).
\section{Conclusions}
We presented a novel method to determine the key element ratios we should look for to identify 
the chemical signature of very massive first stars exploding as PISN. 
Independent on the unknowns related to early cosmic star-formation, metal diffusion and mixing, 
when the mass fraction of heavy elements from PISN represents $\geq 50\%$ of the total, our 
parametric study shows that there is always an under-abundance of the key element ratios
${\rm [(N, Cu, Zn)/Fe] < 0}$. In fact, N, Cu, and Zn are so scarcely produced by PISN that 
even when subsequent generations of Pop~II stars do contribute to the enrichment their 
under-abundance with respect to Fe is preserved. Note that if primordial PISN were fast 
rotators the (under-)production of Cu and Zn would have not been altered, while the internal 
mixing might have induced efficient N production (Takahashi et al. 2018). Furthermore,
due to the activation of the CNO cycle, the N abundance can largely increase in bright 
Red Giant Branch (RGB) stars. With current telescopes, RGB stars represent the majority 
of stellar targets for high-resolution spectroscopic follow-up in nearby dwarf galaxies. 
In conclusion, we can assert that the descendants of PISN, i.e. long-lived stars born in a 
PISN-imprinted ISM, should have ${\rm [(Cu,Zn)/Fe] < 0}$.

Our model results also show that PISN descendants can cover a broad [Fe/H] range across 
the overall parameter space but on average they most likely appear at [Fe/H]$\approx -2$. 
These findings nicely agree with previous studies that employed a completely different 
(cosmological) approach \citep{salvadori07,karlsson08,DB17}, thus strengthening our method
and approximations. At these high [Fe/H], SNIa might have also played a role in the chemical 
evolution. Yet, in our work we demonstrated that an ISM polluted by Pop~II and SNIa typically 
have larger ${\rm [(Cu,Zn)/Fe]}$ ratio than ISM also imprinted by PISN. Furthermore, the 
chemical abundance patterns are quite different, thus allowing us to check and account for 
possible degeneracies.

Among literature stars at [Fe/H]$\approx -2$ we pinpointed BD +80$^\circ$ 245 that has sub-solar 
${\rm [(Cu,Zn)/Fe]}$. Our re-analysis of the new and available spectra for BD +80$^\circ$ 245, 
including NLTE corrections for Cu, demonstrate that this star might be a convincing example of 
a PISN descendants. In fact, the three killing element ratios are sub-solar and its (almost) 
complete chemical abundance pattern agrees with our model results for \fratio$= 10^{-3}$ 
and an ISM equally polluted by PISN and normal Pop~II stars with ${\rm m_{popII}>8\Msun}$.
These core-collapse SN typically evolve on a $\approx 30$~Myr timescale, which implies that 
there is likely a delay between Pop~III and the subsequent star-formation, in agreement with
cosmological simulations of the first star-forming galaxies \citep[e.g.][]{bromm03,wise14,jeon14}.
This positive finding suggests that to search for the rare PISN descendants by performing Zn 
surveys with the HERMES and WEAVE spectrographs is not only feasible but extremely promising. 

It is interesting to note that Hartwig et al. 2019, although not investigating the imprint by PISN, 
found that N, Cu, and Zn are some of the most important elements to discriminate [Fe/H]$<-3$ stars 
that are enriched by a single (mono) or multiple first stars with masses $\leq 140 \Msun$. 
The concordance of these findings by two independent studies aimed at addressing very different 
questions related to primordial stars, strengthen the observational need for additional measurements 
of these key chemical elements.

Our parametric study is extremely general and can therefore be applied to different ``systems" that 
retain the chemical signature of PISN. Not only present-day stars dwelling in different environments 
(dwarf galaxies, stellar halos, bulges) but also high-redshift gas in/and around galaxies that can be
observed as Damped Ly$\alpha$ systems along with the Circum- and Inter- Galactic Medium (CGM, 
IGM). Cosmological simulations of metal-enrichment from the first galaxies have shown that very 
massive first stars exploding as PISN produce much larger metal-enriched bubbles than their lower 
mass companions, and the differences in the IGM morphology are quite remarkable even at $z\approx 4$ 
(Pallottini et al. 2014). Looking for the lack of our key elements in the CGM of high-z galaxies can 
thus be a complementary way to probe the existence of PISN and to understand their persistence
across cosmic times (e.g. Mebane et al. 2017). As for the case of BD +80$^\circ$ 245, once identified 
a system that lacks our killing element ratios, high resolution spectroscopic follow-up (e.g. with HIRES 
on ELT) can fully characterise its abundance pattern. Our work showed that statistical comparison 
between the observed patterns and those predicted by our models is able to reveal the most likely 
mass of the Pop III/II pollutants (Sec.~5.2) and their environment of formation (Sec.~7).

We should conclude by mentioning that several studies have recently focused on the key chemical 
abundance ratios of ``second-generation'' stars, i.e. stars formed out from an environment solely 
imprinted by the first stars (e.g. Hartwig et al. 2018 and Takahashi et al. 2018b for the PISN). 
Although our results for the ``direct descendants" of PISN, i.e. for an ISM imprinted by PISN at 
$100\%$ level, are consistent with these works, we should underline that our simple {\it parametric 
approach} is completely different in both methods and aim. Our goal was in fact to develop new
strategy to identify {\it all descendants} of very massive first stars. These stars, in fact, are clearly 
more common than second-generation stars solely imprinted by PISN. Furthermore, once 
interpreted with cosmological chemical evolution models for the Local Group formation, they can 
provide key information on both the first stars properties and early cosmic star-formation \citep{salvadori15,DB17}. 
By building-up a sample of PISN descendants, therefore, we can not only probe the existence of 
very massive first stars but also provide tight empirical constrain on their unknown initial mass function.
\section*{Acknowledgements}
We thank the anonymous referee for his/her careful reading of the paper and for his/her very positive, 
constructive, and deep comments. SS warmly thank Andrea Ferrara for the incisive expression 
``killing elements'' and acknowledges the Italian Ministry for Education, University, and Research 
through a Levi Montalcini Fellowship. This project has received funding from the European Research 
Council (ERC) under the European UnionÕs Horizon 2020 research and innovation programme 
(grant agreement No 804240).
\' A.S. acknowledges funds from the Alexander von Humboldt Foundation in the framework of the Sofja 
Kovalevskaja Award endowed by the Federal Ministry of Education and Research. 
We are grateful at the Programme National de Cosmologie et Galaxies of CNRS that granted 
us observation time with SOPHIE, under programme  17B.PNCG.BONI.
This research has made use of the Keck Observatory Archive (KOA), 
which is operated by the W. M. Keck Observatory and the NASA Exoplanet Science Institute (NExScI), 
under contract with the National Aeronautics and Space Administration.
The Keck-HIRES observations were acquired during program G401H, PI Mel\'endez. 
\bibliographystyle{mn2e}
\bibliography{salvadori} 
\section{appendix A}
\label{App:A}
\begin{figure*}
  \centering
 \includegraphics[width=0.75\linewidth,clip=,trim={2.0cm 5.0cm 6.5cm 0cm}]{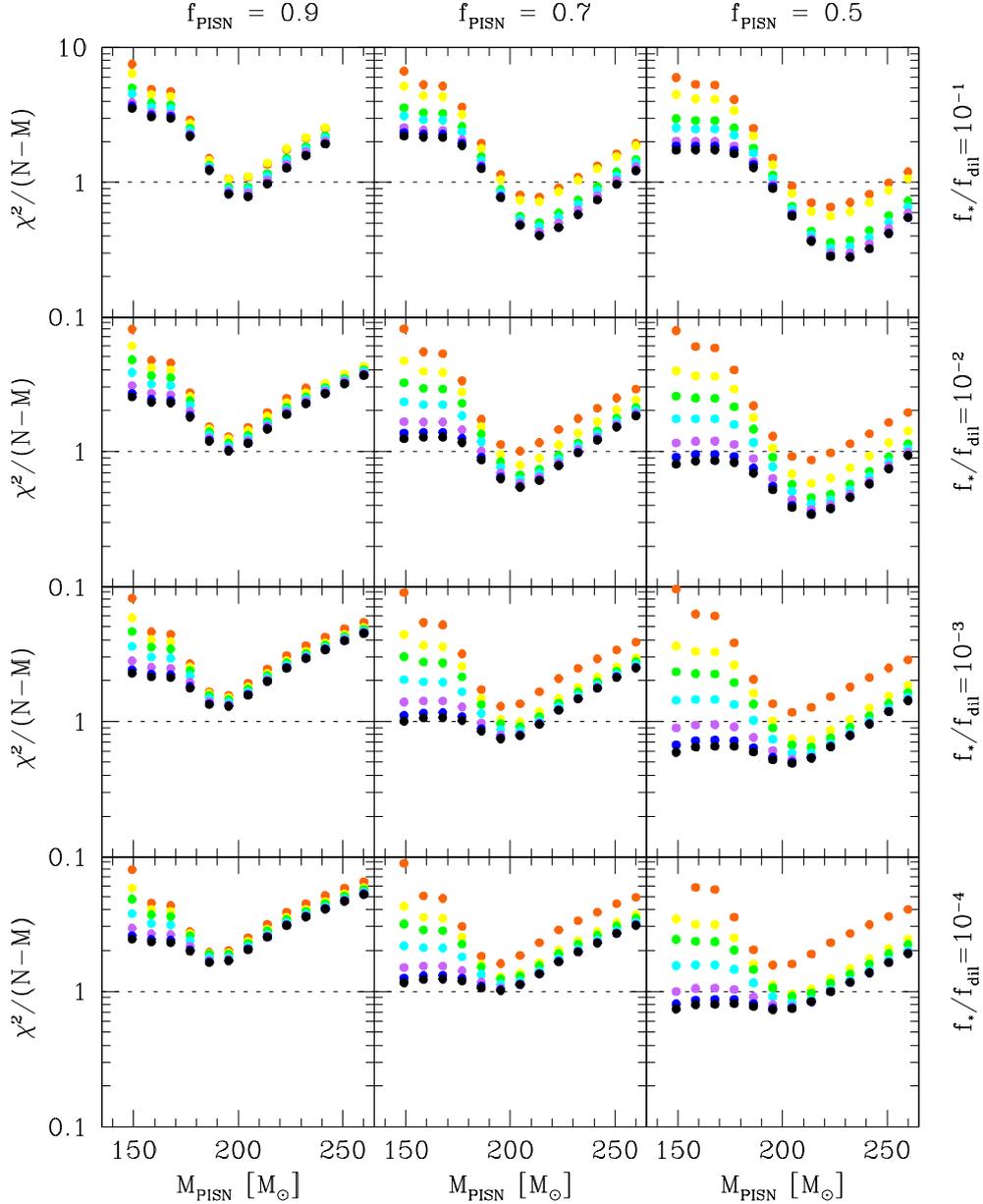}
  \caption{The ${\chi}^2$ distribution normalized to the degree of freedom as a function of the PISN 
  mass and for different combinations of the other free parameters. The panels illustrate how the 
  results change with ${\rm f_{PISN}}$ ({\it columns}) and \fratio ({\it rows}), while different colours 
  in each panel show the time passed since the Pop~II formation, from {\it red} (${\rm t_{popII} 
  \approx 3}$~Myr, uppermost symbols) to {\it black} (${\rm t_{popII} \approx 30}$~Myr, 
  lower-most symbols).}
   \label{Fig:Chi_square}
\end{figure*}
In this Appendix we present the results our ${{\chi}^2}$ analysis to determine the best-fitting model 
of BD +80$^\circ$ 245 across the whole range of model free parameters. These findings are reported
in Fig.~\ref{Fig:Chi_square}, where ${{\chi}^2}_{\nu}$ is shown as a function of ${\rm m_{PISN}}$ (x-axis) 
\fratio (different rows),  ${\rm f_{PISN}}$ (different columns), and ${\rm t_{popII}}$ (different colours). 
We see that for ${\rm f_{PISN} = 0.9}$ (left column) only ${\rm m_{PISN}\approx 200\Msun}$ models 
give a relatively good fit of BD +80$^\circ$ 245, i.e. ${\chi^2}_{\nu} < 2$, a result that is almost 
independent on ${\rm t_{popII}}$. We can understand why this is the case by looking at the left panel
of Fig.~4 where we see that [Fe/H]${\rm_{ISM}\approx -2}$ only for ${\rm m_{PISN}\approx 200\Msun}$, 
and independent of ${\rm t_{popII}}$. In the right panel of Fig.~4, for ${\rm f_{PISN}}=0.5$, we see that 
all models with ${\rm m_{PISN}\leq 200\Msun}$ and ${\rm t_{popII} > 20}~Myr$ give [Fe/H]${\rm_{ISM}\approx -2}$. 
This is because the iron-pollution of the ISM is mainly driven by normal core-collapse SNe, which provide 
enough iron to reach the observed value only if $> 20$~Myr (see Fig.~\ref{Fig:Yx_popII}). 
For these reasons, Fig.~\ref{Fig:Chi_square} shows that the larger the contribution from Pop~II stars, 
the higher the number of models with ${\rm t_{popII}= 30}$~Myr (black points) that provide 
${\chi^2}_{\nu} < 2$. In particular, we find that for ${\rm f_{PISN}}=0.5$
and \fratio$=10^{-4}$ (right-most and lower-most panel) all models with ${\rm t_{popII}= 30}$~Myr
give ${\chi^2}_{\nu} < 2$. These are our best-fitting models shown in Fig.~\ref{Fig:data_predictions}.
Following Ishigaki et al. (2019) we checked that for all of them the probability of the $\chi^2$ distribution 
for ${\rm (N-M)}$ degrees of freedom between ${\rm x^2=\chi^2}$ and infinity provide the largest values.  
Among all these best-fitting models, we find that the one that reproduces the abundance pattern of 
BD +80$^\circ$ 245 at best corresponds to ${\rm m_{PISN}\leq 223\Msun}$, which gives ${\chi^2}_{\nu} =0.99$.
\section{Appendix B}
\label{App:B}
In this Appendix we provide all the tables that illustrate what all the lines used to determine the different 
chemical elements reported in Table 1.
\begin{table}
\centering
\caption{OH lines used for the determination of the oxygen abundance.}
\label{OHlines}
\begin{tabular}{ccrrr} 
\hline
$\lambda$ & log\,$gf$ &$J_{low}$ & $J_{up}$& $E_{low}$ \\
  nm      &           &        &         &cm$^{-1}$  \\
\hline
307.43704&-2.581& 3.5& 4.5&  429.458\\   
308.12544&-1.902&17.5&18.5& 6148.942\\   	
308.32790&-2.066& 4.5& 4.5&  355.900\\   
308.33753&-2.804& 4.5& 3.5&  355.900\\   
308.48950&-1.894&18.5&19.5& 6801.926\\   
308.51993&-1.973& 5.5& 5.5&  544.809\\   
308.53170&-2.865& 5.5& 4.5&  544.809\\   
308.62241&-1.888&20.5&21.5& 7459.105\\   
308.63920&-2.305& 2.5& 1.5&   83.719\\   
308.90082&-1.887&19.5&20.5& 7483.875\\   
308.97356&-1.837& 7.5& 7.5& 1029.092\\   
308.97748&-3.101& 2.5& 3.5&  288.769\\   
308.98130&-3.191& 1.5& 2.5&  187.491\\   
308.98501&-2.328& 2.5& 2.5&  288.769\\   
308.98669&-2.545& 1.5& 1.5&  187.491\\   
308.98958&-2.972& 7.5& 6.5& 1029.092\\   
309.02717&-3.077& 3.5& 4.5&  429.275\\   
309.03683&-2.174& 3.5& 3.5&  429.275\\   
309.04537&-3.462& 0.5& 1.5&  126.291\\   
309.04861&-2.859& 0.5& 0.5&  126.291\\   
309.23976&-1.786& 8.5& 8.5& 1324.291\\   
310.12308&-2.127& 5.5& 4.5&  543.575\\   
310.32691&-2.798& 3.5& 3.5&  429.458\\   
310.33452&-2.354& 3.5& 2.5&  429.458\\   
310.60199&-1.645&12.5&12.5& 2852.485\\   
310.65461&-2.076& 6.5& 5.5&  767.458\\   
310.74582&-2.824& 4.5& 4.5&  608.188\\   
310.75561&-2.257& 4.5& 3.5&  608.188\\   
311.30780&-3.274& 3.5& 2.5&  429.275\\   
311.33656&-1.644&12.5&12.5& 3348.880\\   
311.47735&-1.601&14.5&14.5& 3819.156\\   
312.39487&-1.960& 9.5& 8.5& 1650.790\\   
312.76877&-1.584&15.5&15.5& 4939.902\\   
313.91699&-1.559&17.5&17.5& 6157.854\\   
314.07350&-1.950&10.5& 9.5& 2450.279\\   
315.10035&-1.863&13.5&12.5& 3311.982\\   
316.71693&-1.541&21.5&21.5& 8944.563\\   
317.44843&-1.821&16.5&15.5& 4904.629\\   
317.53019&-1.543&22.5&22.5& 9709.602\\   
319.48479&-1.821&17.5&16.5& 6148.942\\    
320.09601&-1.800&19.5&18.5& 6776.431\\   
325.25948&-1.809&24.5&23.5&10462.394\\   
325.54929&-1.809&23.5&22.5&10484.962\\   
\hline
\end{tabular}
\end{table}

\begin{table}
\centering
\caption{Atomic data for the lines of {Cu}{\sc i}.}
\label{cudata}
\begin{tabular}{ccc}
\hline
$\lambda$& log $gf$  & log $gf$       \\ 
nm    & {\it hfs}&  line          \\ 
\hline                                                                           
324.7511 &   -1.379&     -0.05	    \\ 
324.7513 &   -1.028&     	    \\ 
324.7515 &   -1.379&     	    \\ 
324.7517 &   -0.957&     	    \\ 
324.7520 &   -1.426&     	    \\ 
324.7555 &   -0.288&     	    \\ 
324.7558 &   -1.567&     	    \\ 
\\
327.3927 &   -1.375&     -0.35	   \\ 
327.3929 &   -1.024&     	   \\ 
327.3931 &   -2.074&     	   \\ 
327.3933 &   -1.723&     	   \\ 
327.3971 &   -1.024&     	   \\ 
327.3972 &   -1.375&     	   \\ 
327.3975 &   -1.024&     	   \\ 
327.3976 &   -1.375&     	   \\ 
	 &   	   &               \\ 
510.5504 &   -3.720&     -1.51	   \\ 
510.5513 &   -2.766&     	   \\ 
510.5516 &   -2.813&     	   \\ 
510.5517 &   -3.090&     	   \\ 
510.5521 &   -2.720&     	   \\ 
510.5526 &   -2.398&     	   \\ 
510.5536 &   -2.051&     	   \\ 
510.5563 &   -1.942&     	   \\ 
\\
\hline
\multicolumn{3}{l}{The $gf$ values are from \citet{Warner1968}.}\\
\\
\end{tabular}
\end{table}

\begin{table}
\centering
\caption{Atomic data for the lines of {Zn}{\sc i}.}
\label{zndata}
\begin{tabular}{cccccc}
\hline
$\lambda$& log $gf$  & $J_{low}$ & $J_{up}$ & $E_{low}$ & Ref.   \\ 
  nm     &           &           &          &   cm$^{-1}$ \\
  330.2584&-0.057& 1.0 & 2.0&32501.421& 1\\
  330.2941&-0.534& 1.0 & 1.0&32501.421& 1\\
\\
  481.0528&-0.137& 2.0 &1.0 & 32890.352& 2 \\ 
\hline
\multicolumn{3}{l}{References for $gf$ values:}\\
\multicolumn{3}{l}{1 : \citet{Andersen1973}}\\
\multicolumn{3}{l}{2 : \citet{Kock1968}}\\
\\
\end{tabular}
\end{table}

\begin{table}
\centering
\caption{Atomic  data used to synthesize the  NH 336\,nm band: Ca to Ni.}
\label{nhatoms0}
\begin{tabular}{cccccc}
\hline
$\lambda$& log $gf$  & $J_{low}$ & $J_{up}$ & $E_{low}$   \\ 
  nm     &           &           &          &   cm$^{-1}$ \\
\hline
\multicolumn{2}{l}{Ca {\sc i}}\\
\hline
336.1912 & -0.580& 3.0& 2.0  &     45052.375\\
336.2133 & -1.260& 2.0& 2.0  &     45050.418\\
336.2286 & -2.520& 1.0& 2.0  &     45049.074\\
\hline
\multicolumn{2}{l}{Ti {\sc i}}\\
\hline
335.8271 & -1.190& 2.0 &2.0  &         0.000\\
335.8479 & +0.370& 5.0 &4.0  &     48461.965\\
336.0989 & -1.230& 3.0 &3.0  &       170.134\\
336.1266 & -1.080& 3.0 &3.0  &       170.134\\
336.1831 & -1.460& 3.0 &2.0  &       170.134\\
\hline
\multicolumn{2}{l}{Ti {\sc ii}}\\
\hline
336.0170 & -1.614 &0.5& 1.5 &       9850.900\\
336.1060 & -2.171 &1.5& 0.5 &       9930.690\\
336.1218 & +0.280 &3.5& 4.5 &        225.730\\
\hline
\multicolumn{2}{l}{V {\sc ii}}\\
\hline
336.1496 & -1.200 &4.0 &5.0  &     19112.930 \\
\hline
\multicolumn{2}{l}{Cr {\sc i}}\\
\hline
336.2216 & -0.513 &6.0 &5.0  &     20519.553\\
336.2354 & -1.025 &2.0 &2.0  &     24299.844\\
\hline
\multicolumn{2}{l}{Cr {\sc ii}}\\
\hline
335.8491 & -1.384& 2.5 &1.5   &    19797.881\\
336.0291 & +0.250& 3.5 &3.5   &    25033.699\\
336.1767 & -0.926& 2.5 &3.5   &    25046.760\\
\hline
\multicolumn{2}{l}{Fe {\sc i}}\\
\hline
335.7568 & -1.959& 3.0 &3.0  &     58831.227\\
335.8635 & -2.170& 3.0 &3.0  &     57160.137\\
335.9118 & -2.582& 4.0 &3.0  &     57155.852\\
335.9404 & -2.221& 1.0 &2.0  &     24772.016\\
335.9486 & -3.730& 5.0 &4.0  &      6928.268\\
335.9808 & -1.761& 4.0 &5.0  &     26627.607\\
336.0926 & -2.383& 1.0 &1.0  &     19552.477\\
336.0973 & -3.895& 2.0 &3.0  &     18378.186\\
336.1566 & -3.059& 3.0 &4.0  &     20874.480\\
336.1950 & -1.243& 1.0 &2.0  &     22946.814\\
\hline
\multicolumn{2}{l}{Fe {\sc ii}}\\
\hline
335.7927 & -2.754& 3.5& 2.5 &      33501.254\\
335.8247 & -2.484& 2.5& 3.5 &      31387.947\\
335.8779 & -4.221& 2.5& 2.5 &      13474.411\\
336.0115 & -2.670& 5.5& 5.5 &      32909.906\\
\multicolumn{2}{l}{Co {\sc ii}}\\
\hline
335.8599 & -2.341& 3.0& 2.0 &      17771.711 \\
\hline
\multicolumn{2}{l}{Ni {\sc i}}\\
\hline
335.9101 & +0.019& 5.0& 4.0 &      57829.406\\
336.1237 & -0.282& 4.0& 4.0 &      57810.492\\
336.1551 & -1.390& 2.0& 2.0 &        879.813\\
\\
\hline
\end{tabular}
\end{table}

\begin{table}
\centering
\caption{Atomic  data used to synthetise the  NH 336\,nm band: Y to Er.}
\label{nhatoms1}
\begin{tabular}{cccccc}
\hline
$\lambda$& log $gf$  & $J_{low}$ & $J_{up}$ & $E_{low}$   \\ 
  nm     &           &           &          &   cm$^{-1}$ \\
\hline
\multicolumn{2}{l}{Y {\sc ii}}\\
\hline
336.1986 & -0.230& 2.0& 1.0 &      14832.854\\
\hline
\multicolumn{2}{l}{Zr {\sc ii}}\\
\hline
335.9955 & -0.260& 4.5& 4.5 &      11984.460\\
\hline
\multicolumn{2}{l}{Ce {\sc ii}}\\
\hline
336.0532 & +0.535& 7.5& 7.5 &      11309.972\\
336.1763 & -0.432& 6.5& 6.5 &       3793.634\\
\hline
\multicolumn{2}{l}{Nd {\sc ii}}\\
\hline
335.9766 & -0.260& 8.5& 7.5 &       5085.640\\
\hline
\multicolumn{2}{l}{Gd {\sc ii}}\\
\hline
335.8432 & -0.660& 3.5& 3.5 &       3444.235\\
335.8625 & +0.152& 3.5& 4.5 &        261.841 \\
336.0712 & -0.591& 3.5& 3.5 &        261.841 \\
336.2239 & +0.294& 4.5& 5.5 &        633.273\\
\hline
\multicolumn{2}{l}{Dy {\sc ii}}\\
\hline
335.8602 & -1.359& 8.5& 8.5 &          0.000 \\
335.9462 & +0.567& 8.5& 7.5 &      14895.060 \\
\hline
\multicolumn{2}{l}{Er {\sc ii}}\\
\hline
335.8153 & -0.631& 4.5& 3.5 &       5132.608 \\
\\
\hline
\end{tabular}
\end{table}

\begin{table}
\centering
\caption{Atomic data and hyperfine structure for Sc{\sc ii} and Co{\sc i} used in the synthesis of the NH band}
\label{sc_co}
\begin{tabular}{cccccc}
\hline
$\lambda$& log $gf$ & log($hfs$)  & $J_{low}$ & $J_{up}$ & $E_{low}$  \\ 
  nm     &           &           &        &  &   cm$^{-1}$ \\
\hline 
\multicolumn{2}{l}{Sc{\sc ii}}\\
\hline 
            & -0.770 & \\
   335.9675 &        &-0.626&2.0 & 2.0& 67.720\\
   335.9675 &        &-1.196&2.0 & 2.0& 67.720\\
   335.9677 &        &-1.024&2.0 & 2.0& 67.720\\
   335.9677 &        &-1.038&2.0 & 2.0& 67.720\\
   335.9677 &        &-1.196&2.0 & 2.0& 67.720\\
   335.9679 &        &-1.038&2.0 & 2.0& 67.720\\
   335.9679 &        &-1.049&2.0 & 2.0& 67.720\\
   335.9679 &        &-1.720&2.0 & 2.0& 67.720\\
   335.9680 &        &-1.049&2.0 & 2.0& 67.720\\
   335.9680 &        &-1.222&2.0 & 2.0& 67.720\\
   335.9680 &        &-3.146&2.0 & 2.0& 67.720\\
   335.9681 &        &-1.222&2.0 & 2.0& 67.720\\
   335.9681 &        &-1.398&2.0 & 2.0& 67.720\\
\hline 
            & -0.790\\
   336.1261 &        &-0.594&1.0&1.0&0.000\\
   336.1261 &        &-0.790&1.0&1.0&0.000\\
   336.1266 &        &-0.790&1.0&1.0&0.000\\
   336.1266 &        &-0.794&1.0&1.0&0.000\\
   336.1266 &        &-1.975&1.0&1.0&0.000\\
   336.1269 &        &-0.794&1.0&1.0&0.000\\
   336.1269 &        &-1.049&1.0&1.0&0.000\\
\hline 
   336.1931 &  -0.630&    &1.0& 0.0 & 0.00\\ 
\hline
\hline 
\multicolumn{2}{l}{Co{\sc i}}\\
\hline 
            & 0.127 \\
   336.1487 &  &-1.426&5.5 &4.5&54367.430\\
   336.1494 &  &-1.308&5.5 &4.5&54367.430\\
   336.1494 &  &-1.878&5.5 &4.5&54367.430\\
   336.1503 &  &-1.190&5.5 &4.5&54367.430\\
   336.1503 &  &-1.667&5.5 &4.5&54367.430\\
   336.1503 &  &-2.878&5.5 &4.5&54367.430\\
   336.1516 &  &-1.078&5.5 &4.5&54367.430\\
   336.1516 &  &-1.571&5.5 &4.5&54367.430\\
   336.1516 &  &-2.671&5.5 &4.5&54367.430\\
   336.1533 &  &-0.974&5.5 &4.5&54367.430\\
   336.1533 &  &-1.535&5.5 &4.5&54367.430\\
   336.1533 &  &-2.643&5.5 &4.5&54367.430\\
   336.1552 &  &-0.878&5.5 &4.5&54367.430\\
   336.1552 &  &-1.551&5.5 &4.5&54367.430\\
   336.1552 &  &-2.723&5.5 &4.5&54367.430\\
   336.1575 &  &-0.788&5.5 &4.5&54367.430\\
   336.1575 &  &-1.633&5.5 &4.5&54367.430\\
   336.1575 &  &-2.915&5.5 &4.5&54367.430\\
   336.1601 &  &-0.704&5.5 &4.5&54367.430\\
   336.1601 &  &-1.851&5.5 &4.5&54367.430\\
   336.1601 &  &-3.304&5.5 &4.5&54367.430\\
\\
\hline
\end{tabular}
\end{table}

\begin{table}
\centering
\caption{Molecular data of CH and OH used to synthetise the NH 336\,nm band.}
\label{nhatoms2}
\begin{tabular}{cccccc}
\hline
$\lambda$& log $gf$  & $J_{low}$ & $J_{up}$ & $E_{low}$  \\ 
  nm     &           &           &          &   cm$^{-1}$ \\
\hline
\multicolumn{2}{l}{CH}\\
\hline
336.1977 & -0.682&36.5&37.5 &      21250.174\\
336.1837 & -0.670&37.5&38.5 &      21249.012 \\
\hline
\multicolumn{2}{l}{OH}\\
\hline
335.9900 & -1.401&32.5&31.5 &      17591.371\\
335.9790 & -2.226&18.5&18.5 &      16241.062\\
336.0598 & -1.686&27.5&26.5 &      16029.838\\
336.0075 & -2.094&20.5&19.5 &      14532.071\\
336.0417 & -2.656&13.5&12.5 &      13621.247\\
\\
\hline
\end{tabular}
\end{table}

\begin{table}
\centering
\caption{Molecular  data of NH  used to synthetise the  NH 336\,nm band (first part).}
\label{nh1}
\begin{tabular}{ccccr}
\hline
$\lambda$& log $gf$  & $J_{low}$ & $J_{up}$ & $E_{low}$  \\
  nm     &           &           &          &   cm$^{-1}$ \\
\hline
335.7557 & -1.918& 0.0& 1.0 &         31.574\\
335.7662 & -2.056& 2.0& 1.0 &         32.506\\
335.7736 & -1.302& 3.0& 3.0 &        326.030\\
335.7835 & -2.159& 4.0& 3.0 &        326.909\\
335.8050 & -1.171& 4.0& 4.0 &        488.704\\
335.8053 & -1.590& 1.0& 2.0 &          0.000\\
335.8135 & -2.221& 5.0& 4.0 &        489.460\\
335.8292 & -2.015& 1.0& 2.0 &         97.567\\
335.8300 & -1.072& 5.0& 5.0 &        683.546\\
335.8309 & -1.995& 3.0& 2.0 &         97.718\\
335.8404 & -1.665& 2.0& 3.0 &       3220.365\\
335.8417 & -1.720& 2.0& 2.0 &         98.676\\
335.8503 & -0.992& 6.0& 6.0 &        910.056\\
335.8665 & -2.042& 4.0& 3.0 &        195.513\\
335.8678 & -2.087& 2.0& 3.0 &        195.621\\
335.8684 & -0.925& 7.0& 7.0 &       1168.193\\
335.8783 & -1.418& 3.0& 3.0 &        196.557\\
335.8825 & -1.216& 6.0& 7.0 &       6932.594\\
335.8877 & -2.109& 5.0& 4.0 &        325.748\\
335.8880 & -0.867& 8.0& 8.0 &       1457.516\\
335.8909 & -2.156& 3.0& 4.0 &        326.030\\
335.9008 & -1.240& 4.0& 4.0 &        326.909\\
335.9023 & -2.175& 6.0& 5.0 &        488.267\\
335.9072 & -2.220& 4.0& 5.0 &        488.704\\
335.9077 & -0.816& 9.0& 9.0 &       1777.704\\
335.9158 & -1.117& 5.0& 5.0 &        489.460\\
335.9269 & -0.771&10.0&10.0 &       2128.257\\
335.9285 & -1.024& 6.0& 6.0 &        684.226\\
335.9387 & -1.164& 7.0& 8.0 &       6933.030\\
335.9403 & -0.949& 7.0& 7.0 &        910.637\\
335.9493 & -0.730&11.0&11.0 &       2509.002\\
335.9518 & -0.885& 8.0& 8.0 &       1168.763\\
335.9637 & -0.831& 9.0& 9.0 &       1457.811\\
335.9746 & -0.693&12.0&12.0 &       2919.197\\
335.9776 & -0.783&10.0&10.0 &       1778.187\\
335.9928 & -0.740&11.0&11.0 &       2128.525\\
335.9997 & -0.659&13.0&13.0 &       3358.734\\
336.0042 & -0.820& 9.0& 9.0 &       1167.391\\
336.0046 & -0.872& 8.0& 8.0 &        909.325\\
336.0067 & -0.774&10.0&10.0 &       1456.537\\
\\
\hline
\end{tabular}
\end{table}

\begin{table}
\centering
\caption{Molecular  data of NH  used to synthetise the  NH 336\,nm band (second part).}
\label{nh2}
\begin{tabular}{cccccr}
\hline
$\lambda$& log $gf$  & $J_{low}$ & $J_{up}$ & $E_{low}$  \\ 
  nm     &           &           &          &   cm$^{-1}$ \\
\hline
336.0068 & -1.089& 8.0& 9.0 &       6931.751\\
336.0073 & -0.932& 7.0& 7.0 &        682.958\\
336.0106 & -0.701&12.0&12.0 &       2509.381\\
336.0108 & -0.732&11.0&11.0 &       1776.682\\
336.0165 & -1.003& 6.0& 6.0 &        488.267\\
336.0207 & -0.694&12.0&12.0 &       2127.089\\
336.0275 & -0.627&14.0&14.0 &       3826.364\\
336.0300 & -2.159& 5.0& 6.0 &        489.460\\
336.0308 & -0.666&13.0&13.0 &       2919.347\\
336.0318 & -1.089& 5.0& 5.0 &        325.748\\
336.0330 & -0.660&13.0&13.0 &       2507.714\\
336.0364 & -1.388& 3.0& 4.0 &       3219.414\\
336.0449 & -2.082& 4.0& 5.0 &        326.909\\
336.0483 & -0.628&14.0&14.0 &       2917.983\\
336.0542 & -0.633&14.0&14.0 &       3358.909\\
336.0585 & -1.201& 4.0& 4.0 &        195.513\\
336.0617 & -0.597&15.0&15.0 &       4322.481\\
336.0678 & -0.598&15.0&15.0 &       3357.108\\
336.0703 & -1.992& 3.0& 4.0 &        196.557\\
336.0795 & -0.603&15.0&15.0 &       3826.473\\
336.0901 & -0.570&16.0&16.0 &       3825.097\\
336.0986 & -0.570&16.0&16.0 &       4845.503\\
336.1004 & -1.361& 3.0& 3.0 &         97.718\\
336.1098 & -0.574&16.0&16.0 &       4322.581\\
336.1112 & -1.885& 2.0& 3.0 &         98.676\\
336.1179 & -0.544&17.0&17.0 &       4320.719\\
336.1420 & -0.544&17.0&17.0 &       5395.737\\
336.1426 & -0.548&17.0&17.0 &       4845.563\\
336.1445 & -1.846& 2.0& 2.0 &       3156.788\\
336.1487 & -0.520&18.0&18.0 &       4843.941\\
336.1497 & -1.283& 5.0& 6.0 &       6724.005\\
336.1728 & -1.642& 2.0& 2.0 &         32.506\\
336.1825 & -1.760& 1.0& 2.0 &         33.359\\
336.1833 & -0.497&19.0&19.0 &       5393.767\\
336.1837 & -0.523&18.0&18.0 &       5395.779\\
336.1886 & -0.519&18.0&18.0 &       5971.663\\
336.2137 & -1.224& 6.0& 7.0 &       6724.585\\
336.2275 & -0.474&20.0&20.0 &       5969.806\\
336.2282 & -0.500&19.0&19.0 &       5971.682\\
336.2381 & -0.496&19.0&19.0 &       6573.346\\
\\
\hline
\end{tabular}
\end{table}
\label{lastpage}
\end{document}